\documentclass[11pt, a4paper]{article}
\usepackage{jheparxiv}
\usepackage[utf8]{inputenc}
\usepackage{amsmath}
\usepackage{comment}
\newcommand{\be}{\begin{equation}}
\newcommand{\ee}{\end{equation}}
\usepackage{amsmath,bm}
\usepackage[all]{xy}
\usepackage{tikz-cd}
\usepackage{dsfont}
\usepackage{bm}
\usepackage{amsfonts}
\usepackage{appendix}
\usepackage{amssymb}
\usepackage{latexsym}
\usepackage{float}
\usepackage{mathrsfs}
\usepackage{braket}	
\usepackage{graphicx}
\usepackage{color}
\usepackage{xcolor}
\usepackage{slashed}
\usepackage{mathtools}
\usepackage{sujai}
\usepackage{breqn}
\usepackage[all]{xy}
\usepackage{tikz-feynman} 
\usepackage{dsfont}
\usepackage{cancel}
\usepackage{amsfonts}
\usepackage{amssymb}

\begin{document}
\emergencystretch 3em
\newcommand{\RNum}[1]{\uppercase\expandafter{\romannumeral #1\relax}}
\tikzfeynmanset{compat=1.1.0}
\subheader{\hfill \texttt{}}
\newcommand{\com}[1]{ \textcolor{red}{(#1)}}
\newcommand{\blue}[1]{ \textcolor{blue}{#1}}

\title{Spinning Boundary Correlators from (A)dS$_4$ Twistors}
\author[a]{Mariana Carrillo González,}
\author[a]{and Théo Keseman}
\affiliation[a]{Theoretical Physics Group, Blackett Laboratory \\
Imperial College London, SW7 2AZ, United Kingdom}
\emailAdd{m.carrillo-gonzalez@imperial.ac.uk}
\emailAdd{theo.keseman17@imperial.ac.uk}
\abstract{We develop a twistor-space framework to compute boundary correlators via a boundary limit of nested Penrose transforms in (A)dS$_4$. Starting from correlators of \mbox{(anti-)}self-dual bulk fields, the boundary limit reproduces the correlators of the dual conserved currents; we demonstrate this explicitly for two- and three-point functions. The two-point correlator is rendered finite by working in Euclidean signature. At three points, we obtain compact rational twistor-space representatives obeying a double-copy relation, thereby clarifying the twistor-space origin of the results in \cite{Baumann:2024ttn}. We further extend the analysis to non-conserved currents with integer conformal dimension, dual to massive bulk fields, as well as to the free scalar.}

\maketitle

\newpage
\section{Introduction}

In flat space, many tools have been developed that allow us to perform efficient calculations of scattering amplitudes by using the power of unitarity, causality, and Lorentz invariance \cite{Elvang:2015rqa,deRham:2022hpx,Kruczenski:2022lot}. One such tool is the use of spinor-helicity variables, which have explicit, well-defined little-group transformations and thus serve as perfect building blocks for constructing scattering amplitudes. The use of these variables gives rise to an enormously simple expression for the \(n\)-point maximum-helicity-violating (MHV) scattering amplitudes, which is given by the Parke–Taylor formula \cite{Parke:1986gb}.

Parallel to this effort of moving beyond the cumbersome Feynman-diagram expansion in favor of more compact representations like the Parke–Taylor formula, there was also a drive to understand the hidden geometric and algebraic structures that make such simplicity possible. In this specific case, the remarkable simplicity of the Parke–Taylor expression can be traced to its localization in twistor space: the MHV amplitude has support only on a straight line, the simplest possible holomorphic curve, which fully encodes the kinematic constraints of the process \cite{Witten:2003nn}. Understanding this geometric origin subsequently enabled similar insights and simplifications for more complex helicity configurations, revealing that higher N$^k$MHV amplitudes are localized on higher-degree curves in twistor space and can be systematically constructed from MHV building blocks. While their complexity grows combinatorially—reflecting the number of ways to choose the negative-helicity particles and arrange the higher-degree twistor curves—this remains dramatically simpler than the factorial growth of Feynman diagrams \cite{Cachazo:2004kj,Roiban:2004yf}.

Importantly, the identification of such hidden geometric structures also motivated the development of further simplifications, such as on-shell recursion relations, Mellin-space representations, and other analytic tools that exploit unitarity and factorization properties more directly. For instance, the BCFW recursion relations \cite{Britto:2005fq} provide a remarkably efficient way to construct tree-level amplitudes by shifting external momenta and recursively expressing higher-point amplitudes in terms of lower-point ones. These recursion relations are deeply tied to the analytic structure of amplitudes and their simple-pole behavior, which is already hinted at by their twistor-space localization.

In parallel, Mellin-space techniques emerged as a natural language for conformal-field-theory correlators, particularly in the context of AdS/CFT \cite{Penedones:2010ue}, where they reveal a structure akin to flat-space scattering amplitudes and make factorization and unitarity manifest. Both recursion relations and Mellin amplitudes highlight how the geometric and analytic insights uncovered in the study of flat-space amplitudes can guide the search for similar simplicity and structure in the more intricate setting of curved spacetimes and cosmological correlators.

On the other hand, in curved spacetimes it remains unclear what set of tools is most useful for describing interactions as effectively as in flat space. The observables in this case are correlation functions that, contrary to scattering amplitudes, are field-dependent. Cosmological correlators are notoriously difficult to compute at higher points and loop order \cite{Baumann:2022jpr}, motivating the development of more efficient formalisms to express kinematical data and understand their analytic structure.

In recent years, there has been a large effort to push our understanding of cosmological correlators to the same level as that of scattering amplitudes. Research on cosmological observables now spans analyses of unitarity, analyticity and causality  \cite{Melville:2023kgd,Spradlin:2001nb,Cheung:2022pdk,Grall:2020tqc,Melville:2021lst,Grall:2021xxm,Cespedes:2020xqq,Goodhew:2020hob,Goodhew:2021oqg,Jazayeri:2021fvk,Stefanyszyn:2024msm,Albayrak:2023hie,Goodhew:2024eup,Thavanesan:2025kyc,Benincasa:2020uph,Melville:2024zjq,Serra:2022pzl,Bittermann:2022hhy,Dubovsky:2007ac,Baumgart:2020oby,AguiSalcedo:2023nds,Salcedo:2022aal,Baumann:2021fxj,deRham:2020zyh,CarrilloGonzalez:2023emp,CarrilloGonzalez:2025fqq}. Among the various insights carried over from flat space, soft theorems have found an analogue in cosmological settings, where they have been used to deduce properties of inflationary models \cite{Creminelli:2011mw, Assassi:2012zq, Maldacena:2002vr,Hinterbichler:2012nm,Creminelli:2012ed,Hinterbichler:2013dpa,Berezhiani:2013ewa,Mirbabayi:2014zpa,Avis:2019eav,Jazayeri:2019nbi}.
Inspired by the simplicity of gauge-theory and gravity scattering amplitudes when written using spinor-helicity variables, there have also been proposals for spinor-helicity constructions in de Sitter (dS) and Anti-de Sitter (AdS) \cite{Basile:2024ydc,Maldacena:2011nz,Nagaraj:2018nxq,Nagaraj:2019zmk,Nagaraj:2020sji,Buchbinder:2018nkp,David:2019mos,Binder:2020raz}. 

The recent proposal in \cite{Baumann:2024ttn} introduced a twistor-like\footnote{We refer to the representation in \cite{Baumann:2024ttn} as twistor-like since it considers a \(\mathbb{CP}_3\) while, strictly speaking, the mini-twistor space of flat three-dimensional space is the tangent space to the Riemann sphere, \(\mathrm{T}\mathbb{CP}_1\) \cite{Hitchin:1982gh,Hitchin:1982vry,Jones:1985pla,Ward:1990vs}.} representation for three-dimensional conformal-field-theory (CFT\(_3\)) correlators of conserved currents. Using this representation, it is possible to write embedding-space correlation functions \cite{Costa:2011mg} via nested Penrose transforms of a function of twistor-like variables that encode the scaling dimensions and are explicitly conformally invariant. Here, we will refer to this function as the twistor representative since it is a Čech-cohomology-class representative, as reviewed below. The advantages of this framework are multifaceted. In the traditional embedding-space formalism, the conservation of CFT currents requires solving involved differential equations. By contrast, in the twistor approach, conservation arises naturally as a consequence of demanding that the Penrose transform is well-defined under projective scalings. We show that the twistor formulation naturally aligns with a position space definition of helicity introduced in \cite{Caron-Huot:2021kjy}\footnote{Throughout this work, we use this position-space definition of helicity, rather than the more familiar momentum-space one.}. This is particularly convenient, as it allows the flat-space intuition for helicity to extend straightforwardly to (A)dS, at least for correlators up to three points.

Beyond simplifying such constraints, this framework also seems to be a natural setting for the double copy. The double copy is a powerful idea originating in the study of scattering amplitudes in flat space. It states that amplitudes in gravitational theories can be obtained from those in gauge theory either by squaring colour-ordered amplitudes via a momentum dependent kernel (dubbed the KLT (after Kawai, Lewellen and Tye) double copy) \cite{Kawai:1985xq} or by a systematic replacement of colour structures with kinematic ones, also called BCJ numerators after Bern, Carrasco and Johansson, thanks to the colour-kinematics duality \cite{Bern:2008qj,Bern:2010ue}. In its most famous incarnation, it relates Yang–Mills amplitudes to those of \(\mathcal{N}=0\) supergravity (Einstein gravity coupled to the dilaton and the Kalb–Ramond field), but the duality now extends to a whole web of theories \cite{Adamo:2022dcm,Bern:2022wqg}. The double copy is usually formulated in flat Minkowski spacetime, and its extension to curved spacetimes, such as de Sitter or anti-de Sitter, remains an open and active area of research \cite{carrillogonzalez2018classical,Farnsworth:2023mff,Liang:2023zxo,Alkac:2023glx,Han:2022mze,Prabhu:2020avf,Bahjat-Abbas:2017htu,Ilderton:2024oly,Garcia-Compean:2024uie,Chacon:2024qsq,Chawla:2022ogv}. In the classical context, one way to make progress is through the so-called classical twistor double copy \cite{White:2020sfn,Chacon:2021wbr,Chacon:2021hfe,Chacon:2021lox,Luna:2022dxo,Adamo:2021dfg,Armstrong-Williams:2023ssz,Guevara:2021yud,CarrilloGonzalez:2022ggn}, where one constructs the twistor cohomology-class representative of a gravitational solution from spin-1 and spin-0 twistor representatives. This was was extended to AdS$_3$ in \cite{Beetar:2024ptv}, and the double copy presented here follows the same logic, as we show later.

In this work, we clarify the twistor origin of the representation introduced in \cite{Baumann:2024ttn}. We demonstrate that the twistor-like Penrose transform employed there arises naturally from a boundary limit of nested Penrose transforms in four-dimensional  (A)dS. Upon breaking four-dimensional conformal symmetry down to the (A)dS\(_4\) isometries, we construct correlators involving bulk chiral and anti-chiral fields and take their boundary limits to reconstruct the boundary correlators. Recognising the twistor origin of this description provides a useful perspective on the relationship between solutions in the bulk spacetime and their boundary counterparts, aligning with recent insights from the three-dimensional analysis in \cite{Beetar:2024ptv}. There, a double-copy structure at the level of classical solutions was observed in the context of minitwistor theory, where three-dimensional gauge and gravity solutions are encoded via holomorphic data on minitwistor space. Our results suggest that this classical correspondence admits a natural uplift to the level of correlators. Other related approaches have focused on computing celestial-CFT correlators by viewing four-dimensional Minkowski space as the embedding space of AdS\(_3\) and using traditional AdS\(_3\)/CFT\(_2\) techniques combined with minitwistor-space constructions \cite{Bu:2023cef,Bu:2024cql,Bu:2023vjt,Seet:2024vmh}.

In the same spirit, explicit constructions of bulk-to-boundary propagators and two-point correlators in AdS\(_5\) have also been developed using twistor variables in \cite{Adamo:2016rtr}, where scalar and spinor propagators are realised as cohomology representatives in twistor space. This formalism demonstrates how the twistor space of AdS\(_5\) naturally coincides with the ambitwistor space of its conformal boundary, and how the Penrose transform can be adapted to curved backgrounds to yield boundary two-point functions from bulk twistor data. Taken together, these results illustrate how twistor methods provide a geometrically transparent framework that bridges bulk fields and boundary observables.

In the present work, the use of twistor variables gives rise to complex contour integrals, which are simpler to regulate and evaluate than their real counterpart. This formalism gives rise to twistor space correlators of conserved currents that can be captured by fractions of polynomials directly paralleling the elementary states in classical twistor theory, leading to a tractable and geometrically transparent description. Furthermore, by preserving the  homogeneity of the integrand, we can construct the two-point correlators for fields with general conformal dimension, including those above the unitarity bound. In the process of writing this paper, \cite{Bala:2025qxr} appeared on arXiv which also considers this extension away from conserved currents (see also \cite{Bala:2025gmz}). Their proposal also yields the expected scaling by introducing factors involving the flat space infinity twistor, agreeing with our representative (up to regularisation). As a further demonstration of the framework, we derive a non-trivial Ward–Takahashi identity within our formalism.

The paper is structured as follows. Section~\ref{sec:embeddingspace} reviews the basics of the embedding space formalism. In Section~\ref{sec:ads4twistors}, we derive the embedding-space bispinors used in \cite{Binder:2020raz} from their twistorial origin, establishing our notation. Section~\ref{sec:unitaritybound} presents the computation of bulk-to-bulk, bulk-to-boundary, and boundary-to-boundary propagators in this formalism. In Section~\ref{sec:beyondunitarity}, we extend the method to include non-conserved currents. Section~\ref{sec:3points} applies the construction to three-point boundary correlators, where we illustrate the emergence of the double-copy structure and retrieve the non-regularised and regularised two-point function through the Ward identity. In Appendix~\ref{ap:conventions}, we set out our spinor conventions from six to three dimensions and in Appendix~\ref{ap:bispinors}, we show why the dimensional reduction we performed from (A)dS$_4$ to CFT$_3$ is valid in any signature. In Appendix~\ref{ap:Inm}, we calculate the main integral that is needed for the two-point functions - both for conserved and non-conserved currents. We also relegated the computation of an integral used for several three points functions to Appendix~\ref{ap:generalised Leibniz} and finally show in Appendix~\ref{app:negative helicity} how $\braket{O_1O_2O_3^s}$ can be computed with negative helicity in our formalism. 

\section{Short Review of Embedding Space Formalism}\label{sec:embeddingspace}

Conformal symmetry heavily constrains the form of conformal correlators. The embedding space formalism, provides a transparent way to exploit this fact. The main insight is to embed a $d \geq 3$ dimensional physical space (with coordinates $x_i$) in two dimensions higher (with coordinates $P^I$). While in physical space the conformal generators act non‑linearly, in embedding space they coincide with the linear Lorentz generators of $SO(d+1,1)$. Therefore a Lorentz scalar in embedding space represents a conformally invariant quantity in physical space. We take the $d+2$-dimensional space to be flat and embed the physical space on the projective null cone, i.e. 
\begin{equation}\label{projective null cone}
P^2\,=0 \ , \quad P^I \,\sim r P^I \ .
\end{equation} 
To represent fields in physical space, the corresponding fields in embedding space must live in the tangent space of the cone. 
Irreducible representations of the conformal group are labelled by the conformal dimension $\Delta$ and spin $s$. A primary field is thus a symmetric‑traceless section of the bundle $\mathcal O(-\Delta)$; equivalently, in embedding space it obeys the homogeneity condition
\begin{equation}\label{scaling correlator}
    O_{i,I_1...I_{s}} (r_i P^I)= r_i^{-\Delta_i} O_{i,I_1...I_{s}} (P^I) \ .
\end{equation}
One can also write this expression in terms of spinor variables as in \cite{Binder:2020raz}. Taking $ P^{MN} =\mathbf{\Lambda}^{[M} \mathbf{\Lambda}^{N]}$, where $\mathbf{\Lambda}^{M}$ is a massless spinor in five dimension and $M,N$ are $Sp(4)$ indices, the scaling above reads
\begin{equation}\label{scalingconformaldim}
    O_{i,M_1...M_{2s_i}} (r_i \bold \Lambda^M)= r_i^{-2\Delta_i} O_{M_1...M_{2s_i}} (\bold \Lambda^M) \ .
\end{equation}
To further simplify the construction of correlation functions, we build scalar quantities encoding the correlators. This can be achieved by expressing all correlators through polynomials constructed by contracting indices in Eq. \eqref{scaling correlator} with auxiliary vectors $W^I$ satisfying $W\cdot W = W \cdot P=0$.
The polarised version of Eq. \eqref{scaling correlator} then becomes
\begin{equation}
    O_i(r_i P^I_i, q_i W^I_i)= r_i^{-\Delta} q_i^s O_i (P_i^I, W_i^I) \ .
\end{equation}
Working in physical space the auxiliary vectors correspond to a null $l^i$; hence the polarised spinorial version is constructed by contracting the symmetric traceless field in physical space $f_{a_1...a_{2s}}$ with $2s$ copies of an arbitrary non-zero spinor $l^a$.\\
For 3-points, any parity-even\footnote{The case of parity-odd operators will not be considered here. A twistor-like construction for this was considered in \cite{Bala:2025gmz}.} conformally invariant correlator can be constructed entirely from the following basic invariant structures \cite{Costa:2011mg}
\begin{equation}\label{invariantseven}
    \begin{aligned}
        P_{ij}&\,=-2 P_i \cdot P_j\ ,\\
        H_{ij} &\,=-2((W_i \cdot W_j) (P_i \cdot P_j) -(W_i \cdot P_j)(W_j \cdot P_i))\ ,\\
         V_i &\,= V_{i,jk} = \frac{(W_i \cdot P_j) (P_k \cdot P_i) - (W_i \cdot P_k) (P_j \cdot P_i) }{P_j \cdot P_k}\ .
    \end{aligned}
\end{equation}
 To construct a correlator, one writes the most general scalar built from these objects that respects the scaling of Eq. \eqref{scaling correlator}. The last physical requirement is to set the unitarity bound, which for $d=3$ is $\Delta \geq 1/2$ for scalars and $\Delta \geq s+1$ for spinning fields. In the latter case, the saturation of the bound corresponds to conserved fields. Meanwhile, a scalar with $\Delta=1/2$ corresponds to a free massless scalar, and a scalar with $\Delta=1$ corresponds to a conformally coupled scalar in the alternate quantization. However, conservation not only fixes the scaling dimension, it also constrains the tensor structure. In embedding space, this requires that the correlators satisfy a slightly involved differential equation. As pointed out in \cite{Baumann:2024ttn}, one of the advantages of the twistor formalism is that this differential constraint becomes  automatically satisfied. Additionally, conservation of a spin-$s$ conformal field restricts the helicity to just two values, $h=\pm s$. Dropping the conservation constraint restores the full $2s+1$ helicity spectrum. This is easily seen from the AdS/CFT correspondence: for $s\geq 1$ a conserved boundary current is dual to a massless bulk gauge boson, whereas a non-conserved operator corresponds to a massive bulk field with mass given by\footnote{Taking the Fierz-Pauli mass rather than the Casimir mass which gives the relation for  $m=(\Delta-s)(\Delta+s-d)$, which we take for $s=0, 1/2$ fields.  } $m=(\Delta-(s+d-2)) (\Delta+s-2)$.  While helicity is defined in momentum space, CFT operators with definite helicity can be constructed in coordinate space as shown in \cite{Caron-Huot:2021kjy}. 

With these ingredients, it is now easy to construct the general shape of correlation functions. In the following, we will denote a generic spin $s$ operator by $O^s$, and denote spin 0,1,2 operators by $O,J,T$ respectively. The two-point function is given by 
\begin{equation}\label{2pointgeneral}
    \begin{aligned}
        \braket{O_1^s O_2^s} 
        &\,\propto\frac{H_{12}^s}{P_{12}^{\Delta+s}}\ .
    \end{aligned}
\end{equation}
Additionally, the three-point correlators of mixed spinning and scalars that saturate the unitarity bound are 
\begin{equation}\label{3points ex}
    \begin{aligned}
        \braket{O_1 O_2 O_3^s} 
        &\,\propto \frac{P_{12}^{2s} V_3^s}{(P_{12}P_{23} P_{31})^{(2s+1)/2}}\ ,\\
        \braket{J_1 J_2 O_3 } &\,\propto \frac{V_1 V_2 - H_{12}}{P_{12}^{5/2}P_{23}^{1/2}P_{31}^{1/2}}\ ,\\
    \end{aligned}
\end{equation}
where for $s=1$ in the first line, $J_3$ should either be an Abelian current and the scalars singlets, or $J_3^a$ should be a non-Abelian current and the scalars should be in the fundamental and anti-fundamental representations. In the second line, $J_1$ and $J_2$ should be identical.\\
For equal spin correlators of conserved currents with $s=1$ and $s=2$, the space of possible structures is 2-dimensional and given by
 \begin{equation}
    \begin{aligned}
        \braket{\tilde J_1^{a_1} \tilde J_2^{a_2} \tilde J_3^{a_3}}
        &\, \propto  \sum_{\sigma \in S_2} Tr(T^{a_1} T^{a_{\sigma(2)}} T^{a_{\sigma(3)}}) \frac{V_1 H_{23}+ V_2 H_{31}+  V_3 H_{12}+ V_1 V_2 V_3}{(P_{12}P_{23} P_{31})^{3/2}}\ ,\\
        \braket{\tilde T_1 \tilde T_2 \tilde T_3} &\, \propto \frac{(6V_1^2 H_{23}^2+16 V_2 V_3 H_{31}H_{12}+4 H_{23}V_1^2 V_2 V_3-3V_1^2 V_2^2 V_3^2)+ \text{cyclic}}{(P_{12}P_{23} P_{31})^{5/2}}\ ,
    \end{aligned}
\end{equation}
 or 
\begin{equation}
    \begin{aligned}
        \braket{J_1^{a_1} J_2^{a_2} J_3^{a_3}} &\, \propto \sum_{\sigma \in S_2} Tr(T^{a_1} T^{a_{\sigma(2)}} T^{a_{\sigma(3)}}) \frac{V_1 H_{23}+ V_2 H_{31}+  V_3 H_{12}+5 V_1 V_2 V_3}{(P_{12}P_{23} P_{31})^{3/2}}\ ,\\
        \braket{T_1T_2T_3}&\,  \propto \frac{(-2V_1^2 H_{23}^2+16 V_2 V_3 H_{31}H_{12}+52 H_{23}V_1^2 V_2 V_3+49V_1^2 V_2^2 V_3^2)+ \text{cyclic}}{(P_{12}P_{23} P_{31})^{5/2}}\ ,
    \end{aligned}
\end{equation}
where the spin 1 case above corresponds to coloured-ordered correlators. The first pair encodes the leading bulk Yang–Mills/Einstein interactions, while the second accounts for the subleading bulk $F^3/W^3$ vertices. In this notation, tilded and non-tilded operators distinguish between next-to-leading and leading interactions in the bulk, respectively. More generally, as we will explain in Section \ref{sec:3points}, tilded operators correspond to correlators in which one operator carries opposite helicity to the others, whereas non-tilded operators correspond to correlators where all helicities are the same.\\
Finally, in $d=3$, the structures $H_{ij}$ and $V_i$, are built out of the 6 vectors $P_i$ and $W_i$. However, these live in the (5d) embedding space so there can only be 5 linearly independent vectors. There must be one constraint which in terms of the invariant structures takes the form
\begin{equation}\label{constraint embedding}
     -2H_{12} H_{23} H_{31}= (V_1 H_{23}+V_2 H_{31}+ V_3 H_{12}+ 2 V_1 V_2 V_3)^2\ .
\end{equation}
\section{From (A)dS$_4$ twistors to 3d CFT bispinors}\label{sec:ads4twistors}
Our goal in this section is to show how to obtain the building blocks reviewed in the previous section from a boundary limit of the AdS$_4$ twistor space. We first introduce twistors in the four-dimensional bulk with complex coordinates. Later we shall mainly restrict ourselves to the Lorentzian real slice with $(-,+,+,+)$ signature and an AdS$_4$ bulk spacetime. However, our formalism is consistent with any signature and sign of the curvature as explained below. In what follows, $A,B,..$ will be (dual) twistor indices, i.e. the (anti-)fundamental representation of $SL(4,\mathbb{C})$. $M,N,..$ will refer to the fundamental representation of $Sp(4, \mathbb{C})$, while $\mu, \nu...$ and $i,j,...$ will be tensor indices in four dimensions (4d) and three dimensions (3d) with coordinates denoted $x^{4d}$ and $x$ respectively. Finally, $\alpha, \dot \alpha$ and $a,b$ will denote the massless little group indices for the bulk ($SL(2, \mathbb{C}) \times SL(2, \mathbb{C}))$ and the boundary ($SL(2, \mathbb{C})$), respectively. See Appendix \ref{ap:conventions} for the spinor conventions.

\subsection{(A)dS$_4$ Twistors and Dual Twistors}
The projective twistor space of complexified (A)dS$_4$ is defined through the double fibration
\begin{equation*}
\xymatrix{
 & \mathbb{F} \ar[ld] \ar[rd] & \\
 \mathbb{PT} & & \text{(A)dS}_{4\mathbb{C}}} 
\end{equation*}
together with a choice of infinity twistor $I^{AB}$ which breaks the 4d conformal group. Here, $Z^A$ are homogeneous coordinates on an open subset of $\mathbb{CP}^3$ called the projective twistor space and $\mathbb{F} = \mathbb{CP}^1 \times \text{(A)dS}_{4\mathbb{C}}$ is the correspondence space.  To make the fibration manifest, the twistor coordinates can be written in terms of Weyl spinors as 
\begin{equation}\label{Z Weyl spinors}
Z^A=(\lambda_\alpha,\mu^{\dot \alpha}) \ .
\end{equation}
The incidence relation that gives the correspondence to position space is
\begin{equation}
    \mu^{\dot \alpha} =( x^{4d})^{\dot \alpha  \alpha} \lambda_\alpha \ ,
\end{equation}
and a twistor evaluated on the incidence relation will be denoted by $Z|_X$. Then, the projective  twistor space is the subset $\mathbb{PT} = \mathbb{CP}^3\setminus\mathbb{X_\infty}$ where $\mathbb{X_\infty}= \{ I_{AB}Z_{1}^{[A} Z_{2}^{B]}|_{X} =0\}$, where we take $Z_i=(\lambda_{i,\alpha},\mu^{\dot \alpha}_i)$ such that $\lambda_1$ is not proportional to $ \lambda_2$. Within this setup, the generators $T^A_{\;B}$ of the 4d conformal group ($SO(6,\mathbb{C})$) are linear and holomorphic. They can be explicitly written in terms of twistors as
\begin{equation}\label{4dgenerators}
    T^A_{\;B} = Z^A \frac{\partial}{\partial Z^B} \ .
\end{equation}
The infinity twistor determines where the metric diverges, hence it chooses the structure at infinity. In our case, we take it to satisfy 
\begin{equation}
    I_{AB}I^{AB}=-\frac{1}{4 L^2} \ ,
\end{equation}
which breaks the $SO(6,\mathbb{C})$ symmetries down to $SO(5,\mathbb{C})$. Therefore, the infinity twistor can be used to contract twistors and build scalars that preserve the $SO(5,\mathbb{C})$ symmetry. In order to obtain scalars that are invariant under the full conformal group, one defines the dual twistor
\begin{equation}\label{dual twistors}
     W_A=(\tilde\mu^\alpha,-\tilde\lambda_{\dot\alpha}) \ ,
\end{equation}
which transforms in the anti-fundamental of $SL(4, \mathbb{C})$ (which is not isomorphic to the fundamental one as in the simpler $SL(2, \mathbb{C})$ case). Now, the incidence relation is 
\begin{equation}
    \tilde \mu^{\alpha}= (x^{4d})^{\dot\alpha \alpha}
        \tilde \lambda_{\dot{\alpha}} \ .
\end{equation}
One can construct scalars that preserve the AdS isometries using twistors and dual twistors as follows
\begin{equation}
Z_i \cdot W_j \ , \quad  Z_i \cdot Z_j\, \equiv  Z_i^A I_{AB} Z_j^B \ , \quad W_i \cdot W_j\, \equiv  W_{iA} I^{AB} W_{jB} \ .
\end{equation}
These scalars will be the building blocks for the twistor representatives of the correlation function. We now proceed to construct the spacetime metric. To do so, we consider the embedding of the 4d conformally flat manifold in 6d, which in vector notation is explicitly given by
\begin{equation}\label{YJ}
    Y^J = (x^i,z,\frac{1-x^ix_{i}-z^2}{2},\frac{1+x^ix_{i}+z^2}{2})\ ,
\end{equation}
with signature $(-,+,+,+,+,-)$ and where the constant of proportionality does not matter since $Y^J$ is defined on a projective cone. Here we assumed AdS, but one can easily switch to dS by taking $z\leftrightarrow i t$. In the following, we will consider the corresponding bi-twistors, $Y^{AB}$, satisfying
\begin{equation}\label{null Plucker}
    Y^{AB} Y_{AB}=0 \ .
\end{equation}
These bi-twistors can be written as 
\begin{equation}\label{YAB as twistors}
    Y^{AB}=Z_{1}^{[A} Z_{2}^{B]}|_{X} \ ,
\end{equation}
where again we take $Z_1$ and $Z_2$ to lie on the same twistor line. This expression can be thought of as the massless spinor helicity decomposition of a null vector in 6d \cite{Cheung:2009dc}. In 6d, $Y^{AB}$ is rank 2 rather than rank 4 by Eq. \eqref{null Plucker} (antisymmetric matrices cannot have odd rank), which necessitates the introduction of a second index, the 6d massless little group index. The antisymmetry in Eq. \eqref{YAB as twistors} is then the consequence of their contraction $Y^{AB} \sim \epsilon_{\alpha \beta} p^{A\alpha} p^{B \beta}\sim \epsilon^{\dot \alpha \dot \beta} p^A_{\dot \alpha} p^{B}_{\dot \beta}$, where $\alpha, \dot \alpha$  are chiral and antichiral representation indices of the 4d spinor group $SL(2, \mathbb{C}) \times SL(2, \mathbb{C})$. Geometrically, it describes the image of the Grassmannian $Gr(2,4) $ (the set of planes in 4d passing through the origin) in $\mathbb{CP}^5$. \\
From $Y^{AB}$ and $I_{AB}$, we can construct the metric
\begin{equation}\label{twistormetric}
    ds^2=-\frac{ \epsilon_{ABCD}dY^{AB} dY^{CD}}{(I_{AB}Y^{AB})^2} \ ,
\end{equation}
which is well-defined projectively. $Y^{AB}$ is an antisymmetric $4 \times 4$ matrix, where the homogeneity of $Z^A$ removes a dimension.  We can see that this metric corresponds to AdS$_4$ by choosing coordinates on our spacetime. We will use Poincaré coordinates, $(x^{4d})^\mu= (x^i,z)$, in which case the infinity twistor is given by 
\begin{equation}\label{infinity twistor AdS}
\begin{aligned}
    I_{AB}&\,=\frac{i}{4L}\begin{pmatrix}
        0 & -\delta^{\alpha}_{\;\dot \beta} \\
        \delta_{\dot\alpha}^{\; \beta}&0\\
        \end{pmatrix}
        \ .
\end{aligned}
\end{equation}
This choice will allow us to take a straightforward boundary limit. Using the explicit infinity twistor above together with Eq.~\eqref{YAB as twistors} and applying the incidence relation, we can see that the metric of Eq. \eqref{twistormetric} reduces to
\begin{equation}
    ds^2= \frac{dx^i dx_i + dz^2}{L z^2} \ ,
\end{equation}
where we also used $Y^{AB}I_{AB}= \frac{z}{L}$. 
Finally, for the two-point function, we will also need the flat infinity twistor which we define here as
\begin{equation}
\begin{aligned}
    I^{\text{flat}}_{AB}&\, = \begin{pmatrix}
        \epsilon_{\alpha \beta} &0\\
        0&0\\
    \end{pmatrix}\ .
\end{aligned}
\label{eq:flat_I}
\end{equation}

\subsection{Boundary Limit and Definition of Pseudo-Twistors}
The aim of this section is to recover the 3d pseudo-twistors (to be contrasted from the 3d flat space mini-twistors) used in \cite{Baumann:2024ttn} from the 4d twistors defined in the previous section. Before giving the explicit results, let us comment on the embedding that is taking place at the level of the groups and representations that appear. The embedding is
\begin{equation}\label{embeddingcomplex}
SL(2,\mathbb{C})_\pi \subset_3 SL(2,\mathbb{C})_\lambda \times SL(2,\mathbb{C})_\mu \subset_2 Sp(4,\mathbb{C}) \subset_1 SL(4,\mathbb{C}),
\end{equation}
where $\subset_i$ denotes the order in which the embedding is considered from twistors to the 3d boundary. Here, we consider a complex spacetime and comment on the reality conditions at the end. \\
We start on the right with a (non-projective) twistor $Z^A$,
which transforms in the fundamental representation of $SL(4, \mathbb{C})$. As already mentioned, the first breaking occurs through our choice of infinity twistor. The second embedding $\subset_2$ takes place through the embedding 
\begin{equation}
    \mathbf{4} \xrightarrow{} (\mathbf{2,1})+(\mathbf{1,2}),
\end{equation}
which in practice amounts to writing the twistors using a pair of Weyl spinors as
\begin{equation}
    Z^A=(\lambda_\alpha, \mu^{\dot \alpha}). \label{eq:branching}
\end{equation}
Note that this breaking is still consistent after the first breaking to $Sp(4,\mathbb{C})$ (i.e. $I_{AB} Z^A$ can still be written as a pair of Weyl spinors). Finally, the projection to three dimensions is made explicit by taking $\pi_a = \delta^\alpha_a \lambda_\alpha$ such that the boundary limit of $y^{\dot \alpha \alpha}\lambda_\alpha$ is $x^{a b}\pi_b$. As we shall see, allowing $\pi^a$ to be complex provides a natural regularisation of integrals that would otherwise be divergent.\\
Let us know make the link between our twistors and the bispinors in \cite{Binder:2020raz} for both the bulk and the boundary. In the bulk, this is done by first defining 
\begin{equation}\label{TTbar}
T^M\,=\frac{\delta_A^M Z^A }{\sqrt{z}} \ , \quad \bar T_M\,=\frac{\delta_M^A W_A }{\sqrt{z}} \ ,
\end{equation}
while on the boundary we take 
\begin{equation}\label{pseudotwistors def}
\begin{aligned}
    \mathbf{\Lambda}^M&\,\equiv\lim_{z\to 0} Z^A \delta^M_A \ ,
\end{aligned}
\end{equation}
where $M$ is a fundamental $Sp(4, \mathbb{C})$ index. It is the limit of Eq. \eqref{pseudotwistors def} that we define as pseudo-twistors. While they do not correspond to the usual dimensional reduction to minitwistors, they do realise the conformal group linearly as
\begin{equation}
    T^{3d}_{MN} = \bold{\Lambda}_{(M}\frac{\partial}{\partial \bold{\Lambda}^{N)}} \ ,
\end{equation}
analogously to the 4d case shown in Eq. \eqref{4dgenerators}. Their link with 4d twistors also reveals that they can be decomposed into a pair of Weyl spinors similarly to Eq. \eqref{Z Weyl spinors}, except the indices $\alpha, \dot \alpha$ should be replaced by the single $SL(2, \mathbb{C})$ index $a$. Let us also stress that, contrary to 4d, the pseudo-twistors and dual pseudo-twistors are isomorphic since
\begin{equation}
    \bold \Lambda^M = \Omega^{MN} \bold{\Lambda}_N,
\end{equation}
where $\Omega^{MN}$ is the symplectic form. $\Omega^{MN}$ can be written in terms of the infinity twistor as
\begin{equation}
    \Omega^{MN}= 4i L\delta^M_A I^{AB}\delta_B^N \ .
\end{equation}
Now on the incidence relation, Eq. \eqref{TTbar} and \eqref{pseudotwistors def} decompose to
\begin{equation}\label{eq:TTL_incidecnce}
    \begin{aligned}
        T^M&\,=T^{M\alpha}\lambda_\alpha\ ,\\
        \bar T_M&\,=\bar T_M^{\dot \alpha}\tilde \lambda_{\dot \alpha}\ ,\\
    \mathbf{\Lambda}^M&\,= \pi^a \Lambda^M_a \ ,
    \end{aligned}
\end{equation}
recovering explicitly the bispinors $T^{M\alpha}$, $\bar T_M^{\dot \alpha}$, and $\Lambda^M_a$ of \cite{Binder:2020raz}. For a related construction for $4d$ CFTs, see \cite{Simmons-Duffin:2012juh}. As explained there, $\alpha, \dot \alpha,a$ should be interpreted as 4d/3d tangent space indices, which can also be seen from the twistor incidence relation and the embedding described above.
These bispinors can also be used to reconstruct the 5d embedding space bulk and boundary points 
\begin{align}
 X^{MN}&\,=T^{M \alpha}  T^{ N \alpha} \epsilon_{\alpha \beta}+i\Omega^{MN} \nonumber \\
&\,= \Omega^{M M'} \Omega^{N N'}\bar{T}^{\dot \alpha}_{M'} \bar T^{\dot \beta} _{N'}\epsilon_{\dot \alpha \dot \beta}-i\Omega^{MN}, \label{eq:X_T}\\
 P^{MN}&\,=\Lambda^{Na} \Lambda^M_a, \label{eq:P_Lambda}
\end{align}
or in vector notation in Poincaré coordinates
\begin{align}
        X^I&\,= \frac{1}{z}(x^i,\frac{1-x^ix_{i}-z^2}{2},\frac{1+x^ix_{i}+z^2}{2}) \ ,\\
        P^I&\,= (x^i,\frac{1-x^ix_{i}}{2},\frac{1+x^ix_{i}}{2}) \ ,
\end{align}
where again we considered AdS but it is straightforward to switch to dS. \\
Finally, we should highlight that this construction holds for all choices of reality conditions, as discussed in Appendix~\ref{ap:bispinors}. The real slices and corresponding groups involved in the embedding in Eq.~\eqref{embeddingcomplex} are shown in Table \ref{tab:adsdsequivalence}. 
\begin{table}[h]
  \centering
  \begin{tabular}{@{} l l l l l l l l l @{}}
    \textbf{Bulk} &  $\subset$ & \textbf{3d LLG} & $\subset$ & \textbf{4d LLG } & $\subset$& \textbf{4d Isometry} & $\subset$&  \textbf{Twistor}  \\
    (A)dS$_{4\mathbb{C}}$ &  & $SL(2,\mathbb{C})$ & & $SL(2,\mathbb{C}) \times SL(2,\mathbb{C})$ & & $Sp(4,\mathbb{C})$ & & $SL(4,\mathbb{C})$ \\
    AdS$_{4}$ &  & $SL(2,\mathbb{R})$ & & $SL(2,\mathbb{C})$ & & $Sp(4,\mathbb{R})$ & & $SU(2,2)$ \\
    dS$_{4}$ &  & $SU(2)$ & & $SL(2,\mathbb{C}) $ & & $Sp(2,2,\mathbb{H})$ & & $SU(2,2)$ \\
    EAdS$_{4}$ &  & $SU(2)$ & & $SU(2)_L \times SU(2)_R $ & & $Sp(2,2,\mathbb{H})$ & & $SL(2,\mathbb{H})$ \\
    EdS$_{4}$ &  & $SU(2)$ & & $SU(2)_L \times SU(2)_R $ & & $Sp(4,\mathbb{H})$ & & $SL(2,\mathbb{H})$ \\
  \end{tabular}
  \caption{Summary of isometry groups, massive little groups, and local Lorentz groups for various 4d (A)dS backgrounds and real slices.}
  \label{tab:adsdsequivalence}
\end{table}

\subsection{Scalars and Conformally Invariant Structures}
This subsection rewrites the kinematic data and invariants using little group representations, which simplify manipulations in later sections. It serves as a reference for the key practical identities and expressions employed throughout the remainder of the paper. First note that the bispinors $T^{M \alpha}, \bar{T}^{\dot{\alpha}}_M$ can be contracted to give\footnote{The only difference between $y_{ij}$ and $\tilde y_{ij}$ is the sign in front of $z_j$.}
\begin{subequations}\label{yT}
\begin{align}
        y^{\dot \alpha \alpha}_{ij}&\,\equiv (T_i \cdot \bar T_j)^{\alpha \dot \alpha}= \frac{ (x^{4d})^{\dot \alpha \alpha}_{ij}}{\sqrt{z_i z_j}} \ ,\\
       \tilde y^{\beta \alpha}_{ij}&\,\equiv  (T_i \cdot T_j)^{\alpha \beta}=\frac{1}{\sqrt{z_i z_j}}\begin{pmatrix}
       x_i^0-x_j^0 -(x_i^1-x_j^1)& -i(z_i+z_j)-(x_i^2-x_j^2)\\
      i(z_i+z_j)-(x_i^2-x_j^2)& x_i^0-x_j^0 +(x_i^1-x_j^1)
    \end{pmatrix} \ ,
\end{align}
\end{subequations}
where $(x^{4d})_{ij}=(x^{4d})_i-(x^{4d})_j$ and which are related to the chordal distance $u=((x_{1}-x_{2})^2+(z_1-z_2)^2)/(2z_1 z_2)$ as
\begin{subequations}\label{yytilde}
\begin{align}
        y^{\dot \alpha \beta}_{ij} y_{ij,\dot \alpha \alpha} &\,= -(y_{ij})^2 \delta^\beta_\alpha= -2u \delta^\beta_\alpha \ ,\\
        \tilde y^{ \gamma \beta}_{ij} \tilde y_{ij, \gamma \alpha} &\,= -(\tilde y_{ij})^2 \delta^\beta_\alpha= -2(u+2) \delta^\beta_\alpha \ .
    \end{align}
\end{subequations}
Taking the boundary limits in turn, we have
\begin{subequations}\label{limit1}
    \begin{align}
        w^{\alpha a}_{ij} &\,\equiv \lim_{z_j \xrightarrow[]{}0} \sqrt{z_j} \tilde y_{ij}^{\alpha \beta}\delta_{\beta}^a= \lim_{z_j \xrightarrow[]{}0} \sqrt{z_j}y_{ij}^{\alpha \dot \beta}\delta_{\dot{\beta}}^a,\\
        x_{ij}^{ab} &\,\equiv \lim_{z_i \xrightarrow[]{}0} \sqrt{z_i}  w_{ij}^{\alpha b}\delta_{\alpha}^a= \Lambda^{Ma}_i\Lambda^b_{jM}
    \end{align}
\end{subequations}
where $x_{ij}= x_{i}-x_{j}$. Again
\begin{equation}
    (x_{ij})^{a}_{\;b} (x_{ij})^{b}_{\;c}=  (x_{ij})^2 \delta^a_c \ .
\end{equation}
We can now express Eq. \eqref{invariantseven} in our notation
\begin{equation}\label{structureslittle}
P_{ij}\,= \frac{(x_{ij})^2}{2}\ , \quad H_{ij}\,=-\frac{\braket{l_ix_{ij} l_j}^2}{2} \ , \quad V_i\,= \frac{\braket{l_i \tilde x_{i} l_i}}{2 (x_{jk})^2} \ ,
\end{equation}
where $l^a_i$ are arbitrary polarisation spinors and where 
\begin{equation}\label{xtilde}
   (\tilde x_{i})^a_{\;b} \equiv (x_{ij} x_{jk} x_{ki})^a_{\;b} \ ,
\end{equation}
with $i,j,k$ in cyclic permutations. In order for the correlators to be real, we will require $\sqrt{P_{ij}} $ to be real, that is, $x_i, x_j,x_k$ must be spacelike separated. Here, the constraint of Eq. \eqref{constraint embedding} comes from the fact that not all $x_{ij}$ are independent since $x_{ki}=x_{kj}+x_{ji}$, which can be plugged back in Eq. \eqref{structureslittle} to give Eq. \eqref{constraint embedding}. Finally, note that using Eq. \eqref{2pointgeneral}, the unpolarised two- and three-point functions in spinor notation are completely symmetric in spinor indices. Therefore, we will sometimes assume $l_i=l$ for simplification, as we will see this will be crucial for the non-conserved case.

\section{Propagators of Massless Gauge Fields and Conserved Currents} \label{sec:unitaritybound}
In the following, after briefly summarising the Penrose transform, we show how to obtain the bulk-to-bulk propagators for massless chiral and anti-chiral fields using twistors, the boundary limit of which gives the two-point function when the unitarity bound is saturated, $\Delta=s+1$. 

We will see below that, from an analytic continuation perspective, certain divergent real integrals found in \cite{Baumann:2024ttn} for the two-point functions correspond here to contour integrals around higher-order poles whose residues vanish, and thus evaluate to zero. In the following, we will use the phrase regularization to refer to any prescription that assigns a non-zero and non-divergent value to such integrals. We will see that this is possible by going to Euclidean signature. A similar situation will arise at three-points, where introducing branch cuts with an appropriate choice of contour regularizes the three-point function.

\subsection{Penrose Transform Recap}
The aim of this section is to write the two-point correlator from nested Penrose transforms. For a negative helicity zero-rest-mass spin $s$ field in 4d flat space, the Penrose transform can be written as a contour integral in correspondence space 
\begin{equation}\label{PTlambda}
    \phi_{\alpha_1 ... \alpha_{2s}}^- = \oint D\lambda  \, \lambda_{\alpha_1}... \lambda_{\alpha_{2s}} f^-(\lambda, \mu)|_X \ ,
\end{equation}
where  $D\lambda\equiv \braket{\lambda d\lambda}$ is the standard holomorphic measure on $\mathbb{CP}^1$ and $f^-$ has homogeneity $-2s-2$ in $\lambda$, so that the integral is well-defined projectively. The Penrose transform is an isomorphism between cohomology classes\footnote{We will mostly work with \v{C}ech cohomology classes, but the Dolbeault perspective can also be useful as we will see later. $f(Z)$ is a \v{C}ech cohomology representative (an explicit $p$-cocycle) corresponding to a particular spinning massless free field. In our case, only two charts are needed to cover $\mathbb{CP}^1$, so on the correspondence space the \v{C}ech representatives of $H^1 (\mathbb{CP}^1, \mathcal O (k))$ are locally just given by sections of homogeneity $k$.} $H^1 (\mathbb{PT}, \mathcal O (k))$ and solutions to the zero-rest-mass equation,  $\nabla^{\alpha_1\dot{\alpha}_1}\phi_{\alpha_1 ... \alpha_{2s}}=0$. In the scalar case, the isomorphism is for conformally coupled scalars satisfying $(\nabla^2-R/6)\phi=0$. The same integral can also be written in polarised form by contracting with a chiral 4d polarization spinor $t^\alpha$
\begin{equation}\label{PT-polarised}
    \phi_{2s}^- = \oint D\lambda \,\braket{t \lambda}^{2s} f^-(\lambda, \mu)|_X \ .
\end{equation}
The generalisation to conformally flat spacetimes is obtained by considering the appropriate measure. With a flat infinity twistor $I$, as in Eq.~\eqref{eq:flat_I}, the measure in Eq. \eqref{PT-polarised}, $\braket{\lambda d\lambda}$, can be written as $Z \cdot I \cdot dZ$. In a generic conformally flat spacetime, we simply need to replace the flat infinity twistor with the appropriate one for the curved background, see Section 9 of \cite{Penrose:1986ca}. For the polarizations, we generalize $\braket{t \lambda}$ to $\tau \cdot I \cdot  Z$, where $\tau$ is a polarization-embedding space spinor. Thus, the Penrose transform for generic conformally flat spacetimes in terms of twistors $Z$ is 
\begin{equation}\label{PTZ}
    \phi_s^- = \oint (Z \cdot I \cdot dZ) (\tau \cdot I \cdot  Z)^{2s} f^-(Z)|_X\ ,
\end{equation}
with again the homogeneity requirement $f^-(r Z)= r^{-2s-2} f^-(Z)$. In our case, using the AdS infinity twistor (cf. Eq. \eqref{infinity twistor AdS}), the measure reduces to $z \braket{\lambda d \lambda}$ on the incidence relation, and the polarization-embedding space spinor is
\be
\tau^A = t_{\alpha} T^{M \alpha} \delta^A_M \ . \label{eq:polar_embed}
\ee
Later, we will also need the conjugate polarization spinors given by 
\be
\bar \tau^A = t^{\dot{\alpha}} \bar T^{M}_{\dot \alpha} \delta^A_M \ . 
\ee
When the helicity is flipped, the flat space Penrose transform is given by
\begin{equation}\label{PTlambda+}
    \phi_{\dot \alpha_1 ... \dot \alpha_{2s}}^+ = \oint D\lambda  \;\frac{\partial}{\partial \mu^{\dot \alpha_1}}... \frac{\partial}{\partial \mu^{\dot \alpha_{2s}}} f^+(\lambda, \mu)|_X\ .
\end{equation}
where $f^+$ should now be homogeneous with degree $2s-2$ and the helicity (positive or negative) is fixed by the presence of $\lambda_\alpha$ or $\frac{\partial}{\partial \mu^{\dot \alpha}}$ factors in the transform.
One of the key insights that enabled \cite{Baumann:2024ttn} to formulate a twistor-based interpretation of conformal correlators was the observation that the homogeneity in twistor space aligns with the 3d conformal dimension of conserved negative-helicity fields. For positive helicity, however, this correspondence breaks down, and the correct scaling is no longer manifest, unless the Penrose transform is expressed using the dual twistors of Eq. \eqref{dual twistors}, i.e.
\begin{equation}\label{PTtildelambda}
    \phi_{\dot \alpha_1 ... \dot \alpha_{2s}}^+ = \oint D\tilde \lambda \;\tilde \lambda_{\dot \alpha_1}... \tilde \lambda_{\dot \alpha_{2s}} \tilde f^+(\tilde \lambda, \tilde \mu)|_X\ ,
\end{equation}
with $D\tilde \lambda \equiv [\tilde \lambda d \tilde \lambda]$ and $\tilde f^+(r \tilde \lambda)= r^{-2s-2} \tilde f^+(\tilde \lambda)$. This shows that to work with chiral and anti-chiral fields dual to conserved currents, we should work with both twistors and dual twistors.

\subsection{Bulk-to-Bulk Propagator}
As we illustrate in more detail later, the results of \cite{Baumann:2024ttn}, which we reproduce here, correspond to definite helicity states in the $3d$ CFT. To obtain the bulk twistor origin of these correlators, it is therefore natural to turn to the propagators of (anti-)self-dual fields whose boundary limits are in definite helicity states. These are written locally for bosons as 
\begin{equation}
    \begin{aligned}
        \mathcal{F}_{\alpha_1...\alpha_{2s}} &\,= \nabla_{(\alpha_{s+1}}^{\;\;\dot \alpha_1}... \nabla_{\alpha_{2s}}^{\;\;\dot \alpha_{s}}\varphi_{\alpha_1... \alpha_s) \dot\alpha_1... \dot\alpha_{s}}\ ,\\
        \mathcal{\bar F}_{\dot \alpha_1...\dot \alpha_{2s}} &\,= \nabla_{\;\;(\dot\alpha_{s+1}}^{ \alpha_1}... \nabla_{\;\;\alpha_{2s}}^{\dot \alpha_{s}}\varphi_{|\alpha_1... \alpha_s| \dot\alpha_1... \dot\alpha_{s})} \ ,\\
    \end{aligned}
\end{equation}
and for fermions as
\begin{equation}
    \begin{aligned}
        \mathcal{F}_{\alpha_1...\alpha_{2s}} &\,= \nabla_{(\alpha_{s+1/2}}^{\;\;\dot \alpha_1}... \nabla_{\alpha_{2s}}^{\;\;\dot \alpha_{s-1/2}}\varphi_{\alpha_1... \alpha_{s+1/2}) \dot\alpha_1... \dot\alpha_{s-1/2}}\ ,\\
        \mathcal{\bar F}_{\dot \alpha_1...\dot \alpha_{2s}} &\,= \nabla_{\;\;(\dot\alpha_{s+1/2}}^{ \alpha_1}... \nabla_{\;\;\alpha_{2s}}^{\dot \alpha_{s-1/2}}\varphi_{|\alpha_1... \alpha_{s-1/2}| \dot\alpha_1... \dot\alpha_{s+1/2})}\ ,\\
    \end{aligned}
\end{equation}
where $\varphi$ is a spin $s$ gauge field and for $s=1,2$,  $\mathcal{ F}/ \mathcal{ \bar F}$ are the (anti-)self-dual field strengths and Weyl curvatures respectively. We will also refer to the anti-self-dual and self-dual fields as chiral and anti-chiral fields, respectively. Their linearized equations of motion are given by the zero-rest-mass equation.\\
As usual, it will be easier to work with the polarised versions
\begin{equation}
    \begin{aligned}
        \mathcal{F}^s&\,= t^{\alpha_1}...t^{\alpha_{2s}} \mathcal{F}_{\alpha_1...\alpha_{2s}}\ ,\\
        \mathcal{\bar{F}}^s&\,= \bar t^{\dot \alpha_1}...\bar t^{\dot \alpha_{2s}} \mathcal{\bar F}_{\dot \alpha_1...\dot \alpha_{2s}}\ .\\
    \end{aligned}
\end{equation}
To construct the twistor space correlator, we assume that the scaling of Eq. \eqref{scalingconformaldim} is manifest at the level of the integrand and that twistors and dual twistors are associated with the anti-self-dual and self-dual curvatures, respectively. For concreteness, let us focus on the chiral/anti-chiral case, that is, built out of $Z_1$ and $W_2$. Together with homogeneity, this constrains the form of the (polarised) correlator to be 
\begin{equation}\label{chiral/antichiralnaive}
    \braket{\mathcal{F}^s_{1}(X_1) \bar{\mathcal{F}}^s_{2}(X_2)} \sim \oint\oint  (Z_1 \cdot I \cdot dZ_1) (W_2 \cdot I \cdot dW_2)  \frac{(\tau_1 \cdot I \cdot Z_1)^{2s} (\bar \tau_2 \cdot I\cdot W_2)^{2s}}{(Z_1 \cdot W_2)^{2s+2}}\Bigg|_{X_1,X_2}\, .
\end{equation}
However, this integral vanishes by virtue of the residue theorem\footnote{One could alternatively attempt to evaluate this integral in 4d split signature, that is,  using real twistors on the real line, but this leads to a divergent answer.}. The task is therefore to identify a suitable deformation of the integrand that yields the correct, non-trivial result. To obtain a non-vanishing contribution, the order of the pole must be reduced. Achieving this reduction appears to require the introduction of additional variables, which runs counter to the aim of working entirely within the twistor framework. Nevertheless, in Euclidean signature, one can naturally regularise this integral (i.e., make it non-zero from our point of view). Indeed, now we may exploit the Euclidean conjugates of the twistor variables to factorize the denominator into two terms. The Euclidean conjugation is defined as follows:
\begin{equation}\label{Euclidean Conjugation twistor}
\begin{aligned}
    Z^A&\,=(\lambda_\alpha,\mu^{\dot{\alpha}}) \xrightarrow{} \hat Z^A=(\hat \lambda_\alpha,\hat \mu^{\dot{\alpha}})\ ,\\
\end{aligned}
\end{equation}
where 
\begin{equation}\label{euclidean conjugation}
    \begin{aligned}
        \hat\lambda^\alpha&\,=(-\bar \lambda^1,\bar \lambda^0)\ ,\\
        \hat \mu^{\dot \alpha}&\,=(-\bar \mu^{\dot 1},\bar \mu^{\dot 0})\ .
    \end{aligned}
\end{equation}
For the rest of the two-point calculation, we take $\lambda_\alpha$ in the fundamental representation of $SU(2)$ rather than $SL(2, \mathbb{C})$, as suited for Euclidean signature. With this last step, the simplest way to account for the homogeneity of $\lambda_1, \hat\lambda_1, \tilde \lambda_2$ is to split the pole into a simple pole and a pole of order $2s+1$ such that
\begin{equation}\label{chiral/antichiral}
\begin{aligned}
    \braket{\mathcal{F}^s_{1}(X_1) \bar{\mathcal{F}}^s_{2}(X_2)} =C_{\Delta,s} \int\limits_{{\mathbb{CP}^1_{Z_1}}} \oint &\,\frac{(Z_1 \cdot I \cdot dZ_1) \wedge (\hat Z_1 \cdot I^{\text{flat}} \cdot d\hat Z_1)}{(Z_1 \cdot I^{\text{flat}} \cdot \hat{Z}_1) }(W_2 \cdot I \cdot dW_2)  \\
    &\, \times \ \frac{(\tau_1 \cdot I \cdot Z_1)^{2s} (\bar \tau_2 \cdot I \cdot W_2)^{2s}}{(\hat Z_1 \cdot W_2)(Z_1 \cdot W_2)^{2s+1}}\Bigg|_{X_1,X_2}\ ,
\end{aligned}
\end{equation}
where $C_{\Delta,s}$ is a constant which will play no role in the following. This can also be written more compactly as 
\begin{equation}\label{bulk bulk little group}
    \begin{aligned}
        \boxed{\braket{\mathcal{F}^s_{1}(X_1) \bar{\mathcal{F}}^s_{2}(X_2)} =C_{\Delta,s} (z_1 z_2)^{s+1} \!\!\int\limits_{{\mathbb{CP}^1_{Z_1}}} \!\!\oint \frac{(D\lambda_1 \wedge D\hat\lambda_1)D \tilde \lambda_2}{\braket{\lambda_1 \hat{\lambda}_1}} \frac{\braket{t_1\lambda_{1}}^{2s}[\bar t_2 \tilde \lambda_2]^{2s} }{(\hat Z_1 \cdot W_2) (Z_1 \cdot W_2)^{2s+1}}\Bigg|_{X_1,X_2}}\ .
    \end{aligned}
\end{equation}
This can be seen by using Eq.~\eqref{eq:polar_embed} and \eqref{TTbar}, which implies that $\tau_1 \cdot I \cdot Z_1 \propto z_1 \braket{t_1 \lambda_1}/\sqrt{z_1}$ (similarly $\bar \tau_2 \cdot I \cdot W_2\propto \sqrt{z_2} [t_2 \tilde \lambda_2]$) and by plugging the expression for the infinity twistors on the incidence relation such that 
\begin{equation}
    \begin{aligned}
        (Z_1 \cdot I \cdot dZ_1)&\,\propto z_1 \braket{\lambda_1d \lambda_1}\ ,\\
        (W_2 \cdot I \cdot dW_2) &\,\propto z_2 [\tilde \lambda_2d \tilde \lambda_2]\ ,\\
        \frac{ (\hat Z_1 \cdot I^{\text{flat}} \cdot d\hat Z_1)}{(Z_1 \cdot I^{\text{flat}} \cdot \hat{Z}_1) } &\,\propto \frac{\braket{\hat \lambda_1 d \hat \lambda_1}}{\braket{ \lambda_1 \hat \lambda_1}} \ .
    \end{aligned}
\end{equation}
Our reasoning led us naturally to a mixed \v{C}ech-Dolbeault representative where the integral is a surface integral with respect to the twistor $Z_1$ and a contour integral with respect to $W_2$. This is the form in which the correlator will be computed; however, our nested Penrose transform can also be recast in a purely Dolbeault form as 
\begin{equation}\label{dolbeault}
    \boxed{\begin{aligned}
        \braket{\mathcal{F}^s_{1}(X_1) \bar{\mathcal{F}}^s_{2}(X_2)} =&\,C_{\Delta,s}  (z_1 z_2)^{s+1} \!\!\!\!\!\!\!\!\int\limits_{\mathbb{CP}^1_{Z_1}\times \mathbb{CP}^1_{W_2}} \!\!\!\!\!\!\!\!D\lambda_1 D\tilde\lambda_2  \ \braket{t_1 \lambda_1}^{2s} [t_2 \tilde \lambda_2]^{2s} f|_{X_1,X_2} 
    \end{aligned}}\ ,
\end{equation}
where
\begin{align}
    f=\frac{ \bar\delta_2((\hat Z_1\cdot W_2) (Z_1 \cdot W_2)^{2s+1})}{\braket{\lambda_1 \hat \lambda_1} } \wedge \braket{\hat\lambda_1d \hat\lambda_1}+ \bar\delta_2((\hat Z_1\cdot W_2) (Z_1 \cdot W_2)^{2s+2} )\wedge \langle\hat\lambda_1 dx_1 \tilde\lambda_2]\label{eq:Dolbeault_rep}
\end{align}
with $f\in H^{0,2}(\mathbb{PT}_{1}\times \mathbb{PT}_{2}^{\vee},\mathcal{O}(-2s-2) \otimes \mathcal{O}(-2s-2))$ and $\bar\delta_2(\hat Z_1\cdot W_2)=\bar{\partial}_2(1/\hat Z_1\cdot W_2)$ a closed (0,1)-form on $\mathbb{PT}_{2}^{\vee}$, where $Z_1$ is treated as constant . Note that upon the restriction to the $\mathbb{CP}^1_{Z_1}$ fibre, the contribution $\langle\lambda_1 dx_1 \tilde\lambda_2]$ is not included since it does not point along the Euclidean $\mathbb{CP}^1_{Z_1}$ fibre direction. While it is clear that $f$ is a $\bar\partial_2$-closed $(0,1)$-form on $\mathbb{PT}_{2}^{\vee}$ (since $\langle\lambda_1 dx_1 \tilde\lambda_2]$ is holomorphic w.r.t $\bar \partial_2$), the fact that $f$ is a $\bar\partial_1$-closed $(0,1)$-form also on $\mathbb{PT}_1$ is less obvious, but can be explicitly checked by using the complex structure $\bar{\partial}_1 = \bar{e}_1^0 \, \bar{\partial}_{1,0} + \bar{e}_1^{\dot \alpha} \, \bar{\partial}_{1,\dot{\alpha}},$ with the Euclidean basis for vectors and $(0,1)-$forms given by
\begin{align}
T^{0,1}_{\mathbb{PT}_1} = \mathrm{span} \left\{ 
\bar{\partial}_{1,0} = \langle \lambda_1 \hat{\lambda}_1 \rangle \lambda^\alpha_1 \frac{\partial}{\partial \hat{\lambda}^\alpha_1} , \quad
\bar{\partial}_{1,\dot{\alpha}} = \lambda^\alpha_1 \frac{\partial}{\partial x^{\alpha \dot{\alpha}}_1} 
\right\} , \\
\Omega^{0,1}(\mathbb{PT}_1) = \mathrm{span} \left\{ 
\bar{e}^0_1 = \frac{\langle \hat{\lambda}_1 \, d\hat{\lambda}_1 \rangle}{\langle \lambda_1 \hat{\lambda}_1 \rangle^2} , \quad
\bar{e}_1^{\dot{\alpha}} = \frac{\hat{\lambda}_{1,\alpha} \, dx^{\alpha \dot{\alpha}}_1}{\langle \lambda_1 \hat{\lambda}_1 \rangle}
\right\} .
\end{align}
Going back to the \v{C}ech-Dolbeault representative and using this basis, we can write
\be
f=f_0 \bar{e}^0 + (\bar\partial_{\dot\alpha} \Phi) \bar{e}^{\dot{\alpha}} \ , \quad f_0= \frac{\braket{\lambda_1 \hat \lambda_1}}{\hat Z_1\cdot W_2 (Z_1 \cdot W_2)^{2s+1}} \ , \ \  \Phi=\frac{\log(\hat{Z}_1\cdot W_2)}{(Z_1 \cdot W_2)^{2s+2}} \ .
\ee
In this form, it is easy to see that $f$ is $\bar\partial$-closed on $\mathbb{PT}_1$ by noting that $\bar{\partial}_0\Phi = f_0$. 
We proceed to explicitly show that our ansatz gives rise to the expected coordinate space two-point function. Using Eq. \eqref{yT}, 
\begin{equation}
    (\hat Z_1 \cdot W_2)(Z_1 \cdot W_2)^{2s+1} = (z_1 z_2)^{s+1} \langle \hat \lambda_1 y_{1 2}\tilde \lambda_2] \langle\lambda_1 y_{1  2}\tilde \lambda_2]^{2s+1}\ ,
\end{equation}
we have 
\begin{equation}
    \begin{aligned}
        \braket{\mathcal{F}^s_{1}(X_1) \bar{\mathcal{F}}^s_{2}(X_2)} = C_{\Delta,s} \!\!\int\limits_{{\mathbb{CP}^1_{Z_1}}} \!\!\oint \frac{(D\lambda_1 \wedge D\hat\lambda_1)D \tilde \lambda_2}{\braket{\lambda_1 \hat{\lambda}_1}} \frac{\braket{t_1\lambda_{1}}^{2s}[\bar t_2 \tilde \lambda_2]^{2s} }{\langle \hat \lambda_1 y_{1 2}\tilde \lambda_2] \langle\lambda_1 y_{1  2}\tilde \lambda_2]^{2s+1}}\Bigg|_{X_1,X_2}\ .
    \end{aligned}
\end{equation}
After the contour integral with respect to $\tilde \lambda_2$ we are left with
\begin{equation}\label{FFbar}
    \begin{aligned}
         \braket{\mathcal{F}^s_1(X_1)\mathcal{\bar F}^s_2(\bar  X_2)}
         &\,= C_{\Delta,s} \!\!\int\limits_{{\mathbb{CP}^1_{Z_1}}}  \frac{D\lambda_1 \wedge D\hat\lambda_1}{\braket{\lambda_1 \hat{\lambda}_1}}\frac{\braket{t_1\lambda_1}^{2s} [\bar t_2 y_{12}^T \hat \lambda_1\rangle^{2s} }{\braket{ \lambda_1 y_{1  2}y_{12}^T \hat \lambda_1}^{2s+1}}\\
         &\,= C_{\Delta,s}\frac{I_{2s,0}}{u^{2s+1}}\ ,
    \end{aligned}
\end{equation}
where
\begin{equation}\label{serre}
    \begin{aligned}
    I_{n,0}&\,=\!\!\int\limits_{{\mathbb{CP}^1}} \frac{D\lambda_1 \wedge D\hat\lambda_1}{\braket{\lambda_1 \hat{\lambda}_1}^{n+2}}\braket{t_1\lambda_1}^{n} [\bar t_2 y_{1 2}^T \hat \lambda_1\rangle^{n}\\
    &\,= g(n) \, \langle t_1 y_{12}^T \bar t_2]^n\ ,
    \end{aligned}
\end{equation}
as we prove in Appendix \ref{ap:Inm}. As expected, $I_{n,0}$ depends solely on the unique scalar constructible from its integrand. Finally we have
\begin{equation}\label{chiral antichiral}
    \braket{\mathcal{F}^s(X_1)\mathcal{\bar F}^s(X_2)} = C'_{\Delta,s} \frac{(\tau_1\cdot \bar \tau_2)^{2s}}{u^{2s+1}},
\end{equation}
which reproduces the bispinor expression of \cite{Binder:2020raz} and where we absorbed all the coefficients in $C'_{\Delta,s}$. We note that Eq.~\eqref{serre} is the polarized version of the correspondence between n-index symmetric tensors and a closed, harmonic form in $\mathbb{CP}^1$ representing a Dolbeault cohomology class, $H^{0,1}(\mathbb{CP}^1,\mathcal{O}(-n-2))$, which is an example of Serre duality. Thus, after integrating over $\tilde \lambda_2$, we are left with the so-called Woodhouse representative \cite{Woodhouse:1985id} for the two-point correlator. 

The chiral/chiral and anti-chiral/anti-chiral cases are analogous. As alluded to earlier, these would be obtained by making the $W_2 \xrightarrow{} Z_2$ and $Z_1 \xrightarrow{} W_1$ replacements, respectively. For example, the chiral/chiral correlator is therefore explicitly given by
\begin{equation}\label{chiral/chiral}
\begin{aligned}
    \braket{\mathcal{F}^s_{1}(X_1) {\mathcal{F}}^s_{2}(X_2)} = C_{\Delta,s} \!\!\int\limits_{{\mathbb{CP}^1_{Z_1}}} \!\!\oint &\, \frac{(Z_1 \cdot I \cdot dZ_1) \wedge (\hat Z_1 \cdot I^{\text{flat}} \cdot d\hat Z_1)}{(Z_1 \cdot I^{\text{flat}} \cdot \hat{Z}_1) } (Z_2 \cdot I \cdot dZ_2)  \\
    &\,\frac{(\tau_1 \cdot I \cdot Z_1)^{2s} ( \tau_2 \cdot I \cdot Z_2)^{2s}}{(\hat Z_1 \cdot I\cdot Z_2)(Z_1 \cdot I \cdot Z_2)^{2s+1}}\Bigg|_{X_1,X_2}\ .
\end{aligned}
\end{equation}
Following the same steps as before, this can be written as
\begin{equation}
    \begin{aligned}
        \braket{\mathcal{F}^s_{1}(X_1) \mathcal{F}^s_{2}(X_2)} = C_{\Delta,s}  \!\!\int\limits_{{\mathbb{CP}^1_{Z_1}}} \!\!\oint \frac{(D\lambda_1 \wedge D\hat\lambda_1)D \lambda_2}{\braket{\lambda_1 \hat{\lambda}_1}} \frac{\braket{t_1\lambda_{1}}^{2s}\braket{t_2 \lambda_2}^{2s} }{\braket{ \hat \lambda_1 \tilde y_{1 2}\lambda_2} \braket{\lambda_1 \tilde y_{1  2}\lambda_2}^{2s+1}}\Bigg|_{X_1,X_2}\ .
    \end{aligned}
\end{equation}
By Eq. \eqref{yT}, the denominator is now proportional to $(u+2)^{2s+1}$ and we obtain
\begin{equation}
    \braket{\mathcal{F}^s(X_1)\mathcal{F}^s(X_2)} = C'_{\Delta,s} \frac{(\tau_1\cdot \tau_2)^{2s}}{(u+2)^{2s+1}}\ ,
\end{equation}
which agrees with the bispinor expression of \cite{Binder:2020raz} again. Finally, note that Fubini's theorem does not automatically allow to interchange the order of integration. In our prescription, the contour integral should be performed before the surface integral so that no pole is left on the Riemann sphere. \\
Interestingly, the self-dual/anti-self-dual propagator representative has full conformal invariance since it only involves the contraction between a twistor and a self-dual twistor $Z_1 \cdot W_2$, while the self-dual/self-dual propagator is given by $Z_1\cdot  I \cdot Z_2$, and therefore is only invariant under AdS isometries. This is consistent with the fact that the former survives in the flat space limit while the two-point functions of self-dual curvatures and Weyl tensors with themselves vanishes.

\subsection{Boundary Limits}
We now obtain a regularised version of the twistor-like representation of the boundary-to-boundary propagator from \cite{Baumann:2024ttn}. Our construction arises by taking the appropriate limit of the four-dimensional twistor formula developed in the preceding section. Explicitly, this will be done by verifying
\begin{equation}
    \begin{aligned}
        \braket{O_1^s(P_1) O_2^s(P_2)}= \mathcal{N}_{\Delta,s} \lim_{z_{1}, z_2 \rightarrow{}0} (z_1 z_2)^{-\Delta} \braket{\mathcal{F}^s(X_1)\mathcal{\bar F}^s(X_2)}\ ,
    \end{aligned}
\end{equation}
at the level of the integrand. Here $O^s(P)$ refers to the boundary dual to the bulk fields considered previously. We will take the limits in turn, focusing first on the twistor representation of the bulk-to-boundary propagator. The clearest way to proceed is to start from the bulk correlator written with explicit little group contractions, and then use Eq.~\eqref{limit1} to take the limit. It is also necessary to note that, since the 3d little group of the boundary operator is now $SL(2, \mathbb{C})$, we should adjust for the indices. This means we should act on both $(y_{12})_{\dot \alpha \alpha}$ and on the dummy variable $\tilde \lambda_{2,\alpha}$ with $\delta^{\dot \alpha}_a$ such that $(\delta \tilde\lambda)_a= \delta^{\dot{\alpha}}_a \lambda_{\dot{\alpha}} \equiv \pi_a$, resulting in
\begin{equation}
    \begin{aligned}
    \braket{\mathcal{F}_{1,\alpha_1...\alpha_{2s}}O_{2,a_1...a_{2s}}} =\mathcal{N}_{1,\Delta,s} &\,\lim_{z_2\to 0} \sqrt{z_2}^{-2\Delta} \!\!\int\limits_{{\mathbb{CP}^1_{Z_1}}} \!\!\oint \frac{ (D\lambda_1 \wedge D\hat\lambda_1)D \delta \tilde \lambda_2}{\braket{\lambda_1 \hat{\lambda}_1}} \,\\
    &\,\frac{\lambda_{\alpha_1}...\lambda_{\alpha_{2s}} (\delta \tilde\lambda_2)_{a_1}...(\delta \tilde\lambda_2)_{a_{2s}}}{\langle \hat \lambda_1 (y_{1 2} \delta) (\delta \tilde \lambda_2)] \braket{\lambda_1 (y_{1  2}\delta)(\delta \tilde \lambda_2)}^{2s+1}}\Bigg|_{X_1,X_2}\ ,
    \end{aligned}
\end{equation}
where $\Delta=s+1$. Equivalently, using Eq. \eqref{limit1}, polarising with $t_1^\alpha$ and $l^a$, we have
\begin{equation}
    \begin{aligned}
         \braket{\mathcal{F}^s_1(X_1)O^s_2(P_2)}
        &\,=\mathcal{N}_{1, \Delta,s}\!\!\int\limits_{{\mathbb{CP}^1_{Z_1}}} \!\!\oint  \frac{(D\lambda_1 \wedge D\hat\lambda_1)D \pi_2}{\braket{\lambda_1 \hat{\lambda}_1}}\frac{\braket{t_1 \lambda_1}^{2s} \braket{l \pi_2}^{2s}}{\braket{\hat \lambda_1 w_{1  2}\pi_2}\braket{\lambda_1 w_{1 2}\pi_2}^{2s+1}}\ .\\
    \end{aligned}
\end{equation}
As before, we first integrate over the simple pole $\braket{\hat \lambda_1 w_{1  2}\pi_2}$ 
\begin{equation}
    \begin{aligned}
         \braket{\mathcal{F}^s_1(X_1)O^s_2(P_2)}
        &\,=\mathcal{N}_{1,\Delta,s}\!\!\int\limits_{{\mathbb{CP}^1_{Z_1}}}  \frac{D\lambda_1 \wedge D\hat\lambda_1}{\braket{\lambda_1 \hat{\lambda}_1}}\frac{\braket{t_1 \lambda_1}^{2s} \braket{l w_{12}^T \hat \lambda_1}^{2s}}{\braket{\lambda_1 w_{1 2}w_{12}^T \hat\lambda_1}^{2s+1}}\ .
    \end{aligned}
\end{equation}
This is the same integral as before, which can again only depend on the scalar $\braket{t_1 w_{12}l}=\tau_1 \cdot \sigma_2 $ with $\sigma^M= l^a \Lambda^M_a$. Finally, we read off the bulk-to-boundary propagator to be
\begin{equation}
    \begin{aligned}
         \braket{\mathcal{F}^s_1(X_1)O^s_2(P_2)}
        &\,=\mathcal{N}^{ \prime}_{1,\Delta,s} \frac{(\tau_1 \cdot \sigma_2)^{2s}}{( X_1\cdot P_2)^{2s+1}}\ ,
    \end{aligned}
\end{equation}
as expected. By the first line of Eq. \eqref{limit1}, the $z_2$ limit of both $y_{12}$ and $\tilde y_{12}$ equals $w_{12}$; hence, we could have alternatively taken the boundary limit of the chiral-chiral propagator.\\
Now, taking the boundary limit with respect to $z_1$, we similarly obtain the boundary-to-boundary correlator
 \begin{equation}
    \begin{aligned}
         \braket{O_{1a_1...a_{2s}}(P_1) O_{2b_1...b_{2s}}(P_2)} &\,=\mathcal{N}_{2,\Delta,s}  \!\!\int\limits_{{\mathbb{CP}^1_{\pi_1}}} \!\!\oint \frac{(D\pi_1 \wedge D\hat\pi_1)D\pi_2}{\braket{\pi_1 \hat{\pi}_1}} \frac{\pi_{1a_1}...\pi_{1a_{2s}} \pi_{2b_1}...\pi_{2b_{2s}}}{\braket{\hat \pi_1 x_{1  2}\pi_2}\braket{\pi_1 x_{1 2}\pi_2}^{2s+1}}\Bigg|_{P_{1/2}} ,
    \end{aligned}
\end{equation}
\normalsize or equivalently
\begin{equation}
    \begin{aligned}
         \braket{O^s_1(P_1)O^s_2(P_2)}
        &\,=\mathcal{N}_{2,\Delta,s}\!\!\int\limits_{{\mathbb{CP}^1_{\pi_1}}} \!\!\oint\frac{(D\pi_1 \wedge D\hat\pi_1)D\pi_2}{\braket{\pi_1 \hat{\pi}_1}}\frac{\braket{l \pi_1}^{2s} \braket{l \pi_2}^{2s}}{(\hat{\mathbf{\Lambda}}_1 \cdot \mathbf{\Lambda}_2)(\mathbf{\Lambda}_1 \cdot \mathbf{\Lambda}_2)^{2s+1}}\\
        &\,=\mathcal{N}_{2,\Delta,s} \!\!\int\limits_{{\mathbb{CP}^1_{\pi_1}}} \frac{D\pi_1 \wedge D\hat\pi_1}{\braket{\pi_1 \hat{\pi}_1}}
        \frac{\braket{l \pi_1}^{2s} \braket{l x_{12}\hat \pi_1}^{2s}}{\braket{\pi_1 (x_{1 2})^2 \hat \pi_1}^{2s+1}}\\
        &\,=\mathcal{N}_{2,\Delta,s} \frac{1}{(x_{12})^{4s+2}}\!\!\int\limits_{{\mathbb{CP}^1_{\pi_1}}} \frac{D\pi_1 \wedge D\hat\pi_1}{\braket{\pi_1 \hat{\pi}_1}}
        \frac{\braket{l \pi_1}^{2s} \braket{l x_{12}\hat \pi_1}^{2s}}{\braket{\pi_1 \hat \pi_1}^{2s+1}}\ ,
    \end{aligned}
\end{equation} 
where we used Eq. \eqref{structureslittle}. As explained before, $l_i^a$ are auxiliary spinors and we took $l_1=l_2$ for simplicity (one can also check that this does not change the form of the correlator explicitly). Again, the integral is of the same form as Eq. \eqref{serre}, which gives
\begin{equation}
    \begin{aligned}
         \braket{O^s_1(P_1)O^s_2(P_2)}
         &\,=\mathcal{N}_{2,\Delta,s}^\prime\frac{\braket{l x_{12} l}^{2s}}{(P_1\cdot P_2)^{2s+1}}\\
        &\,=(-2)^{2s}\mathcal{N}_{2,\Delta,s}^\prime  \frac{H_{12}^{s}}{(P_1\cdot P_2)^{2s+1}}\ ,
    \end{aligned}
\end{equation}
using Eq. \eqref{structureslittle} which agrees with the general form of Eq. \eqref{2pointgeneral} when the unitarity bound is saturated.

\subsubsection{Alternative Regularization} \label{sec:alt_reg}
Finally, it is worth noting that the boundary-to-boundary propagator can be regularized in an alternative way by taking
\begin{equation}
    \begin{aligned}
        \braket{O^s_1(P_1)O^s_2(P_2)}
        &\,= \mathcal{N}_{2,\Delta,s} \!\!\oint\!\!\oint D\pi_1 D\pi_2\frac{\braket{l \pi_1}^{2s}\braket{l \pi_2}^{2s}}{\braket{\pi_1 x_{12} \pi_2}^{2s+1}\braket{\pi_1\pi_2}} \left(\frac{\braket{l \pi_2}}{\braket{l x_{12} \pi_2}}\right) \ . 
    \end{aligned}
    \label{eq:altern_reg}
\end{equation}
Contrary to the previous regularization, this does not have a clear twistor origin, but the form presented is absolutely convergent, so that the order of integration does not matter. To recover the embedding space expression for the correlator, one first integrates over the $\braket{\pi_1 \pi_2}$ pole so that
\begin{equation}
    \begin{aligned}
        \braket{O^s_1(P_1)O^s_2(P_2)}
        &\,= \mathcal{N}_{\Delta,s}\oint D\pi_2\frac{\braket{l \pi_2}^{4s}}{\braket{\pi_2 x_{12} \pi_2}^{2s+1}} \frac{\braket{l \pi_2}}{\braket{l x_{12} \pi_2}} \ ,
    \end{aligned}
\end{equation}
and then one integrates over the simple pole $\braket{l x_{12} \pi_2}$, from which we get immediately
\begin{equation}
    \begin{aligned}
        \braket{O^s_1(P_1)O^s_2(P_2)}
        &\,=\mathcal{N}_{\Delta,s}\frac{\braket{l x_{12} l}^{4s+1}}{\braket{l x_{12}^3 l}^{2s+1}} \\
        &\,=\mathcal{N}_{\Delta,s}^\prime\frac{H_{12}^{s}}{(P_{12})^{2s+1}} \ .
    \end{aligned}
\end{equation}

\section{Propagators of Non-Conserved Currents}\label{sec:beyondunitarity}
In this section, we generalise the previous results beyond the case where the unitarity bound is saturated (including the free scalar at $\Delta=1/2$). It is important to note that this generalisation does not have an obvious twistor origin. 

\subsection{Boundary-to-Boundary Propagators} 
\subsubsection{Maximal Helicity}
The generalisation for the boundary-to-boundary propagator is found by preserving the same two assumptions as before; that is we require that the integral should be projectively well defined and we still assume that the scaling of Eq. \eqref{scalingconformaldim} is manifest at the level of the integrand. This suggests the replacement
\begin{equation}
    (\hat{\mathbf{\Lambda}}_1 \cdot \mathbf{\Lambda}_2)(\mathbf{\Lambda}_1 \cdot \mathbf{\Lambda}_2)^{2s+1} \xrightarrow{} (\hat{\mathbf{\Lambda}}_1 \cdot \mathbf{\Lambda}_2)(\mathbf{\Lambda}_1 \cdot \mathbf{\Lambda}_2)^{2\Delta-1} \ .
\end{equation}
By the first assumption, this means we need to add $2(\Delta-s-1)$ factors of $\pi_1$ and $\pi_2$.
In addition, free indices determine the spin, which we want to keep general as before. With that in mind, the simplest possible deformation to the boundary-to-boundary propagator is
\begin{equation}\label{nonconserved unpolarised}
    \begin{aligned}
         \langle O_{1,a_1...a_{2s}}(P_1) &\, O_{2,b_1...b_{2s}}(P_2)\rangle=\\
         &\,\mathcal{N}_{\Delta,s} \!\!\int\limits_{{\mathbb{CP}^1_{\pi_1}}} \!\!\oint \frac{(D\pi_1 \wedge D\hat\pi_1)D\pi_2}{\braket{\pi_1 \hat{\pi}_1}} \frac{\pi_{1,(a_1}...\pi_{1,a_{2s}} \pi_{2,b_1}...\pi_{2,b_{2s})} \braket{\pi_1 \pi_2}^{2(\Delta-s-1)} }{(\hat{\mathbf{\Lambda}}_1 \cdot \mathbf{\Lambda}_2)(\mathbf{\Lambda}_1 \cdot \mathbf{\Lambda}_2)^{2\Delta-1}}\Bigg|_{P_1,P_2} \ .
    \end{aligned}
\end{equation}
Note the symmetrisation of indices, which is now necessary since the two-point correlator is fully symmetric in spinor notation. Although not obvious, this integral is automatically symmetric in its spinor indices when $\Delta = s + 1$. This is why, in the conserved case, we did not have to impose this explicitly. In general, however, this need not hold. In the polarised form, we have
\begin{equation}\label{boundarybeyondpolarised}
    \begin{aligned}
         \braket{O^{s, \Delta}(P_1)O^{s, \Delta}(P_2)}
        &\,=\mathcal{N}_{\Delta,s}\!\!\int\limits_{{\mathbb{CP}^1_{\pi_1}}} \!\!\oint \frac{(D\pi_1 \wedge D\hat\pi_1)D\pi_2}{\braket{\pi_1 \hat{\pi}_1}}\frac{\braket{l \pi_1}^{2s} \braket{l \pi_2}^{2s} \braket{\pi_1 \pi_2}^{2(\Delta-s-1)}}{\braket{\hat \pi_1 x_{1 2}\pi_2}\braket{\pi_1 x_{1 2}\pi_2}^{2\Delta-1}} \\
        &\,=\mathcal{N}_{\Delta,s}\!\!\int\limits_{{\mathbb{CP}^1_{\pi_1}}}  \frac{D\pi_1 \wedge D\hat\pi_1}{\braket{\pi_1 \hat{\pi}_1}}
        \frac{\braket{l \pi_1}^{2s} \braket{l x_{1 2} \hat \pi_1}^{2s} \braket{\pi_1 x_{12}\hat \pi_1}^{2(\Delta-s-1)}}{\braket{\pi_1 x_{1 2} x_{12}\hat \pi_1}^{2\Delta-1}}\\
        &\,=\mathcal{N}_{\Delta,s}\frac{I_{2s,2(\Delta-s-1)}}{(x_{12}^2)^{2\Delta-1}} \ ,
    \end{aligned}
\end{equation} 
where we first integrated around the simple pole exactly as before, and where we defined  
\begin{equation}\label{Inm}
    I_{n,m}=\int_{\mathbb{CP}^1}  \frac{D\pi_1 \wedge D\hat\pi_1}{\braket{\pi_1 \hat{\pi}_1}^{n+m+2}} 
    \braket{l \pi_1}^n\braket{l x_{12}\hat{\pi}_1}^n \braket{\pi_1 x_{12} \hat{\pi}_1}^m \;\;\; n,m \geq 0\ .
\end{equation}
As before, we take the auxiliary polarization spinors to be equal for simplicity. As we show in Appendix \ref{ap:Inm}, 
\begin{equation}
    I_{n,m} =g(n,m) (P_1 \cdot P_2)^{m/2} \braket{l x_{12} l}^n \ ,
\end{equation}
for $m$ even or $n=0,m=-1$.
Substituting into Eq. \eqref{boundarybeyondpolarised}, we arrive at
\begin{equation}
    \braket{O^{s, \Delta}(P_1)O^{s, \Delta}(P_2)} = \mathcal{N}_{\Delta,s}^\prime \frac{H_{12}^s}{(P_1 \cdot P_2)^{\Delta+s}},
\end{equation}
as expected (see Eq. \eqref{2pointgeneral}). From the allowed values of $n$ and $m$, our twistor formulation only works for integer conformal dimension and the free scalar.
\subsubsection{Other Helicities}
As we already mentioned, the operators that don't saturate the unitarity bound are dual to massive spinning fields, so they should contain all possible helicities in $-s,...,s$. Guiding ourselves from twistor space ideas, the positive  helicity operators will be constructed as
\begin{equation}\label{PTlambda-}
    \phi_{ a_1 ...  a_{2s}}^+ = \oint \braket{\lambda d\lambda}  \frac{\partial}{\partial \mu^{a_1}}... \frac{\partial}{\partial \mu^{a_{2s}}} f^+(\lambda, \mu)|_X\ .
\end{equation}
which is a 3d analogue of the 4d Penrose transform for twistors, to be contrasted to the dual twistor realization that we used in the previous section. Thus, the expressions we consider should contain all combinations $\braket{l \pi}^{2s-h} \braket{l\frac{\partial}{\partial\mu }}^h$.\footnote{It is important to emphasise that this is not the standard momentum space definition of helicity. However, it matches the definition taken in \cite{Caron-Huot:2021kjy}, where the link with momentum space is made. In particular, it turns out that this definition aligns well with flat space intuition, as we will discuss in the three-point YM correlator example. } \\
Without regularization, the correlator of two operators of helicities $h_1, h_2$ will then be
\begin{equation}
    \begin{aligned}
         \braket{O^{s,h_1}(P_1)O^{s,h_2}(P_2)}
        &\,=\oint \oint  D\pi_1 D\pi_2 \braket{l \pi_1}^{2s-h_1} \braket{l\frac{\partial}{\partial\mu_1 }}^{h_1}  \braket{l \pi_2}^{2s-h_2} \braket{l\frac{\partial}{\partial\mu_2 }}^{h_2} \frac{\braket{\pi_1 \pi_2}^{m}}{(\bold \Lambda_1 \cdot \bold \Lambda_2)^{q}}\ ,
    \end{aligned}
\end{equation} 
where we now added a helicity label to the boundary operator. Requiring the nested integrals to be well-defined projectively we find that the helicities should be equal, as desired, and  
\begin{equation}
    \begin{aligned}
         \braket{O^{s,h}(P_1)O^{s,h}(P_2)}
        &\,=\oint \oint D\pi_1 D\pi_2 \braket{l \pi_1}^{2s-h} \braket{l\frac{\partial}{\partial\mu_1 }}^{h}  \braket{l \pi_2}^{2s-h} \braket{l\frac{\partial}{\partial\mu_2 }}^{h} \frac{\braket{\pi_1 \pi_2}^{m}}{(\bold \Lambda_1 \cdot \bold \Lambda_2)^{q}}\ ,
    \end{aligned}
\end{equation} 
where $m=2(\Delta-s-1)$ and $q=2(\Delta-h)$. Performing the $2h$ differentiations, we get
\begin{equation}
    \begin{aligned}
         \braket{O^{s,h}(P_1)O^{s,h}(P_2)}
        &\,=(-1)^h q^{(2h)} \oint \oint D\pi_1 D\pi_2 \braket{l \pi_1}^{2s}  \braket{l \pi_2}^{2s} \frac{\braket{\pi_1 \pi_2}^{m}}{(\bold \Lambda_1 \cdot \bold \Lambda_2)^{q+2h}}\ ,
    \end{aligned}
\end{equation} 
where $q^{(2h)}$ is the rising factorial. This is regularised as before to
\begin{equation}
    \begin{aligned}
         \braket{O^{s,h}(P_1)O^{s,h}(P_2)}
        &\,=(-1)^h q^{(2h)} \!\!\int\limits_{{\mathbb{CP}^1_{\pi_1}}} \!\!\oint\frac{(D\pi_1 \wedge D\hat\pi_1)D\pi_2}{\braket{\pi_1 \hat{\pi}_1}}  \frac{ \braket{l \pi_1}^{2s}  \braket{l \pi_2}^{2s}\braket{\pi_1 \pi_2}^{m}}{(\bold{\hat \Lambda}_1 \cdot \bold \Lambda_2)(\bold \Lambda_1 \cdot \bold \Lambda_2)^{q+2h-1}}\\
        &\, =(-1)^h q^{(2h)} \frac{I_{2s,m}}{(x^{2}_{12})^{q+2h-1}}\\
        &\, = \mathcal{N}_{\Delta,s}^{\prime \prime}\frac{H_{12}^s}{(P_1 \cdot P_2)^{\Delta+s}} \ ,
    \end{aligned}
\end{equation}
which gives the correct propagator. This expression is now valid for non-conserved currents and non-conformally-coupled scalars and agrees, up to regularization, with the recent results of \cite{Bala:2025gmz} published while this article was being written.

\subsubsection{Holomorphicity and (Non-)Conservation}
\paragraph{Spinning Case}
From the point of view of twistors, it may seem surprising that the generalisation to the non-conserved case simply introduces factors of $\braket{\pi_1 \pi_2}$ rather than new non-holomorphic data. Just like in the 4d case, one can check that both the conserved and non-conserved representatives that we write are holomorphic. The $\bold \Lambda_2$ holomorphicity is obvious, and the $\bold \Lambda_1$ case can be checked in an analogous manner to the bulk version. Here, we can consider the complex structure $\bar{\partial} = \bar{e}^0 \, \bar{\partial}_0 + \bar{e}^{a} \, \bar{\partial}_{a},$ where the Euclidean basis for vectors and $(0,1)-$forms is now given by
\begin{align}
T^{0,1}_{\mathbb{PS}} = \mathrm{span} \left\{ 
\bar{\partial}_0 \equiv \langle \lambda \hat{\lambda} \rangle \, \lambda_a \frac{\partial}{\partial \hat{\lambda}_a} \ , \quad
\bar{\partial}_a \equiv \lambda^b \frac{\partial}{\partial x^{ab}} 
\right\} \ , \\
\Omega^{0,1}(\mathbb{PS}) = \mathrm{span} \left\{ 
\bar{e}^0 \equiv \frac{\langle \hat{\lambda} \, d\hat{\lambda} \rangle}{\langle \lambda \hat{\lambda} \rangle^2} \ , \quad
\bar{e}^a \equiv \frac{dx^{ab} \, \hat{\lambda}_b}{\langle \lambda \hat{\lambda} \rangle}
\right\} \ .
\end{align}
Even though we are not strictly working with minitwistors from the on-set, the representative is in the projective spinor bundle $\mathbb{PS}=\mathbb{R}^3\times \mathbb{CP}^1$. Writing the representative from Eq.~\eqref{nonconserved unpolarised} as
\be
f=f_0 \bar{e}^0 + (\bar\partial_{a} \Phi) \bar{e}^{a} \ , \quad f_0= \frac{\braket{\pi_1 \hat \pi_1}\braket{\pi_1\pi_2}^{2(\Delta-s-1)}}{(\hat{\mathbf{\Lambda}}_1 \cdot \mathbf{\Lambda}_2)(\mathbf{\Lambda}_1 \cdot \mathbf{\Lambda}_2)^{2\Delta-1}} \ , \ \  \Phi=\frac{\braket{\pi_1\pi_2}^{2(\Delta-s-1)}\log(\hat{\mathbf{\Lambda}}_1 \cdot \mathbf{\Lambda}_2)}{(\mathbf{\Lambda}_1 \cdot \mathbf{\Lambda}_2)^{2\Delta-1}} \ ,
\ee
one can see that $f$ is $\bar\partial$-closed on $\mathbb{PS}_1$ by using that $\bar{\partial}_0\Phi = f_0$. The proof for the other helicities is analogous. 

Nevertheless, in the non-conserved case we observe a new feature arising from the nested transform: the spinor indices are no longer automatically symmetric, so that $\pi_1^a$ contributes to both $O_1(P_1)^{a_1...a_{2s}}$ and $O_2(P_2)^{b_1...b_{2s}}$, rather than only to the operator located $P_1$ (as expected). To check conservation, we consider the unpolarised case and apply the divergence. In our case this can be written as $\pi_{i(a} \frac{\partial }{\partial \mu^{b)}_i}+\hat{\pi}_{i(a} \frac{\partial }{\partial \hat{\mu}^{b)}_i}$. With this, we find that the non-symmetrised version of our integral is divergenceless, so as expected, holomorphicity seems to imply conservation. However, by applying the divergence on either operator, we have
\small
\begin{equation}\label{ddx2}
    \begin{aligned}
      -\pi_{2(a} \frac{\partial }{\partial \mu_2^{b)}}  [\frac{1}{(\hat {\bold \Lambda}_1 \cdot \bold \Lambda_2) (\bold \Lambda_1 \cdot \bold \Lambda_2)^{2\Delta-1}}]&\,= (\pi_{1(a} \frac{\partial }{\partial \mu^{b)}_1}+\hat{\pi}_{1(a} \frac{\partial }{\partial \hat{\mu}^{b)}_1} )[\frac{1}{(\hat {\bold \Lambda}_1 \cdot \bold \Lambda_2) (\bold \Lambda_1 \cdot \bold \Lambda_2)^{2\Delta-1}}] \\
      &\,= \frac{1}{(\hat {\bold \Lambda}_1 \cdot \bold \Lambda_2) (\bold \Lambda_1 \cdot \bold \Lambda_2)^{2\Delta-1}}(\frac{\hat \pi_{1(a} \pi_{2b)}}{\hat {\bold \Lambda}_1 \cdot \bold \Lambda_2}   +(2\Delta-1) \frac{\pi_{1(a} \pi_{2b)}}{ \bold \Lambda_1 \cdot \bold \Lambda_2})\ .
    \end{aligned}
\end{equation}
\normalsize
Therefore, including symmetrization we obtain
\begin{equation}
    \begin{aligned}
        &\,\nabla_{b_1b_2} \braket{O_{1,\Delta}^{(a_{1}...a_{2s}}O_{2, \Delta}^{b_{1}...b_{2s})}}= -  \mathcal{N}_{\Delta,s} \!\!\int\limits_{{\mathbb{CP}^1_{\pi_1}}} \!\!\oint \frac{(D\pi_1 \wedge D\hat\pi_1)D\pi_2}{\braket{\pi_1 \hat{\pi}_1}} \\&\,\frac{\pi_{1}^{(a_1}...\pi_{1}^{a_{2s}} \pi_{2}^{b_1}...\pi_{2}^{b_{2s})} \braket{\pi_1 \pi_2}^{2(\Delta-s-1)} }{(\hat{\mathbf{\Lambda}}_1 \cdot \mathbf{\Lambda}_2)(\mathbf{\Lambda}_1 \cdot \mathbf{\Lambda}_2)^{2\Delta-1}}
        (\frac{\hat \pi_{1(a} \pi_{2b)}}{\hat {\bold \Lambda}_1 \cdot \bold \Lambda_2}   +(2\Delta-1) \frac{\pi_{1(a} \pi_{2b)}}{ \bold \Lambda_1 \cdot \bold \Lambda_2})\Bigg|_{P_1,P_2}\ ,
    \end{aligned}
\end{equation}
which contract to terms proportional to $\braket{\pi_2 \pi_1}^2$, $\braket{\pi_2 \pi_1}\braket{\pi_2 \hat \pi_1}$, $\braket{\pi_2 \pi_1}\braket{\pi_1 \hat \pi_1}$ that do not vanish.
\paragraph{Scalar Case}
Let us now check the scalar equation of motion. For this one needs to apply $\pi_{2}^{(a} \frac{\partial }{\partial \mu_{2b)}}$ to Eq. \eqref{ddx2}, from which we get 
\begin{equation}
    \begin{aligned}
         &\, \pi_{2}^{(a} \frac{\partial }{\partial \mu_{2b)}}  \pi_{2(a} \frac{\partial }{\partial \mu_2^{b)}}  \braket{O^{0,\Delta}(P_1)O^{0,\Delta}(P_2)}\\
   &\,= \frac{\mathcal{N}_{\Delta,0}}{(x_{12}^2)^{2\Delta}} ((2\Delta-1)  I_{0,2\Delta-2}+(\Delta-1)(2\Delta-3) I_{0,2\Delta-4} x_{12}^{2})\ .
    \end{aligned}
\end{equation}
As a sanity check, note that at $\Delta=1/2$, we have 
\begin{equation}
    \nabla^2 \braket{O^{0,1/2}(P_1)O^{0,1/2}(P_2)} \propto I_{0,-3} =0 \ ,
\end{equation}
as expected for the free scalar.

\subsubsection{Alternative Regularization}
The alternative regularization of Section~\ref{sec:alt_reg} can be generalised when the unitarity bound is not saturated in the maximal helicity case and is valid for the free scalar too ($\Delta=1/2$). In a similar manner to Eq.~\eqref{eq:altern_reg}, we take
\begin{equation}
    \begin{aligned}
        \braket{O^s(P_1)O^s(P_2)}
        &\,= \mathcal{N}_{\Delta,s}\oint \oint \braket{\pi_1 d\pi_1}\braket{\pi_2 d\pi_2}\frac{\braket{l \pi_1}^{2s}\braket{l \pi_2}^{2s}}{\braket{\pi_1 x_{12} \pi_2}^{2s+1}\braket{\pi_1\pi_2}} (\frac{\braket{l \pi_2}}{\braket{l x_{12} \pi_2}})^{2(\Delta-s)-1} \\
        &\,=\mathcal{N}_{\Delta,s}\oint \braket{\pi_2 d\pi_2}\frac{\braket{l \pi_2}^{4s}}{\braket{\pi_2 x_{12} \pi_2}^{2s+1}} (\frac{\braket{l \pi_2}}{\braket{l x_{12} \pi_2}} )^{2(\Delta-s)-1} \ .
    \end{aligned}
\end{equation}
Integrating over the pole $\braket{l x_{12} \pi_2}$ and using Stokes' theorem
\begin{align}
&\braket{O^s(P_1)O^s(P_2)}\,=\mathcal{N}_{\Delta,s}^\prime\oint \braket{\pi_2 d\pi_2}\frac{\braket{l \pi_2}^{2(\Delta+s)-1}}{\braket{\pi_2 x_{12} \pi_2}^{2s+1}} (\frac{\braket{l \frac{\partial}{\partial \pi_2}}}{\braket{l x_{12}l}})^{2(\Delta-s-1)}  (\frac{1}{\braket{l x_{12} \pi_2}} ) \nonumber \\
        &\,=(-1)^{2(\Delta-s)}\mathcal{N}_{\Delta,s}^\prime \oint \braket{\pi_2 d\pi_2}\frac{\braket{l \pi_2}^{2(\Delta+s)-1}}{\braket{l x_{12} \pi_2}  } (\frac{\braket{l \frac{\partial}{\partial \pi_2}}}{\braket{l x_{12}l}})^{2(\Delta-s-1)}  (\frac{1}{\braket{\pi_2 x_{12} \pi_2}^{2s+1}} ) \ .
\end{align}
Using the following expression
\begin{equation}
    \begin{aligned}
        \braket{l \frac{\partial}{\partial \pi_2}}^n (\frac{1}{\braket{\pi_2 x_{12} \pi_2}^m})|_{x_{12}l}= \begin{cases} \frac{1}{(x_{12}^2)^{n/2 +m} \braket{lx_{12}l}^m}, \text{    $n$ even } \\
        0, \text{   \;\;\;\;\;\;\;\;\;\;\;\;\;\;\;\;\;\;\;\; \;\;\,$n$ odd }
        \end{cases} \ ,
    \end{aligned}
\end{equation}
for $n=2(\Delta-s-1)$ and $m=2s+1$ and the residue theorem, we obtain
\begin{equation}
    \begin{aligned}
        \braket{O^s(P_1)O^s(P_2)}
        &\,=(-1)^{2(\Delta-s)}\mathcal{N}_{\Delta,s}^\prime \frac{\braket{l x_{12} l}^{4s+1-(2s+1)}}{(x_{12}^2)^{\Delta-s-1+2s+1}} \\
        &\,=(-1)^{2(\Delta-s)}\mathcal{N}_{\Delta,s}^\prime \frac{\braket{l x_{12} l}^{2s}}{(x_{12}^2)^{\Delta+s}} \\
        &\,=\mathcal{N}_{\Delta,s}^{\prime \prime}  \frac{H_{12}^{s}}{(P_{12})^{\Delta+s}}\ .
    \end{aligned}
\end{equation}
\subsection{Bulk-to-Bulk Propagator as a Pochhammer contour?}
If the operators do not saturate the unitarity bound, the bulk-to-bulk propagators become \cite{Binder:2020raz, Costa:2014kfa, Bena:1999py, Bena:1999be, Anguelova:2003kf,Leonhardt:2003qu, Basu:2006ti, Faizal:2011sa} 
\begin{equation}\label{chiral antichiral massive}
\begin{aligned}
    \braket{\mathcal{F}^s(X_1)\mathcal{\bar F}^s(X_2)} &\,\propto \frac{(\tau_1\cdot \bar \tau_2)^{2s}} {u^{\Delta+s}}\phantom{}_2F_1(\Delta-s-1,\Delta+s;2\Delta-2;-\frac{2}{u}) \ ,\\
    \braket{\mathcal{F}^s(X_1)\mathcal{ F}^s(X_2)} &\,\propto \frac{(\tau_1\cdot  \tau_2)^{2s}}{u^{\Delta+s}}\phantom{}_2F_1(\Delta+s-1,\Delta+s;2\Delta-2;-\frac{2}{u}) \ ,
\end{aligned}
\end{equation}
which excludes the $s=0, \Delta=1$ case for the chiral-chiral propagator. Clearly, this is harder to express as an integral over $\mathbb{CP}^1$. Let us mention however that such an integral form indeed exists. As we already noted, a choice of representative entails both a choice of contour as well as a choice of integrand. The Gauss hypergeometric function can then be expressed as an integral over the sphere along a Pochhammer contour $P=ABA^{-1}B^{-1}$, where $A$ is a loop starting at $x=1/2$ circling around $x=1$ anticlockwise and $B$ is a loop starting at $x=1/2$ circling around $x=0$ anticlockwise. 
Starting from the Euler representation of the hypergeometric function
\begin{equation}
   \phantom{a}_2F_1(a,b;c;z)=\frac{1}{B(b,c-b)} \int_0^1 t^{b-1}(1-t)^{c-b-1}(1-zt)^{-a}dt\ ,
\end{equation}
where Re$(c)>$ Re$(b)> 0$ and $z$ is not a real number greater than 1, we can analytically continue the hypergeometric function to
\begin{equation}
    \phantom{a}_2F_1(a,b;c;z)=\frac{1}{B(b,c-b) (1-e^{2 \pi i b}) (1-e^{2 \pi i (c-b)})} \oint_P t^{b-1}(1-t)^{c-b-1}(1-zt)^{-a}dt\ ,
\end{equation}
as a Pochhammer contour integral over $\mathbb{CP}^1$. Let us emphasise that, should a twistor interpretation of the bulk-to-bulk propagator exist, it is not clear whether this is the correct path to take.

\section{3-Point Functions of Conserved Currents}\label{sec:3points}
\subsection{General Formalism}
In this section, we proceed to construct boundary three-point functions of conserved currents and conformally coupled scalars that saturate the unitarity bound. Since bulk three-point functions are not entirely fixed by the (A)dS isometries, we won't explicitly compute them here. Nevertheless, we expect that they can be constructed from nested Penrose transforms as in the two-point case, and that their boundary limit also gives the results that we describe below. In fact, this is the motivation for the construction that we now describe. In the following, we will be able to regularise our integrals (that is, make them non-zero from the complex perspective) by dressing them with logarithmic factors, and it won't be necessary to restrict ourselves to Euclidean signature. Therefore, in the following, we keep the coordinates complex, imposing the reality condition in the end if necessary.\\
The number of consistent independent structures for three-point functions in 3d is \cite{Costa:2011mg}
\begin{equation}
    N_{3d}(s_1,s_2,s_3)= (2s_1+1)(2s_2+1)-p(1+p) \ ,
\end{equation}
where $s_1 \leq s_2 \leq s_3$ are the spins of the three currents and $p= \text{max}(0,s_1+s_2-s_3)$. However, as already mentioned, conservation does not follow immediately from taking these structures and saturating the unitarity bound. As proven in \cite{Caron-Huot:2021kjy}, imposing conservation lowers that number to at most 4 for any spin. In the even case, which we shall restrict ourselves to, the naively
\begin{equation}
    N_{3d}^{even}(s_1,s_2,s_3)= 2s_1 s_2+s_1+s_2-\frac{p(1+p)}{2} \ ,
\end{equation}
possible structures go down to at most 2: either all three helicities are equal or one of them is opposite to the other two.\\
Let us start with the simplest case, which corresponds to the case where all the helicities are negative and in the same direction. As usual, the spin is completely determined by the number of free indices (which, for convenience, we contract with the auxiliary spinor $l^a$). Then we know that the representative of the nested Penrose transforms can only depend on $\mathbf{\Lambda}_i\cdot\mathbf{\Lambda}_j$ since it must be a scalar. With the simplest representative, we obtain
\begin{equation}\label{samehelicity}
    \boxed{\braket{O_1^{s_1} O_2^{s_2} O_3^{s_3}}= \# \oint D \pi_{123} \frac{\braket{l \pi_1}^{2s_1} \braket{l \pi_2} ^{2s_2}\braket{l \pi_3}^{2s_3}} {(\mathbf{\Lambda}_1\cdot\mathbf{\Lambda}_2)^{n_3} (\mathbf{\Lambda}_2\cdot\mathbf{\Lambda}_3)^{n_1}  (\mathbf{\Lambda}_3\cdot\mathbf{\Lambda}_1)^{n_2}}} \ .
\end{equation}
Here we take
\begin{equation}
    n_i= s_j+s_k-s_i+1 \ ,
\end{equation}
for the integral to be projectively well-defined and $D \pi_{ij...} \equiv \braket{\pi_i d\pi_i} \braket{\pi_j d\pi_j} ...$. Eq.~\eqref{samehelicity} will be non-zero\footnote{If $n_i<0$, one has to include a logarithm term to obtain the correlator as shown in Appendix \ref{app:negative helicity}.} if $n_i>0$. From now on, the spins should not be ordered, but they should sum to an integer. For that reason, we restrict ourselves to bosonic correlators. Note that this also implies the usual scaling with respect to the conformal dimension $f(r_i \mathbf{\Lambda}_i)= r_i^{-2\Delta_i}f(\mathbf{\Lambda}_i)$. \\
Let us now consider the case where one of the operators, say $O^{s_3}_3$, carries opposite helicity with $s_3\geq1$. As discussed in Section \ref{sec:embeddingspace}, we will distinguish this situation from the previous one by using tilded operators. Naively, applying the same reasoning, we should consider 
\begin{equation}\label{oppositehelicitynolog}
\begin{aligned}
    \braket{\tilde O_1^{s_1} \tilde O_2^{s_2} \tilde O_3^{s_3}}&\,=\#\oint D \pi_{123} 
     \frac{\braket{l \pi_1}^{2s_1} \braket{l  \pi_2}^{2s_2}}{(\mathbf{\Lambda}_1\cdot\mathbf{\Lambda}_2)^{s_1+s_2+s_3+1}} \braket{l \frac{\partial}{\partial \mu_3}}^{2s_3}
    \frac{1}{ (\mathbf{\Lambda}_2\cdot\mathbf{\Lambda}_3)^{\tilde n_1}  (\mathbf{\Lambda}_3\cdot\mathbf{\Lambda}_1)^{\tilde n_2}}\ ,
\end{aligned}
\end{equation}
where
\begin{equation}
    \begin{aligned}
        \tilde n_1&\,=1-s_1+s_2-s_3 \ ,\\
        \tilde n_2&\,= 1+s_1-s_2-s_3\ ,
    \end{aligned}
\end{equation}
are fixed by homogeneity. For the case of interest ($s_3 \geq 1$), $\tilde n_1$ and $\tilde n_2$ cannot be both strictly positive, but they can be both negative, in which case this integral vanishes automatically. It can, however, be regularised by taking  
 $\tilde n_1 >0$ and $\tilde n_2\leq0$ as
\begin{equation}\label{oppositehelicity1log}
\begin{aligned}
    \boxed{ \braket{\tilde O_1^{s_1} \tilde O_2^{s_2} \tilde O_3^{s_3}}=\#\oint D \pi_{123} 
     \frac{\braket{l \pi_1}^{2s_1} \braket{l  \pi_2}^{2s_2}}{(\mathbf{\Lambda}_1\cdot\mathbf{\Lambda}_2)^{s_1+s_2+s_3+1}} \braket{l \frac{\partial}{\partial \mu_3}}^{2s_3}\frac{\log(\mathbf{\Lambda}_3\cdot\mathbf{\Lambda}_1)}{ (\mathbf{\Lambda}_2\cdot\mathbf{\Lambda}_3)^{\tilde n_1}  (\mathbf{\Lambda}_3\cdot\mathbf{\Lambda}_1)^{\tilde n_2}} }\ ,
\end{aligned}
\end{equation}
where the integral would vanish without the logarithmic term. This can be seen by noticing that after applying the $\mu_3$ derivatives, we obtain a pole in $\mathbf{\Lambda}_3$ which is at least the degree of the polynomial in $\pi_3$ on the numerator plus two. Finally, if $\tilde n_1 \leq 0$ and $\tilde n_2 \leq 0$,
\begin{equation}\label{oppositehelicity}
\begin{aligned}
    \boxed{\braket{\tilde O_1^{s_1} \tilde O_2^{s_2} \tilde O_3^{s_3}}= \# \oint D \pi_{123} 
     \frac{\braket{l \pi_1}^{2s_1} \braket{l  \pi_2}^{2s_2}}{(\mathbf{\Lambda}_1\cdot\mathbf{\Lambda}_2)^{s_1+s_2+s_3+1}} \braket{l \frac{\partial}{\partial \mu_3}}^{2s_3}
    \frac{ \log(\mathbf{\Lambda}_2\cdot\mathbf{\Lambda}_3)\log(\mathbf{\Lambda}_3\cdot\mathbf{\Lambda}_1)}{ (\mathbf{\Lambda}_2\cdot\mathbf{\Lambda}_3)^{\tilde n_1}  (\mathbf{\Lambda}_3\cdot\mathbf{\Lambda}_1)^{\tilde n_2}} }\ .
\end{aligned}
\end{equation}
To be precise, the logarithm is scale invariant up to a shift that vanishes after differentiation, such that the integral is still well-defined projectively. 
Although logarithmic factors in the integrand could suggest a more intricate analytic structure with branch cuts $L$, we can select the contour on the slit sphere $S^2 \setminus L$ so that, after performing an integration by parts, every total derivative contribution is finite and single-valued on the contour. By Cauchy’s theorem these terms vanish, and no logarithmic terms remain in the resulting integral. Note that representatives involving logarithms have been previously considered when constructing self-dual Coulomb fields \cite{Bailey:1985}.

\paragraph{Link with the real distributional representation}
Our representatives can naturally be seen as the analytic continuation of the distributional representatives $\delta^{[n]}(x)$ (the $n-$th derivative of the delta function) used in \cite{Baumann:2024ttn}. The complex analogue of the delta function is just a simple pole; therefore, for $n\geq0$, we have
\begin{equation}
    \delta^{[n]}(x) \xrightarrow{} \frac{1}{z^n} \ ,
\end{equation}
up to an irrelevant constant. For $n\leq -1$, the delta function representative of \cite{Baumann:2024ttn} is defined as
\begin{equation}
    \delta^{[n]}(x)= \frac{1}{2(|n|-1)!} \text{sign}(x) x^{|n|-1}\ .
\end{equation}
The sign function can be expressed in terms of the Heaviside step function \(H(x)\) as
$ \operatorname{sign}(x) = 2H(x) - 1 $. In hyperfunction theory, real functions with discontinuities can be represented as differences of boundary values of holomorphic functions defined on the upper and lower half-planes. The complex logarithm $\log(z)$ has a branch cut along the real axis, and its boundary values on the upper and lower half-planes differ by a constant jump. This allows us to write the sign function as a hyperfunction $
\operatorname{sign}(x) = (1 - log(z)/\pi i,\ -1 - log(z)/\pi i)$, where the pair $(F_+, F_-) = (1 - \log(z)/\pi i,\ -1 - \log(z)/\pi i)$ are holomorphic functions in the upper and lower half-planes, respectively. The value of the sign function on the real axis is recovered as the jump of the logarithm across its branch cut $\operatorname{sign}(x) = F_+(x + i0) - F_-(x - i0)$. Thus, away from the positive real axis, the logarithm provides a natural analytic continuation of $\delta^{[-1]}(x)$. Therefore for $n\leq -1$, we have
\begin{equation}
    \delta^{[n]}(x) \xrightarrow{} z^{-n-1} \log(z) \ .
\end{equation}
The link with the representation used in \cite{Baumann:2024ttn} can be made even more explicit by defining 
\begin{equation}
    \begin{aligned}
         M_n(z)&\,=\begin{cases}
             \frac{z^n}{n!} (log(z)-H_n)\ ,\;\; n\geq 0\\
             (-1)^n \frac{(|n|-1)!}{z^{|n|}}, \;\; n<0 \ 
         \end{cases}\\
         H_n&\,= \sum_{k=1}^n \frac{1}{k}\ ,\\
         H_0&\,=0 \ ,
    \end{aligned}
\end{equation}
which, similarly to $\delta^{[n]}(x)$, defines an Appell sequence normalised as $\partial_xf_n(x)= f_{n-1}(x)$. Then both the same and opposite helicity 3-points can be equivalently written as
{\small
\begin{equation}
\begin{aligned}
    \boxed{ \braket{\tilde O_1^{s_1} \tilde O_2^{s_2} \tilde O_3^{s_3}}=\# \oint D \pi_{123} 
     \braket{l \pi_1}^{2s_1} \braket{l  \pi_2}^{2s_2} \braket{l  \pi_3}^{2s_3} M_{-n_1}(\mathbf{\Lambda}_2\cdot\mathbf{\Lambda}_3) M_{-n_2}(\mathbf{\Lambda}_3\cdot\mathbf{\Lambda}_1)M_{-n_3}(\mathbf{\Lambda}_1\cdot\mathbf{\Lambda}_2)}
\end{aligned}
\end{equation}
}
and 
{\small
\begin{equation}\label{oppositehelicityM}
\begin{aligned}
    \boxed{ \braket{\tilde O_1^{s_1} \tilde O_2^{s_2} \tilde O_3^{s_3}}=\# \oint D \pi_{123} 
     \frac{\braket{l \pi_1}^{2s_1} \braket{l  \pi_2}^{2s_2}}{(\mathbf{\Lambda}_1\cdot\mathbf{\Lambda}_2)^{s_1+s_2+s_3+1}} \braket{l \frac{\partial}{\partial \mu_3}}^{2s_3} M_{-\tilde n_1}(\mathbf{\Lambda}_2\cdot\mathbf{\Lambda}_3) M_{-\tilde n_2}(\mathbf{\Lambda}_3\cdot\mathbf{\Lambda}_1)}\ ,
\end{aligned}
\end{equation}
}
irrespective of the sign of $n_3$, $\tilde n_1$ and $\tilde n_3$. These can be evaluated following \cite{Baumann:2024ttn} using
{\small
\begin{equation}\label{equivalent B.15}
    \begin{aligned}
        \int D\pi_3 M_{-n}(\mathbf{\Lambda}_2\cdot\mathbf{\Lambda}_3) M_{n-2}(\mathbf{\Lambda}_3\cdot\mathbf{\Lambda}_1)&\,= (-1)^{n-1} (\frac{\braket{l x_{31} \pi_1}}{\braket{\pi_2 x_{23} l}})^{n-1}\int D\pi_3 M_{-1}(\mathbf{\Lambda}_2\cdot\mathbf{\Lambda}_3) M_{-1}(\mathbf{\Lambda}_3\cdot\mathbf{\Lambda}_1)  ,
    \end{aligned}
\end{equation}
}
which is the analogue of their Eq. B.15.
\paragraph{Selection Rules}
We conclude this subsection with a comment on the relation to 4d selection rules. The flat 4D angular momentum selection rules require that the total angular momentum $ J = S + L $ satisfies the inequality 
\begin{equation}\label{selection rules}
    |j_1 - j_2| \leq J \leq j_1 + j_2\ .
\end{equation}
Interestingly, the twistor representatives appear to automatically comply with these constraints: when regularised, it ``adds'' the correct units of $L$ so that Eq. \eqref{selection rules} is satisfied. For example, the configuration $(s_1, s_2, s_3) = (0, 0, 2)$, equivalently in the negative helicity case $(n_1, n_2, n_3) = (3, 3, -1)$, is only allowed when two units of angular momentum are added, as is the case for the GR vertex $\kappa h_{\mu \nu} T^{\mu \nu}_\phi$. Using just Eq.~\eqref{samehelicity}, the correlator would be zero. However, as we show in Appendix~\ref{app:negative helicity}, regularising the result by a logarithmic factor, we obtain the correct (non-zero) correlator consistent with the selection rules. Although this gives the correct answer, the link with angular momentum remains unclear, and understanding this aspect would be an interesting direction for further investigation.

\subsection{Examples}
We now show how this formalism recovers the well-known examples listed in Section \ref{sec:embeddingspace} by simply performing the nested contour integrals using the residue theorem.
\subsubsection{$s_1=s_2=s_3=0$}
The simplest example is given by the three-point function of three conformally-coupled scalars. In this case, $s_3=0$, so we should use Eq.~\eqref{samehelicity}, which yields
\begin{equation}\label{3scalars}
\begin{aligned}
    \braket{O_1 O_2 O_3}&\,=\# \oint D \pi_{123} \frac{1} {(\mathbf{\Lambda}_1\cdot\mathbf{\Lambda}_2) (\mathbf{\Lambda}_2\cdot\mathbf{\Lambda}_3)  (\mathbf{\Lambda}_3\cdot\mathbf{\Lambda}_1)}\\
    &\,= \# \oint D \pi_{123} \frac{1} {\braket{\pi_1 x_{12} \pi_2} \braket{\pi_2 x_{23} \pi_3}\braket{\pi_3 x_{31} \pi_1}} \ ,
\end{aligned}
\end{equation}
Since the three poles are symmetric, it does not matter which pole we integrate over first, and we obtain
\begin{equation}
\begin{aligned}
    \braket{O_1 O_2 O_3}&\,= \# \oint D \pi_{12} \frac{1} {\braket{\pi_1 x_{12} \pi_2} \braket{\pi_2 x_{23} x_{31} \pi_1}}\\
    &\,=\# \oint D \pi_{1} \frac{1} {\braket{\pi_1 \tilde x_{1} \pi_1} } \ .
\end{aligned}
\end{equation}
Decomposing the pole into its roots as $\braket{\pi_1 \tilde x_{1} \pi_1}= \braket{\pi_1 \pi_+}\braket{\pi_- \pi_1}$ (cf. Eq.~\eqref{xtilde} for the definition of $\tilde x_1^{3d}$), we can integrate around one of those, say $\pi_+$, to obtain
\begin{equation}
\begin{aligned}
    \braket{O_1 O_2 O_3}
    &\,= \#\frac{1} {\braket{\pi_- \pi_1} }|_{\pi_+}\\
    &\,\propto \frac{1}{ (P_{12} P_{23} P_{31})^{1/2}} \ .
\end{aligned}
\end{equation}
Thus, we have recovered the scalar version of Eq.~\eqref{3points ex}.

\subsubsection{$s_1=s_2=0$, $s_3 \geq 1$}\label{sec:001}
We now turn on the spin of one operator. Since the correlator involves two scalars, it is still degenerate in the sense that distinguishing correlators by same/opposite helicity has no meaning. Here, taking the spinning operator to have positive or negative helicity should be a matter of convention.  Indeed, we know that the correlator, given in Eq. \eqref{3points ex}, is completely fixed by conformal symmetry. To illustrate our formalism, we will compute this correlator using both Eq. \eqref{samehelicity} (cf. Appendix \ref{app:negative helicity}) and Eq. \eqref{oppositehelicity}.
\paragraph{Positive helicity}
We start with $s_3=s=1$ with positive helicity since it is the simplest case, giving
\begin{equation}\label{001-}
    \begin{aligned}
        \braket{O_1 O_2 J_3} 
        &\,=\#  \oint D\pi_{123} \braket{l \frac{\partial}{ \partial \mu_3} }^{2} \frac{\log(\braket{\pi_2 x_{23} \pi_3}) \log(\braket{\pi_3 x_{31} \pi_1})}{\braket{\pi_1 x_{12} \pi_2}^{2}} \\
        &\,=\#  \oint D\pi_{123}  \frac{1}{\braket{\pi_1 x_{12} \pi_2}^{2}}\biggl(-\frac{\braket{l \pi_2}^2}{\braket{\pi_2 x_{23} \pi_3}^2}\log(\braket{\pi_3 x_{31} \pi_1})\\
        &\,\;\;\;\;\;\;\;\;\;\;\;\;\;\;\;\;-2  \frac{\braket{l \pi_1}\braket{l \pi_2}}{\braket{\pi_2 x_{23} \pi_3} \braket{\pi_3 x_{31} \pi_1}}-\frac{\braket{l \pi_1}^2}{\braket{\pi_3 x_{31} \pi_1}^2} \log(\braket{\pi_2 x_{23} \pi_3}) \biggl) \ .
    \end{aligned}
\end{equation}
We now write the double pole as a simple pole by using
\begin{equation}
    \frac{1}{\braket{\pi_3 x_{31} \pi_1}^{2}}= - \frac{\braket{l' \frac{\partial}{\partial \pi_3}}}{\braket{l' x_{31} \pi_1}} (\frac{1}{\braket{\pi_3 x_{31} \pi_1}}) \ , \label{eq:derivative_trick}
\end{equation}
where $l'$ is arbitrary. We will choose $l'=l$ and consider a contour that encloses the simple pole, but not the branch cut from the logarithm. Then, using Stokes' theorem, we obtain
\begin{equation}
    \begin{aligned}
        \braket{O_1 O_2 J_3}   
        &\,=\#  \oint D\pi_{123}  \frac{(\braket{l \pi_2} \braket{l x_{31}\pi_1}+ \braket{l \pi_1} \braket{\pi_2 x_{23} l})^2}{\braket{\pi_1 x_{12} \pi_2}^{2} \braket{\pi_2 x_{23} \pi_3} \braket{\pi_3 x_{31} \pi_1}\braket{\pi_2 x_{23} l} \braket{l x_{31}\pi_1}}\\
        &\,=\#  \oint D\pi_{12}  \frac{(\braket{l \pi_2} \braket{l x_{31}\pi_1}+ \braket{l \pi_1} \braket{\pi_2 x_{23} l})^2}{\braket{\pi_1 x_{12} \pi_2}^{2} \braket{\pi_2 x_{23} x_{31}  \pi_1} \braket{\pi_2 x_{23} l} \braket{l x_{31}\pi_1}}\\
        &\,= \# \frac{1}{(x_{23})^2} \oint D\pi_{1}   \frac{(\braket{l x_{23} x_{31}\pi_1} + (x_{23})^2\braket{l \pi_1})^2}{\braket{\pi_1 \tilde x_{1} \pi_1}^{2}} \ ,
    \end{aligned}
\end{equation}
where we integrated over the simple pole $\braket{\pi_3 x_{31} \pi_1}$ in the first step and over $\braket{\pi_2 x_{23} x_{31}  \pi_1}$ in the second step. Using $x_{31}= - x_{12} -x_{23}$, this further simplifies to
\begin{equation}\label{00snewway}
    \begin{aligned}
        \braket{O_1 O_2 J_3}  
        &\,= \# \frac{1}{(x_{23})^2} \oint D\pi_{1}  \frac{\braket{l x_{23} x_{12}\pi_1}^2}{\braket{\pi_1 \tilde x_{1} \pi_1}^{2}}.
    \end{aligned}
\end{equation}
As before, we decompose the pole into the two roots $\pi_{\pm}$ and we rewrite the pole as the derivative of a simple pole as in Eq.~\eqref{eq:derivative_trick}. It is now helpful to take $l'= x_{12} x_{23} l$ so that once we use Stokes' theorem, the derivative does not affect the numerator of Eq. \eqref{00snewway}. By the residue theorem, we obtain
\begin{equation}
    \begin{aligned}
        \braket{O_1 O_2 J_3}  
        &\,= \# \frac{1}{(x_{23})^2}    \frac{\braket{l x_{23} x_{12}\pi_1}^2}{\braket{\pi_+ x_{12} x_{23} l}} \braket{l x_{23} x_{12}\frac{\partial}{ \partial \pi_1}} (\frac{1}{\braket{\pi_- \pi_1}^{2}})|_{\pi_+} \ .
    \end{aligned}
\end{equation}
By definition $\braket{l' \pi_+}\braket{\pi_- l'}= \braket{l'\tilde x_1 l'}$ for any $l'$, hence the correlator simplifies to
\begin{equation}
    \begin{aligned}
        \braket{O_1 O_2 J_3}  
        &\, = \# \frac{1}{(x_{23})^2} \frac{\braket{l x_{23} x_{12} \tilde x_{1} x_{12} x_{23} l} }{\braket{\pi_- \pi_+}^3}\ ,\\
    \end{aligned}
\end{equation}
which upon using the little group version of the conformal structures in Eq. \eqref{structureslittle} gives
\begin{equation}
    \begin{aligned}
        \braket{O_1 O_2 J_3}  
        &\,  \propto \frac{P_{12}^2 V_3}{(P_{12} P_{23} P_{31})^{3/2}} \ .\\
    \end{aligned}
\end{equation}
We now consider the general spin $s$ case
\begin{equation}\label{00sgeneral}
\begin{aligned}
    \braket{O_1 O_2 O_3^s} 
        &\, =\#\oint D\pi_{123} \braket{l \frac{\partial}{ \partial \mu_3} }^{2s} \frac{\braket{\pi_2 x_{23} \pi_3}^{s-1} \braket{\pi_3 x_{31} \pi_1}^{s-1}}{\braket{\pi_1 x_{12} \pi_2}^{s+1}} \log(\braket{\pi_2 x_{23} \pi_3}) \log(\braket{\pi_3 x_{31} \pi_1}) \ .
\end{aligned}
\end{equation}
To solve this, we use the following integral
\begin{equation}\label{generalised leibniz}
    \begin{aligned}
    J_s&\,=\oint D\pi_{3} \braket{l \frac{\partial}{ \partial \mu_3} }^{2s} \braket{\pi_2 x_{23} \pi_3}^{s-1} \braket{\pi_3 x_{31} \pi_1}^{s-1} \log(\braket{\pi_2 x_{23} \pi_3}) \log(\braket{\pi_3 x_{31} \pi_1})\\ 
        &\,  \propto\oint D\pi_3  \frac{(\braket{l \pi_2} \braket{l x_{31}\pi_1}+ \braket{l \pi_1} \braket{\pi_2 x_{23} l})^{2s}}{\braket{\pi_2 x_{23} \pi_3} \braket{\pi_3 x_{31} \pi_1}\braket{\pi_2 x_{23} l}^s \braket{l x_{31}\pi_1}^s}\ ,
    \end{aligned}
\end{equation}
which is derived in Appendix \ref{ap:generalised Leibniz}. Then the general correlator is simply given by
\begin{equation}
\begin{aligned}
    \braket{O_1 O_2 O_3^s} 
        &\, =\#\oint D\pi_{123} \frac{(\braket{l \pi_2} \braket{l x_{31}\pi_1}+ \braket{l \pi_1} \braket{\pi_2 x_{23} l})^{2s}}{\braket{\pi_1 x_{12} \pi_2}^{s+1} \braket{\pi_2 x_{23} \pi_3} \braket{\pi_3 x_{31} \pi_1}\braket{\pi_2 x_{23} l}^s \braket{l x_{31}\pi_1}^s}\ .
\end{aligned}
\end{equation}
This is evaluated in the same way as in the spin 1 case, i.e., we integrate over the simple poles such that
\begin{equation}
    \begin{aligned}
        \braket{O_1 O_2 O_3^s}   &\,= \# \frac{1}{(x_{23})^{2s}} \oint D\pi_{1}   \frac{\braket{l x_{23} x_{12}\pi_1}^{2s}}{\braket{\pi_1 \tilde x_{1} \pi_1}^{s+1}}\\
        &\, = \# \frac{1}{(x_{23})^{2s}} \frac{(\braket{l x_{23} x_{12}\pi_+} \braket{\pi_- x_{12} x_{23} l})^s }{\braket{\pi_- \pi_+}^{2s+1}}\\
        &\,\propto \frac{P_{12}^{2s} V_3^s}{(P_{12}P_{23} P_{31})^{(2s+1)/2}} \ ,
    \end{aligned}
\end{equation}
where we used the generalization of Eq.~\eqref{eq:derivative_trick} 
\begin{equation}
    \frac{1}{\braket{\pi \pi_+}^n}=\frac{1}{(n-1)!\braket{\pi_+ l'}^{n-1}}\braket{l' \frac{\partial}{\partial \pi}}^{n-1}\frac{1}{\braket{\pi \pi_+}}\ .\label{eq:derivative_trick_gen_s}
\end{equation}

\subsubsection{$s_1=s_2=1$, $s_3=0$}
With two spinning particles, we have the first case where the same/opposite helicity cases could be non-degenerate a priori. However, for $s_1=s_2=1$, the correlator in Eq.~\eqref{3points ex} is completely fixed by conformal symmetry. We could compute $\braket{J_1^+ J_2^+O}$ or $\braket{J_1^+ J_2^-O}$ using Eq. \eqref{samehelicity} or Eq. \eqref{oppositehelicity} with the appropriate logarithmic factors, but it is simpler to consider the negative, equal-helicity correlator given by
\begin{equation}\label{1+1+0}
\begin{aligned}
    \braket{J_1 J_2 O_3} &\,=\# \oint D \pi_{123} \frac{\braket{l \pi_1}^{2} \braket{l \pi_2}^{2}}{\braket{\pi_1 x_{12} \pi_2}^3\braket{\pi_2 x_{23} \pi_3}\braket{\pi_3 x_{31} \pi_1}}\\
    &\,=\#\oint D \pi_{12} \frac{\braket{l \pi_1}^{2} \braket{l \pi_2}^{2} }{\braket{\pi_1 x_{12} \pi_2}^3 \braket{\pi_2 x_{23} x_{31} \pi_1}}  \\
   &\,=\#\oint D \pi_{1}  \frac{\braket{l \pi_1}^{2} \braket{l x_{23} x_{31}\pi_1}^{2}}{\braket{\pi_1 \pi_+}^3 \braket{\pi_- \pi_1}^3}\ , 
\end{aligned}
\end{equation}
where we integrated over the simple poles both times. Performing this last integral around either root, we obtain
\begin{equation}
\begin{aligned}
    \braket{J_1 J_2 O_3} 
   &\,=\#[\frac{\braket{lx_{23}x_{31}l}^2}{\braket{\pi_- \pi_+}^3}-\frac{3}{2 \braket{\pi_- \pi_+}^3} \left(\frac{(x_{23})^2(x_{31})^2}{(x_{12})^2}\braket{l x_{12}l}^2+\braket{lx_{23}x_{31}l}^2\right)]\\
   &\,\propto \frac{V_1 V_2 - H_{12}}{P_{12}^{5/2}P_{23}^{1/2}P_{31}^{1/2}}\ ,
\end{aligned}
\end{equation}
agreeing with Eq. \eqref{3points ex}. Above, we used the identity
\begin{equation}
    \begin{aligned}
        \braket{l x_{23} x_{31}l}^2 &\,=(\braket{l x_{12} l}^2\frac{(x_{23})^2(x_{31})^2}{(x_{12})^2}-\frac{\braket{l \tilde x_1 l} \braket{l \tilde x_2 l}}{(x_{12})^2}),\\
        &\,= -2\frac{(x_{23})^2(x_{31})^2}{(x_{12})^2}(H_{12}+2V_1V_2) \ .
    \end{aligned}
\end{equation}

\subsubsection{$s_1=s_2=s_3$ with Equal Helicities}
We now compute the equal-spin correlator and show that Eqs. \eqref{samehelicity} and \eqref{oppositehelicity} yield genuinely different results. For spin-1 and spin-2, this non-standard helicity-based dichotomy aligns precisely with the distinct bulk interactions: same-helicity correlators emerge from next-to-leading-order vertices ($F^3$, $W^3$) whereas opposite-helicity correlators arise from the leading Yang–Mills and Einstein interactions (YM, GR) \cite{Caron-Huot:2021kjy}, just as the scattering amplitudes in flat space\footnote{In the flat space case, one can use dimensional analysis and spinor-helicity methods to easily find the correspondence between different interactions and the helicities of the scattered states. In curved spacetime, it is important to note that when using the standard momentum-space definition of helicity, the split into negative- and positive-helicity components does not align with the leading- and next-to-leading-order interaction hierarchy as it does in flat space. For example, the Yang–Mills vertex includes an opposite-helicity contribution that vanishes only in the flat-space limit \cite{Baumann:2020dch}.} \cite{Elvang:2015rqa}.\\
We start with the same helicity/next-to-leading-order case where the correlator takes the form
\begin{equation}\label{+++}
    \begin{aligned}
        \braket{O^s_1 O^s_2 O^s_3}&\,=\#\oint D\pi_{123} \frac{\braket{l\pi_1}^{2s} \braket{l\pi_2}^{2s} \braket{l\pi_3}^{2s}}{\braket{\pi_1 x_{12}\pi_2}^{s+1} \braket{\pi_2 x_{23}\pi_3}^{s+1} \braket{\pi_3 x_{31}\pi_1}^{s+1}}\ .
    \end{aligned}
\end{equation}
We first integrate around the $\braket{\pi_3 x_{31}\pi_1}^{s+1}$ pole, giving 
\begin{equation}\label{3pointsequal spin}
    \begin{aligned}
        \braket{O^s_1 O^s_2 O^s_3}
        &\,=\#\oint D\pi_{12} \frac{\braket{l\pi_1}^{2s}\braket{l\pi_2}^{2s}\braket{lx_{31}\pi_1}^s \braket{\pi_2 x_{23} l}^s}{\braket{\pi_1 x_{12}\pi_2}^{s+1} \braket{\pi_2 x_{23}x_{31}\pi_1}^{2s+1}}\\
        &\,=\# \oint D\pi_{1}  \braket{l\pi_1}^{2s} \braket{lx_{12}\pi_1}^s\braket{lx_{31}\pi_1}^s \braket{l \frac{\partial}{\partial \pi_2}}^s(\frac{\braket{\pi_2 x_{23} l}^s}{\braket{\pi_2 x_{23}x_{31}\pi_1}^{2s+1}})|_{x_{12}\pi_1} \ ,
    \end{aligned}
\end{equation}
where we have assumed that $s$ is an integer. It would be interesting to understand how to deal with fermions, but for bosons, one just needs to expand this last derivative, which we do now for $s=1$ and $s=2.$
\paragraph{Conserved Currents Correlator} 
In the case of s = 1, we obtain
\begin{equation}\label{pi1JJJ}
    \begin{aligned}
        \braket{J_1 J_2 J_3}
        &\,=\#\oint D\pi_{1} \braket{l\pi_1}^2 \braket{lx_{31}\pi_1}\braket{lx_{12}\pi_1} (\frac{\braket{l x_{23}l}}{\braket{\pi_1 \tilde x_1 \pi_1}^3}-3\frac{\braket{l x_{23}x_{31} \pi_1} \braket{\pi_1 x_{12} x_{23} l}}{\braket{\pi_1 \tilde x_1 \pi_1}^4})\ .
    \end{aligned}
\end{equation}
Although this is not immediately obvious from the expression, the second term in Eq.~\eqref{pi1JJJ} is not merely a fourth-order pole: it is the sum of a fourth-order pole and a third-order pole. Therefore, applying the residue theorem naively would lead to a wrong answer. Seeing that the numerator contains factors of $\braket{\pi_1 \tilde x_1 \pi_1}$ can be done by Fierzing
\begin{equation}\label{fierz 3points}
    \braket{l x_{23}x_{31}\pi_1}\braket{\pi_1 x_{12}x_{23}l}=\braket{lx_{23}l}\braket{\pi_1 \tilde x_1 \pi_1}- (x_{23})^2\braket{\pi_1 x_{12}l}\braket{l x_{31} \pi_1}\ .
\end{equation}
By using this and again Eq.~\eqref{eq:derivative_trick_gen_s}, we can simplify the integral to
\begin{equation}\label{JJJ integrals}
    \begin{aligned}
        \braket{J_1J_2J_3}&\,=\# \oint \braket{\pi d \pi} \braket{l \pi}^2 \braket{l x_{12} \pi} \braket{l x_{31}\pi}(  -2\frac{ \braket{l x_{23}l}}{\braket{\pi \tilde x_1 \pi}^3} +3 (x_{23})^2 \frac{\braket{l x_{12} \pi} \braket{l x_{31}\pi}}{\braket{\pi \tilde x_1 \pi}^4})\ .
    \end{aligned}
\end{equation}
One could evaluate these integrals as we did before, by residues and simplification. However, there is a faster route which we will now explain. Restricting to the parity-even sector, the current correlator can be built from the three structures of Eq. \eqref{structureslittle}, with $P$ of homogeneity $-\Delta$ and $H_{ij}, V_{i}$ of homogeneity $s$. For $s=1, \Delta=2$,  we arrive at the four-dimensional basis
\begin{equation}\label{basis 1}
 \mathcal{B}_{s=1, \Delta=2}=\frac{1}{(-P_{12}P_{23} P_{31})^{3/2}} \{V_1 H_{23}, V_2 H_{31}, V_3 H_{12}, V_1 V_2 V_3\} \ .    
\end{equation}
Note that by integrating over $\pi_3$ and $\pi_2$, we broke the permutation symmetry between $\{1\}$ and $\{2,3\}$ and hence we don't consider a basis with permutation invariant elements\footnote{For colour-ordered correlators as those coming from Yang-Mills interactions, the correlators are invariant under cyclic permutations. Therefore, doing this at two rather than four random points suffices as well for this computation.}, so we should have
\begin{equation}
    \begin{aligned}
        \braket{J_1 J_2 J_3}&\,= \frac{\alpha_1 V_1 H_{23}+\alpha_2 V_2 H_{31}+ \alpha_3 V_3 H_{12}+\alpha_4 V_1 V_2 V_3}{(-8P_{12}P_{23} P_{31})^{3/2}}\ .\\
    \end{aligned}
\end{equation}
We can now compute Eq. \eqref{JJJ integrals} numerically at four random, distinct (spacelike-separated) points $x_{i,12}, x_{i,23}$ to obtain the coefficients $\alpha_i, \beta_i$. Doing this for various random choices, we found
\begin{equation}\label{JJJF3}
    \braket{J_1J_2J_3} \propto \frac{V_1 H_{23}+ V_2 H_{31}+  V_3 H_{12}+5 V_1 V_2 V_3}{(P_{12}P_{23} P_{31})^{3/2}} \ ,
\end{equation}
as expected for the 3-point function coming from an $F^3$ bulk interaction. Note that during the numerical evaluations, any non-zero $l^a$ can be taken since it is arbitrary (which we checked explicitly by varying the polarisation spinor).

\paragraph{Conserved Stress-Energy Tensor Correlator}
By substituting again Eq. \eqref{fierz 3points} in Eq. \eqref{3pointsequal spin} with $s=2$, we  obtain
\begin{equation}
    \begin{aligned}
        \braket{T_1 T_2 T_3}
        &\,=\#\oint D\pi_{1} \braket{l\pi_1}^4 \braket{lx_{31}\pi_1}^2\braket{lx_{12}\pi_1}^2 (12\frac{ \braket{l x_{12}l}^2}{ \braket{\pi_1 \tilde x_1 \pi_1}^5}\;+\\
        &\,40 (x_{23})^2\frac{\braket{l x_{23}l} \braket{lx_{31}\pi_1}\braket{lx_{12}\pi_1} }{\braket{\pi_1 \tilde x_1 \pi_1}^6} +30 (x_{23})^4\frac{ \braket{lx_{31}\pi_1}^2\braket{lx_{12}\pi_1}^2 }{\braket{\pi_1 \tilde x_1 \pi_1}^7}) \ .
    \end{aligned}
\end{equation}
We evaluate this integral exactly as we did with the spin $1$ case by listing all the terms consistent with $s=2, \Delta=3$. Since the operators have identical spin and dimension, the final result should be invariant under all permutations, but the integral over $\pi_2$ and  $\pi_3$ broke this symmetry, which leads to the four-dimensional basis
\begin{equation}\label{basis 2}
 \mathcal{B}_{s=2, \Delta=3}=\frac{1}{(-P_{12}P_{23} P_{31})^{5/2}} \{V_{(i}^2 H_{jk)}^2, V_{(i} V_j H_{jk}H_{ki)}, H_{(ij}V_k^2 V_i V_{j)}, V_i^2 V_j^2 V_k^2\} \ .    
\end{equation}
Naively, one should include $\frac{H_{12}H_{23}H_{31}}{(-P_{12}P_{23} P_{31})^{5/2}}$ which is also a consistent term. However in $d=3$, this would make our basis over-complete because of the constraint of Eq. \eqref{constraint embedding}. Evaluating the integral at various random sets of four space-like separated points, we find
\begin{equation}\label{TTTW}
     \braket{T_1T_2T_3} \propto \frac{(-2V_1^2 H_{23}^2+16 V_2 V_3 H_{31}H_{12}+52 H_{23}V_1^2 V_2 V_3+49V_1^2 V_2^2 V_3^2)+ \text{cyclic}}{(P_{12}P_{23} P_{31})^{5/2}}\ ,
\end{equation}
which is associated to the $W^3$ bulk interaction. 

\subsubsection{$s_1=s_2=s_3$ with Opposite Helicities}
We now analyze the equal-spin correlator in the opposite-helicity configuration, taking the operator at $P_3$ to have positive helicity and those at $P_1, P_2$ negative helicity, following Eq. \eqref{oppositehelicity}. As expected (see Ref.~\cite{Caron-Huot:2021kjy}), this yields a distinct result tied to the leading bulk interaction for the $s=1$ and $s=2$ cases, which we exhibit explicitly. We have
\small \begin{equation}\label{++-}
\begin{aligned}
    &\,\braket{\tilde O_1^{s} \tilde O_2^{s} \tilde O_3^{s}}=\\
   \#&\,\oint D\pi_{123}\frac{\braket{l \pi_1}^{2s} \braket{l  \pi_2}^{2s} }{\braket{\pi_1 x_{12} \pi_2}^{3s+1}} 
     \braket{l \frac{\partial}{\partial \mu_3}}^{2s}
    [\braket{\pi_2 x_{23} \pi_3}^{s-1}\log(\pi_2 x_{23} \pi_3) \braket{\pi_3 x_{31} \pi_1}^{s-1}\log(\pi_3x_{31}\pi_1)] \ .
\end{aligned}
\end{equation}
\normalsize
The same $2s$ derivatives appeared in the $\braket{O_1O_2O^s_3}$ correlator of Eq. \eqref{00sgeneral} and was evaluated in Appendix \ref{ap:generalised Leibniz}, so we obtain
\begin{equation}
    \begin{aligned}
        \braket{\tilde O_1^{s} \tilde O_2^{s} \tilde O_3^{s}}   
        &\,=\# \oint D\pi_{123}  \frac{\braket{l \pi_1}^{2s}\braket{l \pi_2}^{2s}(\braket{l \pi_2} \braket{l x_{31}\pi_1}+ \braket{l \pi_1} \braket{\pi_2 x_{23} l})^{2s}}{\braket{\pi_1 x_{12} \pi_2}^{3s+1} \braket{\pi_2 x_{23} \pi_3} \braket{\pi_3 x_{31} \pi_1}\braket{\pi_2 x_{23} l}^s \braket{l x_{31}\pi_1}^s}\ .
    \end{aligned}
\end{equation}
Again, we integrate over the simple poles
\begin{equation}
    \begin{aligned}
        \braket{\tilde O_1^{s} \tilde O_2^{s} \tilde O_3^{s}}   
        &\,=\# \oint D\pi_{12}  \frac{\braket{l \pi_1}^{2s}\braket{l \pi_2}^{2s}(\braket{l \pi_2} \braket{l x_{31}\pi_1}+ \braket{l \pi_1} \braket{\pi_2 x_{23} l})^{2s}}{\braket{\pi_1 x_{12} \pi_2}^{3s+1} \braket{\pi_2 x_{23} x_{31} \pi_1}\braket{\pi_2 x_{23} l}^s \braket{l x_{31}\pi_1}^s}\\
        &\,= \# \frac{1}{(x_{23}^2)^{s}}\oint D\pi_{1}  \frac{\braket{l \pi_1}^{2s}  \braket{l x_{23} x_{31}\pi_1}^{2s} (\braket{l x_{23} x_{31} \pi_1} +x_{23}^2 \braket{l \pi_1})^{2s}}{\braket{\pi_1 \tilde x_{1} \pi_1}^{3s+1}}\ .
    \end{aligned}
\end{equation}
As in Eq. \eqref{00snewway}, we can simplify the integrand further by using $x_{31}=x_{32}+x_{21}$, so that
\begin{equation}
    \begin{aligned}
        \braket{\tilde O_1^{s} \tilde O_2^{s} \tilde O_3^{s}}   
        &\,= \# \frac{1}{(x_{23}^2)^{s}}\oint D\pi_{1}  \frac{\braket{l \pi_1}^{2s}  \braket{l x_{23} x_{31}\pi_1}^{2s}  \braket{l x_{23} x_{12}\pi_1}^{2s}}{\braket{\pi_1 \tilde x_{1} \pi_1}^{3s+1}}\ .\\
    \end{aligned}
\end{equation}
We evaluate this integral in the same way as for Eq. \eqref{JJJF3} and Eq. \eqref{TTTW} for spin one and two. Since we haven't changed the dimensions of the operators, we expand the correlators with the same bases $\mathcal{B}_{s=1,\Delta=2}$ and $\mathcal{B}_{s=2,\Delta=3}$ and match the coefficients to obtain this time
\begin{equation}
    \begin{aligned}
        \braket{\tilde J_1 \tilde J_2 \tilde J_3}
        &\,\propto  \frac{V_1 H_{23}+ V_2 H_{31}+  V_3 H_{12}+ V_1 V_2 V_3}{(P_{12}P_{23} P_{31})^{3/2}}\ ,\\
        \braket{\tilde T_1 \tilde T_2 \tilde T_3} &\,\propto \frac{(6V_1^2 H_{23}^2+16 V_2 V_3 H_{31}H_{12}+4 H_{23}V_1^2 V_2 V_3-3V_1^2 V_2^2 V_3^2)+ \text{cyclic}}{(P_{12}P_{23} P_{31})^{5/2}} \ ,
    \end{aligned}
\end{equation}
which are indeed the three-point functions corresponding to a YM and a GR bulk interaction, respectively.

\subsection{Ward-Takahashi Identity} 
It is known that the Ward-Takahashi identity for the YM three points should obey the non-trivial relation \cite{Baumann:2020dch}
\begin{equation}
    \begin{aligned}
          \nabla^{x_3}_\mu \braket{\tilde J_1 \tilde J_2 \tilde J^\mu_3}\propto \delta^3(x_3 -x_1) \braket{(\delta \tilde J_1) \tilde J_2}+\delta^3(x_3 -x_2) \braket{\tilde J_1 (\delta \tilde J_2)}\ .
    \end{aligned}
\end{equation}
Let us check if this can be seen from our formalism\footnote{ We thank Guilherme Pimentel for suggesting this calculation.}. In particular, we need to obtain the regularized two-point correlator from Section~\ref{sec:unitaritybound} on the RHS to have a non-zero result. To derive this, we integrate over $x_3$ and apply the divergence theorem to a sphere at $x_2$ (excluding $x_1$), and check whether
\begin{equation}
      \int dS^{ab}_{x_3} \oint D\pi_1 D\pi_2 D\pi_3  \frac{\braket{l\pi_1}^2 \braket{l\pi_2}^2}{(\mathbf{\Lambda}_1\cdot\mathbf{\Lambda}_2)^{4}}\frac{\partial}{\partial \mu^{a}_3} \frac{\partial}{\partial \mu^{b}_3}  (M_0 (\mathbf{\Lambda}_3\cdot\mathbf{\Lambda}_1)M_0 (\mathbf{\Lambda}_1\cdot\mathbf{\Lambda}_2)) \propto \braket{\tilde J_1 \tilde J_2} \ ,
\end{equation}
using the representation of Eq. \eqref{oppositehelicityM}. We consider the a sphere with unit normal vector $n^{ab}=\frac{x^{ab}_{23}}{\epsilon}$ where $\epsilon$ is the radius, such that the surface element is $dS^{ab}_{x_3}=\epsilon^2 n^{ab}d\Omega$. Then using that
\begin{equation}
    \begin{aligned}
        &\,\braket{\frac{\partial}{\partial \mu_3} x_{23} \frac{\partial}{\partial \mu_3} } (M_0(\mathbf{\Lambda}_3\cdot\mathbf{\Lambda}_1) M_0(\mathbf{\Lambda}_2\cdot\mathbf{\Lambda}_3))= \braket{\pi_1 x_{23} \pi_1} M_{-2}(\mathbf{\Lambda}_3\cdot\mathbf{\Lambda}_1) M_0(\mathbf{\Lambda}_2\cdot\mathbf{\Lambda}_3)\\
        &\,-2 \braket{\pi_1 x_{23} \pi_2} M_{-1}(\mathbf{\Lambda}_3\cdot\mathbf{\Lambda}_1) M_{-1}(\mathbf{\Lambda}_2\cdot\mathbf{\Lambda}_3)+ \braket{\pi_2 x_{23} \pi_2} M_0(\mathbf{\Lambda}_3\cdot\mathbf{\Lambda}_1) M_{-2}(\mathbf{\Lambda}_2\cdot\mathbf{\Lambda}_3) \ ,
    \end{aligned}
\end{equation}
we obtain that the $\pi_3$ integral evaluates to
\begin{align}
     \oint D\pi_3 \braket{\frac{\partial}{\partial \mu_3} x_{23} \frac{\partial}{\partial \mu_3} }   (M_0 (\mathbf{\Lambda}_3\cdot\mathbf{\Lambda}_1)M_0 (\mathbf{\Lambda}_1\cdot\mathbf{\Lambda}_2&))=\frac{\epsilon}{\braket{\pi_2 x_{23} x_{31} \pi_1}} \Bigg(-\braket{\pi_1 x_{23} \pi_1} \frac{\braket{\pi_2 x_{23} p}}{\braket{p x_{31} \pi_1}}  \nonumber \\
    &-2 \braket{\pi_1 x_{23} \pi_2} -\braket{\pi_2 x_{23} \pi_2} \frac{\braket{p x_{31} \pi_1}}{\braket{\pi_2 x_{23} p}}\Bigg) \ ,
\end{align}
where we used Eq. \eqref{equivalent B.15} and where $p$ is an arbitrary spinor. The first two terms scale as $\epsilon^2$ and $\epsilon$ respectively, so they will vanish in the limit $\epsilon\rightarrow 0$. Thus, it suffices to evaluate the last one. Now, we exchange the integration order between the sphere and the contour integrals to obtain the twistor space representation of the two-point correlator.  Taking $p=\pi_2$, we obtain
\begin{equation}
\oint D\pi_1 D\pi_2\int d\Omega\frac{\braket{l\pi_1}^2 \braket{l\pi_2}^2}{(\mathbf{\Lambda}_1\cdot\mathbf{\Lambda}_2)^{3}\braket{\pi_2 n x_{21} \pi_1}}= \# \oint D\pi_1 D\pi_2 \frac{\braket{l\pi_1}^2 \braket{l\pi_2}^2}{(\mathbf{\Lambda}_1\cdot\mathbf{\Lambda}_2)^{4}}\ .
\end{equation}
Note that to obtain this result, the sphere integral needs to be regularized. We take
\be
 \int d\Omega\frac{1}{\braket{\pi_2 n x_{21} \pi_1}}= \lim_{\epsilon\rightarrow 0} \int d\Omega\frac{1}{n\cdot q+ i\epsilon}=-i\frac{2\pi^2}{|\braket{\pi_2 x_{21} \pi_1}|}
\ee
where $q^{ab}=\pi_2^{(a}(x_{21} \pi_1)^{b)}$ is a constant vector from the point of view of the sphere integral. For $q$ complex, the integral vanishes, but choosing reality conditions such that $\pi_i$ and hence $q$ are real (i.e. Lorentzian AdS) we can use the Sokhotski–Plemelj theorem applied to the real line. Then, the principal value of the integral vanishes, but we obtain a non-zero contribution from the delta function. Thus, we have reached the integral for for the two-point twistor representation without regularization. \\
To obtain the regularized version we note that we shouldn't exchange integration order since Fubini's theorem isn't satisfied, similar to the construction of the regularized version described in Section~\ref{sec:unitaritybound}. Performing the $\pi_2$ contour integral first we get
\begin{equation}
     \int d\Omega \oint D\pi_1 D\pi_2\frac{\braket{l\pi_1}^2 \braket{l\pi_2}^2}{(\mathbf{\Lambda}_1\cdot\mathbf{\Lambda}_2)^{3}\braket{\pi_2 n x_{21} \pi_1}}= \frac{1}{(x_{12}^2)^3}\int d\Omega\oint D\pi_2\frac{\braket{lx_{12} n\pi_2}^2 \braket{l\pi_2}^2}{\braket{\pi_2 n \pi_2}^{3}}\ .
\end{equation}
Note in particular the similarity with the regularisation that was found above, where the two-points was also given by a contour integral nested in a sphere integral. This last integral is finally evaluated to
\begin{equation}
    -\frac{8\pi}{3} \frac{\braket{lx_{12}l}^2}{(x_{12}^2)^3}\ ,
\end{equation}
where we Fierzed the factor $\braket{lx_{12} n\pi_2} \braket{l\pi_2}$ and evaluated each contribution separately using the residue theorem. This is indeed proportional to the conserved two-point function. \\

As another example, we can apply the same reasoning to $\braket{OOJ}$, which also satisfies a non-trivial Ward identity. For simplicity, we consider the negative helicity representation of Eq. \eqref{twistorlog} which has only one logarithmic factor. We consider again a sphere  $S^2_{x^3}$ centered at $x_{2}$ excluding $x_1$. After applying the divergence, we have that
\begin{equation}
    \int dx_3 \nabla_\mu^{x_3} \braket{O_1 O_2 J_3} = \# \int d\Omega \oint D \pi_{123} \frac{\braket{\pi_3 n \pi_3}}{(\mathbf{\Lambda}_3 \cdot \mathbf{\Lambda}_1) \braket{\pi_2 n \pi_3}^2 (\mathbf{\Lambda}_1 \cdot \mathbf{\Lambda}_2)} \frac{\braket{l x_{12} \pi_2}}{\braket{l x_{13} \pi_3}}\ ,
\end{equation}
where $l^a$ is an arbitrary spinor. Since $x_{13}= x_{12} + \epsilon n$, we obtain
\begin{equation}
\begin{aligned}
    \int dx_3 \nabla_\mu^{x_3} \braket{O_1 O_2 J_3} &\,= \# \frac{1}{x_{12}^2}\int d\Omega \oint D \pi_{23} \frac{\braket{\pi_3 n \pi_3}}{\braket{\pi_2 n \pi_3}^2 \braket{\pi_3 \pi_2}} \frac{\braket{l x_{12} \pi_2}}{\braket{l x_{12} \pi_3}} \\
     &\,= \# \frac{1}{x_{12}^2}\int d\Omega \oint D \pi_{3} \frac{1}{\braket{\pi_3 n \pi_3}}  \\
     &\,= \# \frac{1}{x_{12}^2}\ ,
\end{aligned}
\end{equation}
which is proportional to the scalar two-point function with $\Delta=1$ as expected. One can compare this to $\braket{JJO}$, which is a homogeneous solution of the Ward identity. In this case, performing the same argument, we see that the divergence of the correlator scales as $\epsilon$ and therefore vanishes automatically. It is very interesting to note that in twistor space, the representatives with logarithmic terms, initially introduced for regularisation, turn out to be in 1-to-1 correspondence with the correlators satisfying a non-trivial Ward identity.
\subsection{Double Copy}
It is not immediately clear how the correlator $\langle \tilde T_1 ,\tilde T_2 ,\tilde T_3 \rangle$ relates to $\langle \tilde J_1 ,\tilde J_2 ,\tilde J_3 \rangle$, regardless of whether one works in position space, embedding space, or momentum space. However, as pointed out in \cite{Baumann:2024ttn}, the twistor formalism offers a natural framework in which such a relation becomes manifest. By rewriting their distributional representatives (involving derivatives of delta functions) as integrals over Schwinger parameters, they observed that squaring these variables connects correlators of equal spin. The existence of such relations at the level of CFT three–point functions was previously observed in different scenarios, see for instance \cite{Farrow:2018yni,Lipstein:2019mpu,Bzowski:2017poo, Albayrak:2020fyp, Jain:2021qcl, Alday:2021odx, Lee:2022fgr}. \\
Upon complexifying the domain, all distributions become rational functions with simple poles, as seen in the examples above. In this setting, the double copy can be carried out directly on the representatives by multiplying them and dividing by the scalar kernel. This kernel is simply the representative for the conformally coupled scalar in the alternate quantization. For example, the next–to–leading order \(s=1\) correlator, divided by the scalar correlator, yields the next–to–leading order \(s=2\) correlator
\begin{equation}
  \frac{
    \Bigl[\frac{1}{(\boldsymbol{\Lambda}_1\!\cdot\!\boldsymbol{\Lambda}_2)^2\,
                   (\boldsymbol{\Lambda}_2\!\cdot\!\boldsymbol{\Lambda}_3)^2\,
                   (\boldsymbol{\Lambda}_3\!\cdot\!\boldsymbol{\Lambda}_1)^2}
    \Bigr]^2
  }{
    \frac{1}{
      (\boldsymbol{\Lambda}_1\!\cdot\!\boldsymbol{\Lambda}_2)\,
      (\boldsymbol{\Lambda}_2\!\cdot\!\boldsymbol{\Lambda}_3)\,
      (\boldsymbol{\Lambda}_3\!\cdot\!\boldsymbol{\Lambda}_1)
    }
  }
  =
  \frac{1}{
    (\boldsymbol{\Lambda}_1\!\cdot\!\boldsymbol{\Lambda}_2)^3\,
    (\boldsymbol{\Lambda}_2\!\cdot\!\boldsymbol{\Lambda}_3)^3\,
    (\boldsymbol{\Lambda}_3\!\cdot\!\boldsymbol{\Lambda}_1)^3
  }\ ,
\end{equation}
which can be written as
\be
\langle 1^-1^-1^-\rangle \;\otimes_{0}\;\langle 1^-1^-1^-\rangle
= \langle 2^-2^-2^-\rangle\ .
\ee
Here \(\otimes_0\) denotes pointwise multiplication of twistor–space representatives, followed by division by the scalar correlator. This aligns with the twistor double copy for classical solutions \cite{White:2020sfn,White:2024pve}. One should note that logarithmic regularization factors should be omitted in the double copy. For example, the leading–order double copy between Yang–Mills and gravity is
\begin{equation}
  \frac{
    \bigl[\tfrac{1}{(\boldsymbol{\Lambda}_1\!\cdot\!\boldsymbol{\Lambda}_2)^4}\bigr]^2
  }{
    \tfrac{1}{
      (\boldsymbol{\Lambda}_1\!\cdot\!\boldsymbol{\Lambda}_2)\,
      (\boldsymbol{\Lambda}_2\!\cdot\!\boldsymbol{\Lambda}_3)\,
      (\boldsymbol{\Lambda}_3\!\cdot\!\boldsymbol{\Lambda}_1)
    }
  }
  =
  \frac{
    (\boldsymbol{\Lambda}_2\!\cdot\!\boldsymbol{\Lambda}_3)\,
    (\boldsymbol{\Lambda}_3\!\cdot\!\boldsymbol{\Lambda}_1)
  }{
    (\boldsymbol{\Lambda}_1\!\cdot\!\boldsymbol{\Lambda}_2)^7
  }\ ,
\end{equation}
that is, 
\be
\langle 1^-1^-1^+ \rangle \;\otimes_{0}\;\langle 1^-1^-1^+ \rangle
= \langle 2^-2^-2^+ \rangle\ .
\ee
Note that it is crucial to align the correct helicities together, i.e. $\langle 1^-1^-1^+ \rangle \;\otimes_{0}\;\langle 1^-1^+1^- \rangle
\neq \langle 2^-2^-2^+ \rangle$, which is the same as for amplitudes. Both cases generalize to arbitrary spin, as can be verified using the general representatives in Eqs.~\eqref{+++} and \eqref{++-}. Our notation also suggests a straightforward extension to mixed–spin correlators and non–identical single copies. For instance:
\[
\langle J_1 J_2 O_3 \rangle \;\otimes_0\; \langle O_1 O_2 J_3 \rangle
=
\begin{cases}
\langle J_1 J_2 J_3 \rangle, & J_3 \text{ aligned with } J_1,J_2\ ,\\
\langle \tilde J_1 \tilde J_2 \tilde J_3 \rangle, & J_3 \text{ anti–aligned with } J_1,J_2\ .
\end{cases}
\]
So, in the aligned case we have
\begin{equation}
    \langle 1^-1^-0 \rangle \;\otimes_{0}\;\langle 001^- \rangle
= \langle 1^-1^-1^- \rangle \ ,
\end{equation}
or explicitly 
\begin{equation}
  \frac{
    \bigl[\tfrac{1}{(\boldsymbol{\Lambda}_1\!\cdot\!\boldsymbol{\Lambda}_2)^3 (\boldsymbol{\Lambda}_2\!\cdot\!\boldsymbol{\Lambda}_3)(\boldsymbol{\Lambda}_3\!\cdot\!\boldsymbol{\Lambda}_1)}\bigr] \bigl[\tfrac{1}{(\boldsymbol{\Lambda}_2\!\cdot\!\boldsymbol{\Lambda}_3)^2(\boldsymbol{\Lambda}_3\!\cdot\!\boldsymbol{\Lambda}_1)^2}\bigr]
  }{
    \tfrac{1}{
      (\boldsymbol{\Lambda}_1\!\cdot\!\boldsymbol{\Lambda}_2)\,
      (\boldsymbol{\Lambda}_2\!\cdot\!\boldsymbol{\Lambda}_3)\,
      (\boldsymbol{\Lambda}_3\!\cdot\!\boldsymbol{\Lambda}_1)
    }
  }
  = \frac{
    1
  }{
    (\boldsymbol{\Lambda}_1\!\cdot\!\boldsymbol{\Lambda}_2)^2(\boldsymbol{\Lambda}_2\!\cdot\!\boldsymbol{\Lambda}_3)^2\,
    (\boldsymbol{\Lambda}_3\!\cdot\!\boldsymbol{\Lambda}_1)^2
  }
  \ ,
\end{equation}
where we used the representatives of Eq. \eqref{1+1+0}, Eq. \eqref{twistorlog} and Eq. \eqref{+++}. In the anti-aligned case, we have 
\begin{equation}
    \langle 1^-1^-0 \rangle \;\otimes_{0}\;\langle 001^+ \rangle
= \langle 1^-1^-1^+ \rangle \ ,
\end{equation}
which is explicitly
\begin{equation}
  \frac{
    \bigl[\tfrac{1}{(\boldsymbol{\Lambda}_1\!\cdot\!\boldsymbol{\Lambda}_2)^3 (\boldsymbol{\Lambda}_2\!\cdot\!\boldsymbol{\Lambda}_3)(\boldsymbol{\Lambda}_3\!\cdot\!\boldsymbol{\Lambda}_1)}\bigr] \bigl[\tfrac{1}{(\boldsymbol{\Lambda}_1\!\cdot\!\boldsymbol{\Lambda}_2)^2}\bigr]
  }{
    \tfrac{1}{
      (\boldsymbol{\Lambda}_1\!\cdot\!\boldsymbol{\Lambda}_2)\,
      (\boldsymbol{\Lambda}_2\!\cdot\!\boldsymbol{\Lambda}_3)\,
      (\boldsymbol{\Lambda}_3\!\cdot\!\boldsymbol{\Lambda}_1)
    }
  }
  = \frac{
    1
  }{
    (\boldsymbol{\Lambda}_1\!\cdot\!\boldsymbol{\Lambda}_2)^4
  }\ ,
\end{equation}
using Eq. \eqref{1+1+0}, Eq. \eqref{00sgeneral} and Eq. \eqref{++-}.
Hence, while such relations are obscure in embedding space, they become manifest in twistor space. 

\section{Conclusions and Open Questions} 

The formalism presented here provides a complete description of all regularised 2- and 3-points, even, conserved correlators in 3d CFT. It rests on the combination of two ideas. On the one hand, the Penrose transform is used to trivialize the constraint imposed by conservation; while on the other hand, the kinematics is trivialised by considering only inner products of twistors and dual twistors (equivalent on the boundary), giving rise to a nested Penrose transform. In contrast to the generic single Penrose transform, the representatives can then be written without the use of any external input, which usually fixes the specific form of the spacetime field. Naively, the space of possible representatives is infinite since any projectively well-defined representative can be written. We found, however, that restricting the representatives to purely rational functions with or without a logarithmic factor not only suffices to cover all possible even boundary correlators, but are in 1-to-1 correspondence with distinct bulk interactions. Indeed, when more than one solution is allowed by conformal symmetry and conservation, the representative with the logarithm gives the leading order bulk interaction, and the purely rational representative gives the next-to-leading order. Therefore, the type of interaction giving rise to a correlator is given by fixing a basis of representatives in twistor space, which in our examples is aligned with the helicity basis given in \cite{Caron-Huot:2021kjy}. Even though the logarithm was initially introduced for regularisation purposes, its presence also turns out to be equivalent to a non-trivial solution of the Ward identity. Finally, the twistor representation also puts all spins on the same footing, as the representatives do not become more involved with increasing spin: they remain a single term whose exponent is controlled by the configuration of spin. \\
For the two-point correlators, we derived this formalism from the point of view of 4d twistors, which describe self-dual and anti-self-dual fields in the bulk. 4d twistors and dual twistors are not equivalent anymore, which gives a natural origin to the self-dual/self-dual propagators in curved spacetime in addition to the usual self-dual/anti-self-dual propagators that are also present in flat space. This was done in the usual complex setting of twistors which streamlined the computation. It also enabled us to obtain a fully regularised construction. Interestingly, this revealed a mixed Čech–Dolbeault representation as the most natural formulation. Taking the boundary limit, the same formalism yielded a twistor origin of conformal field theory correlators. We verified that our nested representatives are indeed holomorphic in the conserved case, as expected from twistor theory. On the boundary, the formalism also extends to non-conserved operators, in agreement with the results of \cite{Bala:2025qxr} (modulo regularisation). These results place the twistor-like structures identified in \cite{Baumann:2024ttn} within a broader framework rooted in the geometry of (A)dS twistors and their boundary limits.

Looking ahead, several open questions remain. A more systematic understanding of the regularisation procedures we employed would be highly valuable. In the case of the two-point function, the mixed Čech–Dolbeault representative we used was the simplest natural choice, but required us to work in Euclidean signature. For instance, we could have split the pole in a more complex ways than by just isolating a simple pole. For the three-point function, by contrast, no choice of reality condition was necessary, but regularisation required the introduction of branch cuts on the Riemann spheres. In the latter case, it would be worthwhile to understand more precisely the connection with angular momentum that emerged in the context of the selection rules. Although the use of branch cuts can be interpreted as an analytic continuation of the results in \cite{Baumann:2024ttn}, it would be interesting to uncover a more intrinsic or geometrical interpretation of both regularizations. The Ward identity, in particular, provides a map between the two approaches, leading to closely related integrals, both of which involve integration over a sphere.

We have here primarily focused on bosonic fields for simplicity. Fermionic representatives typically introduce square roots in the denominators, altering the analytic structure of the correlators, and this deserves further investigation. Another natural next step is to extend our construction of non-conserved boundary-to-boundary propagators to bulk-to-bulk propagators for massive fields, allowing Witten diagrams to be fully formulated within this twistor framework. This would be particularly valuable, given the well-known complexity of Witten diagrams in both position and momentum space.

Furthermore, it would be desirable to also express higher-point correlators in this formalism. The four-point case, in particular, is expected to reveal new structural features: beyond pairwise contractions, one can now contract four bulk twistors using the fully antisymmetric invariant tensor $\epsilon_{ABCD}$, suggesting richer geometric content. While twistors are most naturally defined in four dimensions, it would be very interesting to explore whether this framework can be extended to higher dimensions using ambitwistor space. This could lead to a simple and neat connection between bulk and boundary correlator since, at least for a 4d boundary, the ambitwistors of the boundary are the twistors of the bulk \cite{Adamo:2016rtr,Bailey:1998zif}. Additionally, in $d>3$, the boundary three-point correlators of conserved currents allow for more structures than in the present $d=3$ case and it would be compelling to understand how they arise from twistor space.

Finally, we observed that at three points, the double copy emerges naturally in our construction. It is formally identical to the classical twistor double copy and extends straightforwardly to other arbitrary different single copies, provided that the helicities are suitably aligned. Since expressing a correlator-level double copy becomes significantly more difficult beyond three points in position or momentum space, it would be compelling to test whether a simpler relation arises in the twistor setting for more involved examples. We also note that a twistor space scalar kernel has been proposed in \cite{Adamo:2024hme} for AdS$_4$ correlators, within the formulation of \cite{Adamo:2015ina}. Understanding the relationship between their construction and the results presented here would be very insightful.
\section*{Acknowledgements}
We thank Guilherme Pimentel and Facundo Rost for many useful discussions and collaborations in related topics. MCG is supported by the Imperial College Research Fellowship. TK is supported by an STFC studentship.
\newpage

\appendix

\section{Conventions} 
This appendix sets the spinor conventions in three, four, five, and six dimensions.\label{ap:conventions}
\subsection{3d}
We take the three-dimensional sigma matrices to be
\begin{equation}
    \begin{aligned}
    \sigma^0&\,=\begin{pmatrix}
        1&0\\
        0&1
    \end{pmatrix},
    \sigma^1=\begin{pmatrix}
        1&0\\
        0&-1
    \end{pmatrix}, \sigma^2=\begin{pmatrix}
        0&1\\
        1&0
    \end{pmatrix}\ , \\
    \epsilon^{ab}&\,= \begin{pmatrix}
        0 && 1\\
        -1 &&0
    \end{pmatrix} = -\epsilon_{ab}\ , \\
    (\sigma^i)^{ab}&\,=(\sigma^0,\sigma^1, \sigma^2)\ ,\\ (\sigma^i)^{ac} (\sigma^j)_{cb}&\,+ (\sigma^j)^{ac} (\sigma^i)_{cb}  =-2 \eta^{ij} \delta^{a}_{b}\ ,\\
    \end{aligned}
\end{equation}
with signature $-++$. In this convention, our 3d coordinates $x^i$ become in spinor notation 
\begin{equation}
    \begin{aligned}
    x^{ab}&\,=(\sigma^i)^{ab} x_{i}
    =\begin{pmatrix}
        -x^0+x^1&& x^2\\
        x^2&& -x^0-x^1
    \end{pmatrix}\ ,\\
    x^{ac} x_{cb }&\,= -x^2 \delta^{a}_{b}\ ,
    \end{aligned}
\end{equation}
where $x^2$ is the squared norm of the vector. We also take NW-SE contractions such that
\begin{equation}
    \braket{\lambda_1 v \lambda_2} \equiv \lambda_1^a v_{a}^{\;b} \lambda_{2b}\ .
\end{equation}

\subsection{4d}
We take the four-dimensional sigma matrices to be
\begin{equation}
    \begin{aligned}
    (\sigma^A)^{\dot \alpha \alpha}&\,=(\sigma^i,\sigma^3),\;\;\;
    (\Tilde \sigma^A)_{\alpha \dot\alpha}=(\sigma^0,-\sigma^1,-\sigma^2,-\sigma^3)\ ,
    \end{aligned}
\end{equation}
with
\begin{equation}
    \begin{aligned}
        \sigma^3=\begin{pmatrix}
        0&-i\\
        i&0
    \end{pmatrix}\ ,
    \end{aligned}
\end{equation}
such that
\begin{equation}
    \begin{aligned}
    \sigma^A \Tilde \sigma^B+ \sigma^B \Tilde \sigma^A  =-2 \eta^{A B} \delta^{\dot \alpha}_{\dot \beta}\ , 
    \end{aligned}
\end{equation}
where the signature is $-+++$ and where $A$ is a tetrad index. In this convention, our 4d coordinates $x^A=(x^i_{3d},z)$ become in spinor notation 
\begin{equation}
    \begin{aligned}
    (x^{4d})^{\dot\alpha \alpha}&\,=(\sigma^A)^{\dot\alpha \alpha} x^{4d}_{A}
    =\begin{pmatrix}
        -x^0+x^1&& x^2 -i z\\
        x^2+iz&& -x^0-x^1
    \end{pmatrix}\ ,\\
    (x^{4d})^{\dot \alpha \alpha} (x^{4d})_{\dot \beta \alpha }&\,= -x^2 \delta^{\dot \alpha}_{\dot \beta}\ ,
    \end{aligned}
\end{equation}
where $(x^{4d})^2$ is the squared norm of the vector, but minus the determinant of $(x^{4d})_{\dot \alpha \alpha}$. We will also take NW-SE contractions such that
\begin{equation}
    \braket{\lambda_1 v \lambda_2} \equiv \lambda_1^\alpha v_{\alpha}^{\;\beta} \lambda_{2\beta}\ ,
\end{equation}
and similarly for the dotted indices and the boundary little group indices. Finally, the chordal distance in terms of 5d and 4d coordinates is
\begin{equation}
    \begin{aligned}
        u=\frac{(X_1-X_2)^2}{2}=\frac{(x_{1}-x_{2})^2+(z_1-z_2)^2}{2z_1 z_2}\ .
    \end{aligned}
\end{equation}
We take again numerically
\begin{equation}
    \begin{aligned}
        \epsilon^{\alpha \beta}&\,= \begin{pmatrix}
        0 && 1\\
        -1 &&0
    \end{pmatrix} \\
    &\,= -\epsilon_{\alpha \beta}=\epsilon^{\dot \alpha \dot \beta}=-\epsilon_{\dot \alpha \dot \beta}\ . \\
    \end{aligned}
\end{equation}
\subsection{5d}
The five-dimensional gamma matrices are taken to be
\begin{equation}
    \begin{aligned}
        (\Gamma^0)^M_{\;N}&\,=\begin{pmatrix}
            0&1&0&0\\
            -1&0&0&0\\
            0&0&0&-1\\
            0&0&1&0\\
        \end{pmatrix},\;
        (\Gamma^1)^M_{\;N}=\begin{pmatrix}
            0&-1&0&0\\
            -1&0&0&0\\
            0&0&0&-1\\
            0&0&-1&0\\
        \end{pmatrix},\;
        (\Gamma^2)^M_{\;N}=\begin{pmatrix}
            1&0&0&0\\
            0&-1&0&0\\
            0&0&1&0\\
            0&0&0&-1\\
        \end{pmatrix}\ ,\\
        (\Gamma^3)^M_{\;N}&\,=\begin{pmatrix}
            0&0&0&-1\\
            0&0&1&0\\
            0&1&0&0\\
            -1&0&0&0\\
        \end{pmatrix},\;
        (\Gamma^4)^M_{\;N}=\begin{pmatrix}
            0&0&0&1\\
            0&0&-1&0\\
            0&1&0&0\\
            -1&0&0&0\\
        \end{pmatrix}\ ,
    \end{aligned}
\end{equation}
which obey the Clifford algebra
\begin{equation}
    \{\Gamma^I,\Gamma^J\}^M_{\;N}= 2 \eta^{IJ} \delta^M_N\ .
\end{equation}
Explicitly, the 5d bispinors are then
\begin{equation}
    \begin{aligned}
        T^{M\alpha}=\frac{1}{\sqrt{z}}\begin{pmatrix}
           1&&0\\
           0&&1\\
           -x^0+x^1&&-iz+x^2\\
           iz+x^2&&-x^0-x^1\\
       \end{pmatrix},\;\;
     {\bar T}_M^{\dot{\alpha}}=\frac{1}{\sqrt{z}}\begin{pmatrix}
        -x^0+x^1&&iz+x^2\\
        -iz+x^2&&-x^0-x^1\\
         -1&&0\\
        0&&-1\\
        \end{pmatrix}\ .
    \end{aligned}
\end{equation}
Indices are raised and lowered using the symplectic form $\Omega_{MN}$ using again the NW-SE convention
\begin{equation}
    (S \cdot T )= S^M \Omega_{MN} T^N\ ,
\end{equation}
with 
\begin{equation}
    \Omega_{MN}=\begin{pmatrix}
        0&&0&&1&&0\\
        0&&0&&0&&1\\
        -1&&0&&0&&0\\
        0&&-1&&0&&0\\
    \end{pmatrix}\ .
\end{equation}

\subsection{6d}
We take the six-dimensional sigma matrices to be 
\begin{equation}
    \begin{aligned}
        (S^0)_{AB}&\,=i \sigma^2 \otimes \sigma^3=\pm(\tilde S^0)^{AB},\\
        (S^1)_{AB}&\,=i \sigma^3 \otimes \sigma^2=\pm(\tilde S^1)^{AB},\\
        (S^2)_{AB}&\,=-i \sigma^3 \otimes \sigma^1=\pm(\tilde S^2)^{AB},\\
        (S^3)_{AB}&\,= \sigma^3 \otimes \sigma^0=\pm(\tilde S^3)^{AB},\\
        (S^4)_{AB}&\,=-i \sigma^0 \otimes \sigma^3=\pm(\tilde S^4)^{AB},\\
        (S^5)_{AB}&\,=-i \sigma^1 \otimes \sigma^3=\pm(\tilde S^5)^{AB},
    \end{aligned}
\end{equation}
which satisfy the Clifford algebra
\begin{equation}
    (S^\mu)_{AB} (\tilde S^\nu)^{BC}+(S^\nu)_{AB} (\tilde S^\mu)^{BC}=-2 \eta^{\mu\nu} \delta^C_A\ .
\end{equation}
Single indices cannot be lower or raised, but pairs of indices are lowered/ raised with respect to the $SL(4,\mathbb{C})$ invariant tensor $\frac{1}{2} \epsilon^{ABCD}$. 

\section{Unified Coset Construction for AdS, dS, EAdS and EdS and their Bispinors} \label{ap:bispinors}

This appendix gives a single complex construction that uniformly describes the four constant--curvature 4D slices: AdS$_4$, dS$_4$, Euclidean AdS$_4$ (the hyperbolic space $H^4$),
and Euclidean dS$_4$ (the sphere $S^4$) and their corresponding bispinors. We will use the coset construction of the manifold,
\begin{equation}
\mathcal{M}\approx G/H \ ,
\end{equation}
where $G$ is the spin isometry group and $H$ the local Lorentz stabilizer (the group that fixes a point in the manifold, that is, the local Lorentz group):
\begin{equation}
\begin{array}{c|c|c}
\text{Slice} & G\ \ \text{(isometry)} & H\ \ \text{(stabilizer)}\\ \hline
\text{AdS}_4 & \mathrm{Spin}(3,2)\ \cong Sp(4,\mathbb R) & \mathrm{Spin}(3,1)\cong SL(2,\mathbb C)\\
\text{dS}_4 & \mathrm{Spin}(4,1)\cong Sp(2,2,\mathbb{H}) & \mathrm{Spin}(3,1)\cong SL(2,\mathbb C)\\
\text{EAdS}_4=H^4 & \mathrm{Spin}(4,1)\cong Sp(2,2,\mathbb{H}) & \mathrm{Spin}(4)\cong SU(2)_L\times SU(2)_R\\
\text{EdS}_4=S^4 & \mathrm{Spin}(5)\cong Sp(4,\mathbb{H}) & \mathrm{Spin}(4)\cong SU(2)_L\times SU(2)_R
\end{array}
\end{equation}
All formulas below will be written at the complexified level. A parameter $s=\pm1$ will fix the curvature (AdS vs dS), while the choice of $H$ (Lorentzian vs Euclidean) fixes the index types and Lorentz blocks; all other structures are common to the four cases. \\
Since $Sp(4)$ has a natural action of $Sp(4) \times Sp(4)$, we will consider $O_M{}^{\hat M}$, a $4\times4$ matrix with a left index $M$ and a right index $\hat M$, both in the fundamental of $G$. In all cases, both left and right copies of $G$  preserve non--degenerate antisymmetric forms (symplectic forms) $\Omega_{MN}$ and $\hat\Omega_{\hat M\hat N}$ respectively
\begin{equation}
\label{eq:coset-constraints}
\Omega_{MN}\,O^{M}{}_{\hat M}\,O^{N}{}_{\hat N}=\hat\Omega_{\hat M\hat N} \ ,
\qquad
\hat\Omega_{\hat M\hat N}\,O_{M}{}^{\hat M}\,O_{N}{}^{\hat N}=\Omega_{MN} \ .
\end{equation}
Note that for symplectic groups the dual and anti-fundamental are identified via $\Omega$.\\
We can now decompose the right index $\hat M$ into two doublets. For Lorentzian slices (AdS, dS) we have $\hat M\to(\alpha,\dot\alpha)$ of $SL(2,\mathbb C)$ and for Euclidean slices (EAdS, EdS): $\hat M\to(\alpha,\tilde\alpha)$ of $SU(2)_L\times SU(2)_R$. For simplicity of notation, in the following we will work with only dotted indices, but it should be understood as a tilde one in the Euclidean cases. We make this decomposition explicit by choosing the $H$-covariant block form
\begin{equation}
\label{eq:Omega-hat}
\hat\Omega_{\hat M\hat N}=
\begin{pmatrix}
is\,\varepsilon_{\alpha\beta} & 0\\
0 & \,i\,\varepsilon_{\dot\alpha\dot\beta}\ 
\end{pmatrix} \ ,
\qquad s=\pm1 \ ,
\end{equation}
where the overall $i$ is a convention to make $\hat\Omega$ anti--Hermitian and the relative sign $s$ is the
only invariant element (it cannot be removed by an $H$-basis change). The coset is defined by the quadratic constraints in Eq.~\eqref{eq:coset-constraints} and Eq.~\eqref{eq:Omega-hat}. Last, we can parametrize the coset representative as
\begin{equation}
\label{eq:M-blocks}
O_M{}^{\hat M}=\frac{1}{\sqrt2}\,\big(T_M{}^{\alpha}\ \ \ \bar T_M{}^{\dot\alpha}\big) \ .
\end{equation}
Lowering the left index with $\Omega$ and using \eqref{eq:coset-constraints}--\eqref{eq:M-blocks} we obtain
\begin{align}
\label{eq:TT}
\Omega_{MN}\,T^{M}{}_{\alpha}\,T^{N}{}_{\beta}&=+\,2is\,\varepsilon_{\alpha\beta} \ ,\\
\label{eq:TbTb}
\Omega_{MN}\,\bar T^{M}{}_{\dot\alpha}\,\bar T^{N}{}_{\dot\beta}&=+\,2\,i\,\varepsilon_{\dot\alpha\dot\beta} \ ,\\
\label{eq:Tcross}
\Omega_{MN}\,T^{M}{}_{\alpha}\,\bar T^{N}{}_{\dot\beta}&=0 \ .
\end{align}
Thus, $T_M^{\alpha}$ and $\bar T_M^{\dot\alpha}$ are nothing but the bispinors from \eqref{eq:TTL_incidecnce} and when considering the AdS case it corresponds to those of \cite{Binder:2020raz}. We still have to choose the value of $s$. To do this, we use Eq.~\eqref{eq:X_T},
\begin{equation*}
 X^{MN}\,=T^{M \alpha}  T^{ N \alpha} \epsilon_{\alpha \beta}+i\Omega^{MN} \ ,
\end{equation*}
together with \eqref{eq:TT}--\eqref{eq:Tcross} which imply
\begin{equation}
\label{eq:X2}
X \cdot X = \frac{X^{MN}X_{MN}}{4} = s \ ,
\end{equation}
where we used $\epsilon^{\alpha \beta}\epsilon_{ \alpha \beta}= -2$ and $\Omega^{MN} \Omega_{MN}=4$. Thus $s=+1$ gives $X\!\cdot\!X=+1$ (the de Sitter/sphere unit hyperboloid), and $s=-1$ gives $X\!\cdot\!X=-1$ (the AdS/hyperbolic unit hyperboloid).

\section{Evaluation of $I_{n,m}$}\label{ap:Inm}
In this appendix we show how to compute the integrals that appear in the regularized two-point functions. We start with 
\begin{equation}
    I_{n,0}=\int_{\mathbb{CP}^1}  \frac{D\pi \wedge D\hat\pi}{\braket{\pi \hat{\pi}}^{n+2}} 
    \braket{A \pi}^n\braket{B\hat{\pi}}^n,
\end{equation}
where $A^a, B^a$ are unconstrained constant spinors. The integrand and the measure are both invariant under $SU(2)$ transformations, which we denote by $g_{a}^{\;b}$. To see this, note that $SU(2)$ acts in the same way on both $\pi_a$ and $\hat \pi_a$, due to the way the Euclidean conjugation was defined in Eq. \eqref{euclidean conjugation}.\footnote{This can be checked by taking a patch where $\pi^a=(1,z)$ and applying the standard $SU(2)$ transformations $z \xrightarrow{} z'=\frac{az+\bar b}{-bz + \bar a}$ and $\bar z \xrightarrow{} \bar z'=\frac{a \bar z-\bar b}{-b \bar z + \bar a}$.} Under an $SU(2)$ action on $\pi_a$ and $\hat \pi_a$, the integrand therefore changes to 
\begin{equation}
\begin{aligned}
     \frac{D\pi' \wedge D\hat\pi'}{\braket{\pi' \hat{\pi'}}^{n+2}} 
    \braket{A \pi'}^n\braket{B\hat{\pi'}}^n&\,= \frac{D(g\pi) \wedge D(g\hat\pi)}{\braket{\pi g^{-1} g\hat{\pi}}^{n+2}} 
    \braket{A g\pi}^n\braket{B g\hat{\pi}}^n\\
    &\,= \frac{D(\pi) \wedge D(\hat\pi)}{\braket{\pi \hat{\pi}}^{n+2}} 
    \braket{A'\pi}^n\braket{B'\hat{\pi}}^n,\\
\end{aligned}
\end{equation}
where $A'=Ag$ and $B'=Bg$ and therefore $\braket{A'B'}=\braket{AB}$. Now since $A, B$ are fixed, the integral can only depend on the invariant quantity $\braket{AB}$. This explains why $A$ and $B$ must have the same power (beside it being fixed by homogeneity already). Finally, the reasoning we employed is only true if the space over which we integrate has $SU(2)$ isometry which is true for $\mathbb{CP}^1$. Therefore, this shows why retaining the assumption of homogeneity is key when the unitarity bound is not saturated: without homogeneity, the integral would be over $\mathbb{C}^2$ which does not have $SU(2)$ isometry. 
We have established that\footnote{This can also be understood as a special instance of Serre duality \cite{Boels:2006ir}.}
\begin{equation}
    I_{n,0}=f(\braket{AB}).
\end{equation}
Now scaling the constant spinor $A\xrightarrow{} \alpha A$, implies that $I_{n,0} \xrightarrow{} \alpha^n I_{n,0}$ and therefore
\begin{equation}
    I_{n,0}= g(n)\braket{AB}^n,
\end{equation}
for some coefficient $g(n)$. Now consider the general case
\begin{equation}
    I_{n,m}=\int_{\mathbb{CP}^1}  \frac{D\pi \wedge D\hat\pi}{\braket{\pi \hat{\pi}}^{n+m+2}} 
    \braket{A \pi}^n\braket{B\hat{\pi}}^n \braket{\pi C \hat{\pi}}^m,
\end{equation}
where we added a new spinor $C^{a b}$. Again, this integrand is $SU(2)$ invariant and, by homogeneity, the domain also has $SU(2)$ invariance. Now there are three possible invariant quantities, namely
\begin{equation}\label{invariants}
    \begin{aligned}
        S\equiv\braket{AB}, \; \; T\equiv \text{det}(C), \;\;U= \braket{ACB},
    \end{aligned}
\end{equation}
where $T$ and $U$ are only $SU(2)$ invariant if $C^{ab}=C^{(ab)}$ as expected since there is no non-trivial antisymmetric representation of $SU(2)$. By the same scaling argument as before we know that $S,T,U$ should appear in combinations with total scalings $A^n B^n C^m$, which partially constrains the integral to be 
\begin{equation}
    I_{n,m}= f( S^{n-j}U^j T^{\frac{m-j}{2}}),
\end{equation}
since det$(C)$ scales as $C^2$. An integral of polynomials outputs polynomials, therefore none of the powers of $S,T,U$ should be negative, constraining $j$ to the range $0\leq j \leq min(m,n)$ and $m-j$ to be even, otherwise the integral vanishes, so
\begin{equation}\label{general Jnm}
    I_{n,m}= \sum_{j}^{min(n,m)} g_j(n,m) S^{n-j}U^j T^{\frac{m-j}{2}},
\end{equation}
where the sum is over $j$ even (odd) for $m$ even (odd) respectively. This means one would still have to compute the integral explicitly to know the coefficients $g_j(n,m)$, or at least their relative value. However, we actually only need to calculate
\begin{equation}
    I_{n,m}=\int_{\mathbb{CP}^1}  \frac{D\pi \wedge D\hat\pi}{\braket{\pi \hat{\pi}}^{n+m+2}} 
    \braket{l \pi}^n\braket{l x_{12}\hat{\pi}}^n \braket{\pi x_{12} \hat{\pi}}^m,
\end{equation}
so that $A^a=l^a, B^a=(l x_{12})^a, C^{ab}=x_{12}^{ab}$. With these specific values, $U=0$. Hence, with these values, the integral is completely constrained up to an overall factor to 
\begin{equation}
    I_{n,m}\propto  \begin{cases} S^{n} T^{\frac{m}{2}}, \text{  \;\;\;\;\;\;\;\;\;\;\;\;\;  $m$ even } \\
        0, \text{   \;\;\;\;\;\;\;\;\;\;\;\;\;\;\;\;\;\;\;\, \;\,$m$ odd }
        \end{cases}
\end{equation}
which we have checked explicitly for different values of $n,m$. Finally, allowing for $l_1 \neq l_2$, our integral does not lead to the correlator for the non-conserved case ($m\neq 0$). This is because now $U\neq 0$ so there is another invariant built out of $x_{12}^2$ and $\braket{l_1 l_2}$.
\subsection{$m<0$}
When either $m$ or $n$ is negative, $I_{n,m}$ diverges because of poles on the sphere. Since we are interested in unitary correlators, this generally does not matter, except in the scalar case, where $m=-1$ corresponds to the free scalar theory. In this case, one should regularise the integral by noting that
\begin{equation}
    I_{0,m} = \int_{\mathbb{CP}^1}  \frac{D\pi \wedge D\hat\pi}{\braket{\pi \hat{\pi}}^{2}}\frac{\braket{\pi x_{12} \hat{\pi}}^m}{\braket{\pi \hat{\pi}}^{m}}
\end{equation}
is a sphere integral with normal vector $n^{ab} = \pi^{(a}\hat \pi^{b)}$ and is therefore proportional to $\int dS_2 (n \cdot x_{12})^m$. Choosing a coordinate system in which $x_{12}$ is aligned to the $z$ axis of the sphere, the integral reduces to
\begin{equation}
    I_{0,m} \propto |x_{12}|^m \int_0^{2\pi} cos(\theta)^m d\theta \ ,
\end{equation}
which for $m<0$, can be analytically continued to (with $z=cos (\theta)$)
\begin{equation}
    I_{0,m} \propto |x_{12}|^m \int_0^{1} (z^m +e^{i \pi m} z^m)dz= |x_{12}|^m\frac{1+e^{i\pi m}}{m+1} \ .
\end{equation}
This vanishes for any odd $m$, except $m=-1$ where the limit goes to $\frac{i \pi}{|x_{12}|}$.\\

Similarly, for $n>0$, we can rewrite $\braket{l \pi} \braket{l x_{12} \hat \pi}= (n \cdot D)$. Since we are interested in the case where $U=0$ (defined in Eq. \eqref{invariants}), we have $ D \cdot x_{12}=0$, and we can align $D$ in the x axis, i.e. $n \cdot D = \# |D| sin(\theta) cos(\theta)$, giving
\begin{equation}
    \begin{aligned}
        I_{n,m}=\# |x_{12}|^m |D|^n \int_0^{\pi} d\theta sin(\theta)^{n+1} cos(\theta)^m \int_0^{2\pi} d\phi cos(\phi)^n\ .
    \end{aligned}
\end{equation}
For $n$ even (integer spin), the second integral is an overall non-zero coefficient. For $m$ even positive and $n$ positive, the first integral is the beta function $\frac{1+ (-1)^m}{2}B(\frac{n+2}{2}, \frac{m+1}{2})$. When the sphere integral does not converge, the beta function gives its analytic continuation and therefore we obtain the analytic continuation of $I_{n,m}$ to integer conformal dimensions below the unitarity bound for integer spin. This includes in particular the case of partially massless fields for two-point functions.
\section{Evaluation of $J_s$ }\label{ap:generalised Leibniz}
We will now compute the integral in Eq.~\eqref{generalised leibniz} from Section \ref{sec:001}. We show for $s \in \mathbb{N}^*$ that
\begin{equation}
\begin{aligned}
    J_s&\,=\oint D\pi_3 \braket{l \frac{\partial}{ \partial \mu_3} }^{2s} (A^{s-1} \log(A) B^{s-1} \log(B)) \\
    &\,=(-1)^s (2s)! ((s-1)!)^2 \oint D\pi_3 \frac{(C B'-A' D)^{2s}}{ABC^s D^s}\ ,
\end{aligned}
\end{equation}
where we defined
\begin{equation}
    \begin{aligned}
        A'&\,= \braket{l \frac{\partial}{ \partial \mu_3} }A\ ,\\
        B'&\,= \braket{l \frac{\partial}{ \partial \mu_3} }B\ ,
    \end{aligned}
\end{equation}
and 
\begin{equation}
    \begin{aligned}
        C&\,= \braket{l \frac{\partial}{ \partial \pi_3} }A\ ,\\
        D&\,= \braket{l \frac{\partial}{ \partial \pi_3} }B\ .
    \end{aligned}
\end{equation}
In our context $A= \braket{\pi_2 x_{23} \pi_3}$ and $B= \braket{\pi_3 x_{31} \pi_1}$, so $A'= \braket{l \pi_2},\; B'=- \braket{l \pi_1}, $ $C= \braket{lx_{23} \pi_2},\; D= \braket{lx_{31} \pi_1}$. Using the generalised Leibniz rule, we have
\begin{equation}
    \begin{aligned}
        J_s&\,= \oint D\pi_3 \sum_{i_1+i_2+i_3+i_4=2s}f\ ,\\
        f&\,=\begin{pmatrix}
            2s\\
            i_1, i_2,i_3,i_4
        \end{pmatrix} \braket{l \frac{\partial}{ \partial \mu_3} }^{i_1} (A^{s-1}) \braket{l \frac{\partial}{ \partial \mu_3} }^{i_2}  (B^{s-1})\braket{l \frac{\partial}{ \partial \mu_3} }^{i_3}(\log(A)) \braket{l \frac{\partial}{ \partial \mu_3} }^{i_4}(\log(B))\ .
    \end{aligned}
\end{equation}
In the following, we will take $i_4=2s-i_1-i_2-i_3$ to be fixed.
As in the $s=1$ case done in Section \ref{sec:001}, we want to use Stokes' theorem on the log terms, therefore we split the sum as follows 
\begin{equation}
    J_s= J_{A0}+J_{0B}+J_{AB}+J_{00}\ ,
\end{equation}
where the integrand of $J_{A0}$ is proportional to $\log(A)$, the integrand of $J_{0B}$ is proportional to $\log(B)$,  the integrand of $J_{AB}$ is proportional to $\log(A) \log(B)$, and the integrand of $J_{00}$ has no logarithmic terms. Imposing $i_3=0,\; i_4=2s-i_1-i_2 > 0, \; i_2 \geq0$, the first term simplifies to 
\begin{equation}
     J_{A0}= \oint D\pi_3 \log(A) \sum_{i_1=0}^{2s-1} \sum_{i_2=0}^{2s-i_1-1} \frac{f}{\log(A)}\ .
\end{equation}
However, since the derivatives on $A^{s-1}$ and $B^{s-1}$ truncate for integer spin, the sum simplifies further to
\begin{equation}\label{JA0}
    \begin{aligned}
        J_{A0}&\,= \oint D\pi_3 \log(A) \sum_{i_1=0}^{s-1} \sum_{i_2=0}^{min(s-1,2s-i_1-1)} f_{A0}\ ,\\
         f_{A0}&\,= (-1)^{2s-i_1-i_2-1}\frac{(2s)!}{2s-i_1-i_2} \begin{pmatrix}
    s-1\\i_1
\end{pmatrix} \begin{pmatrix}
    s-1\\i_2
\end{pmatrix}  (A')^{i_1}(B')^{2s-i_1} \frac{A^{s-1-i_1}}{B^{s+1-i_1}}\ .
    \end{aligned}
\end{equation}
Similarly, imposing $i_3> 0,\; i_4=2s-i_1-i_2 = 0$, the second term simplifies to
\begin{equation}
    \begin{aligned}
        J_{0B}&\,= \oint D\pi_3 \log(B) \sum_{i_1=0}^{s-1} \sum_{i_2=0}^{min(s-1,2s-i_1-1)} \frac{f_{0B}}{\log (B)}\ ,\\
         f_{0B}&\,= (-1)^{2s-i_1-i_2-1}\frac{(2s)!}{2s-i_1-i_2} \begin{pmatrix}
    s-1\\i_1
\end{pmatrix} \begin{pmatrix}
    s-1\\i_2
\end{pmatrix}  (B')^{i_2}(A')^{2s-i_1} \frac{B^{s-1-i_2}}{A^{s+1-i_2}}\ .
    \end{aligned}
\end{equation}
Imposing $i_3=i_4=0$ implies that $i_2=2s-i_1$. However, the sums truncate again to $i_1= s-1$ and $i_2=s-1$ which is incompatible with the previous range and therefore 
\begin{equation}
    J_{AB}= 0\ .
\end{equation}
Finally, imposing $i_3>0, \; i_4>0$, the last term simplifies to
\begin{equation}
    \begin{aligned}
        J_{00}&\,=\oint D\pi_3 \sum_{i_1=0}^{2s-2} \sum_{i_2=0}^{2s-2-i_1} \sum_{i_3=1}^{2s-1-i_1-i_2} f_{00}\ ,\\
        f_{00}&\,=(-1)^{i_3+i_4} \frac{(2s)!}{i_3+i_4} \begin{pmatrix}
            s-1\\i_1
        \end{pmatrix} \begin{pmatrix}
            s-1\\i_2
        \end{pmatrix} (A')^{i_1+i_3} (B')^{i_2+i_4} A^{s-1-i_1-i_3} B^{s-1-i_2-i_4}\ .
    \end{aligned}
\end{equation}
Now, we evaluate these integrals in turn. We rewrite the pole in Eq. \eqref{JA0} with
\begin{equation}
    \frac{1}{B^n}= \frac{1}{(n-1)!} (- \frac{\braket{l \frac{\partial}{ \partial \pi_3} }}{D})^{n-1}\frac{1}{B}\ ,
\end{equation}
so that we can use Stokes' theorem to obtain
\begin{equation}
\begin{aligned}
     J_{A0}&\,= \oint D\pi_3 \frac{1}{B} \sum_{i_1=0}^{s-1} \sum_{i_2=0}^{min(s-1,2s-i_1-1)} \tilde f_{A0}\ ,\\
         \tilde f_{A0}&\,= \frac{(-1)^{2s-i_1-i_2-1} (2s)!}{(2s-i_1-i_2)(s-i_1)!} \begin{pmatrix}
    s-1\\i_1
\end{pmatrix} \begin{pmatrix}
    s-1\\i_2
\end{pmatrix}  \frac{(A')^{i_1}(B')^{2s-i_1}}{D^{s-i_1}}  \braket{l \frac{\partial}{ \partial \pi_3} }^{s-i_1}(A^{s-1-i_1} \log(A))\ ,
\end{aligned}
\end{equation}
which kills the remaining logarithmic terms since
\begin{equation}\label{derivative AnlogA}
    \braket{l \frac{\partial}{ \partial \pi_3} }^{n}(A^{n-1} \log(A)) = \frac{(n-1)! C^n}{A}\ .
\end{equation}
Furthermore the $i_2$ sum can be evaluated using that
\begin{equation}
    \sum_{i_2=0}^{min(s-1,2s-i_1-1)} \frac{(-1)^{i_2}}{(2s-i_1-i_2)} \begin{pmatrix}
    s-1\\i_2
\end{pmatrix}= (-1)^{s-1} \frac{(s-1)! (s-i_1)!}{(2s-i_1)!}\ ,
\end{equation}
resulting in 
\begin{equation}\label{JA0 single sum}
    \begin{aligned}
         J_{A0}&\,=\oint D\pi_3 \frac{(-1)^{s} (2s)! ((s-1)!)^2}{AB} \sum_{i_1=0}^{s-1} \frac{(-1)^{i_1}}{(2s-i_1)!i_1!} (A')^{i_1}(B')^{2s-i_1}  \left(\frac{C}{D}\right)^{s-i_1}\ .
    \end{aligned}
\end{equation}
To evaluate $J_{0B}$, note that because $\sum_{i_1=0}^{s-1} \sum_{i_2=0}^{min(s-1,2s-i_1-1)}= \sum_{i_2=0}^{s-1} \sum_{i_1=0}^{min(s-1,2s-i_1-1)}$, we have $J_{A0}=J_{0B}\; (A \xleftrightarrow[]{} B)$ so 
\begin{equation}\label{J0B single sum}
    \begin{aligned}
         J_{0B}&\,=\oint D\pi_3 \frac{(-1)^{s} (2s)! ((s-1)!)^2}{AB} \sum_{i_1=0}^{s-1} \frac{(-1)^{i_1}}{(2s-i_1)!i_1!} (B')^{i_1}(A')^{2s-i_1}  \left(\frac{D}{C}\right)^{s-i_1}\ .
    \end{aligned}
\end{equation}
To evaluate $f_{00}$, we observe that most terms will not contribute inside the integral. Indeed each term will be proportional to
\begin{equation}
    \begin{aligned}
         A^{s-1-i_1-i_3} B^{s-1-i_2-i_4}&\,=  A^{s-1-i_1-i_3} B^{s-1-i_2-(2s-i_1-i_2-i_3)}\\
         &\,=  A^{s-1-i_1-i_3} B^{-s-1+i_1 +i_3}\\
         &\,=\frac{A^{i-2}}{B^{i}}\ ,
    \end{aligned}
\end{equation}
where $i=s-i_1-i_3+1$. By the residue theorem, only the $i=1$ term will contribute. This collapses this sum to the single term
\begin{equation}
    J_{00}= \oint D\pi_3 \frac{(2s)!}{s^2} \frac{(A' B')^s}{AB}\ ,
\end{equation}
which coincides with $i_1=s$ in either Eq. \eqref{JA0 single sum} or \eqref{J0B single sum}, giving
\begin{equation}
\begin{aligned}
    J_s&\,= \oint D\pi_3 \frac{(-1)^{s} (2s)! ((s-1)!)^2}{AB} \sum_{i_1=0}^{2s} \frac{(-1)^{i_1}}{(2s-i_1)!i_1!} (A')^{i_1}(B')^{2s-i_1}  \left(\frac{C}{D}\right)^{s-i_1}\\
    &\,=(-1)^{s} (2s)! ((s-1)!)^2 \oint D\pi_3 \frac{(CB'-A'D)^{2s}}{ABC^s D^s} \ .
\end{aligned}
\end{equation}

\section{$\braket{O_1O_2 O_3^s}$ with Negative Helicity}\label{app:negative helicity}
We now show that we can alternatively use Eq. \eqref{samehelicity} to obtain the same result. As anticipated, the result needs to be regularised since $n_3=1-s \leq0$ for $s \geq 1.$ which is done by dressing the representative with another logarithmic factor
\begin{equation}\label{twistorlog}
    \braket{O_1 O_2 O_3^{s}}= \# \oint D\pi_{123} \frac{\braket{l \pi_3}^{2s} (\mathbf{\Lambda}_1\cdot\mathbf{\Lambda}_2)^{s-1} } { (\mathbf{\Lambda}_2\cdot\mathbf{\Lambda}_3)^{s+1}  (\mathbf{\Lambda}_3\cdot\mathbf{\Lambda}_1)^{s+1}}  \log(\frac{\mathbf{\Lambda}_1\cdot\mathbf{\Lambda}_2}{\braket{\pi_1 \pi_2}})\ .
\end{equation}
As before, we can remove the logarithmic term by integrating by parts and considering a contour which does not enclose the branch cut. First we do the contour integral around the pole $(\mathbf{\Lambda}_3\cdot\mathbf{\Lambda}_1)^{s+1}$.  To do this, we write it again as the derivative of a simple pole. However, it is now advantageous to use $l'= \pi_2$
\begin{equation}
    \frac{1}{\braket{\pi_3 x_{31} \pi_1}^{s+1}}= \frac{1}{s!}\left(- \frac{\braket{\pi_2 \frac{\partial}{\partial \pi_1}}}{\braket{\pi_2 x_{31} \pi_3}} \right)^s\left(\frac{1}{\braket{\pi_3 x_{31} \pi_1}}\right) \ .
\end{equation}
Because $l'$ is arbitrary, it does not need to be fixed and importantly, the integral is still well-defined projectively. Using Stokes' theorem we have
\small \begin{equation}
    \begin{aligned}
        \braket{O_1 O_2 O_3^s} &\,=\#\oint D\pi_{23} \frac{\braket{l \pi_3}^{2s}}{\braket{\pi_2 x_{23} \pi_3}^{s+1}} \left(\frac{ \braket{\pi_2 \frac{\partial}{\partial \pi_1}}}{\braket{\pi_2 x_{13} \pi_3}}\right)^s   \left[\braket{\pi_1 x_{12} \pi_2}^{s-1} \log\left(\frac{\braket{\pi_1 x_{12} \pi_2}}{\braket{\pi_1\pi_2}}\right)\right]\Bigg|_{\pi_1=x_{13}\pi_3}
    \end{aligned}
\end{equation}
\normalsize
which removes the logarithm since
\begin{equation}
    \braket{\pi_2 \frac{\partial}{\partial \pi_1}}^s[\braket{\pi_1 x_{12} \pi_2}^{s-1} \log(\frac{\pi_1 x_{12} \pi_2}{\braket{\pi_1\pi_2}})]=(s-1)!\frac{\braket{\pi_2 x_{12}\pi_2}^s}{\braket{\pi_1 x_{12}\pi_2}}\ .
\end{equation}
Hence, the correlator becomes 
\begin{equation}
    \begin{aligned}
        \braket{O_1 O_2 O_3^s}
        &\,=\#(s-1)!\oint D\pi_{23} \frac{\braket{l \pi_3}^{2s} \braket{\pi_2 x_{12}\pi_2}^s }{\braket{\pi_2 x_{23} \pi_3}^{s+1} \braket{\pi_2 x_{13} \pi_3}^s \braket{\pi_3 x_{31} x_{12}\pi_2}} \\
        &\,=\#(-1)^{s}(s-1)!(x_{12})^{2s} \oint D\pi_{3} \frac{\braket{l \pi_3}^{2s}}{\braket{\pi_3 \tilde x_3 \pi_3}^{s+1}}\ ,
    \end{aligned}
\end{equation}
where integrate over the simple pole.
The last integration is completely analogous to before and results in 
\begin{equation}\label{00sfirstway}
    \begin{aligned}
        \oint D\pi_{3} \frac{\braket{l \pi_3}^{2s}}{(\braket{\pi_3 \pi_+ }\braket{\pi_-\pi_3})^{s+1}} 
        &\, =\#\frac{\braket{l \pi_3}^{2s}}{\braket{\pi_+l}^{s}} \braket{l \frac{\partial}{ \partial \pi_3}}^s\left(\frac{1}{\braket{\pi_- \pi_3}^{s+1}}\right)\Bigg|_{\pi_3=\pi_+}\\
        &\, =\# \frac{(\braket{l \pi_+} \braket{\pi_- l})^s}{\braket{\pi_- \pi_+}^{2s+1}}\ .
    \end{aligned}
\end{equation}
Therefore, plugging in Eq. \eqref{structureslittle}, we recover Eq. \eqref{3points ex}. 

\section{Proof of Eq. \eqref{equivalent B.15}}
Proving Eq. \eqref{equivalent B.15} can be done by using the chain rule and integrating by parts as we did in the other sections. For conciseness, we call $f_2 \equiv\mathbf{\Lambda}_2\cdot\mathbf{\Lambda}_3 $ and $f_1 \equiv\mathbf{\Lambda}_3\cdot\mathbf{\Lambda}_1 $. First note that
\begin{equation}
    \begin{aligned}
        M_{-2}(f_2)&\,= \frac{d}{df_2} M_{-1}(f_2)\\
        &\,= \frac{\braket{l \frac{\partial}{\partial \pi_3}}}{\braket{\pi_2 x_{23} l}} M_{-1}(f_2) \ , 
    \end{aligned}
\end{equation}
and therefore 
\begin{equation}
    \begin{aligned}
        \oint D\pi_3 M_{-n}(f_2) M_{n-2}(f_1)&\,= \oint D\pi_3 (\frac{\braket{l \frac{\partial}{\partial \pi_3}}}{\braket{\pi_2 x_{23} l}})^{n-1} M_{-1}(f_2) M_{n-2}(f_1)\\
        &\,= (-1)^{n-1} \frac{1}{\braket{\pi_2 x_{23} l}^{n-1}} \oint D\pi_3 M_{-1}(f_2) \braket{l \frac{\partial}{\partial \pi_3}}^{n-1} M_{n-2}(f_1)
    \end{aligned}
\end{equation}
Using the chain rule and the fact that $f_1$ is linear in $\pi_3$, we see that the second term must be proportional to $\braket{l \frac{\partial f_1}{\partial \pi_3}}^{n-1} (f_1)^{-1}$ (we already used that fact, see Eq. \eqref{derivative AnlogA}). Substituting for $\braket{l \frac{\partial f_1}{\partial \pi_3}}$, we then obtain 
\begin{equation}
    \begin{aligned}
        \oint D\pi_3 M_{-n}(f_2) M_{n-2}(f_1)&\,= \oint D\pi_3 (\frac{\braket{l \frac{\partial}{\partial \pi_3}}}{\braket{\pi_2 x_{23} l}})^{n-1} M_{-1}(f_2) M_{n-2}(f_1)\\
        &\,= (-1)^{n-1} (\frac{\braket{lx_{31} \pi_1}}{\braket{\pi_2 x_{23} l}})^{n-1} \oint D\pi_3 M_{-1}(f_2) M_{-1}(f_1) \\
         &\,= (-1)^{n-1} (\frac{\braket{l x_{31} \pi_1}}{\braket{\pi_2 x_{23} l}})^{n-1} \frac{1}{\braket{\pi_2 x_{23} x_{31} \pi_1}}\ .
    \end{aligned}
\end{equation}
\newpage
\bibliography{refs.bib}

\providecommand{\href}[2]{#2}\begingroup\raggedright\begin{thebibliography}{100}

\bibitem{Baumann:2024ttn}
D.~Baumann, G.~Mathys, G.~L. Pimentel and F.~Rost, \emph{{A New Twist on
  Spinning (A)dS Correlators}},
  \href{https://arxiv.org/abs/2408.02727}{{\ttfamily 2408.02727}}.

\bibitem{Elvang:2015rqa}
H.~Elvang and Y.-t. Huang, \emph{{Scattering Amplitudes in Gauge Theory and
  Gravity}}.
\newblock Cambridge University Press, 4, 2015.

\bibitem{deRham:2022hpx}
C.~de~Rham, S.~Kundu, M.~Reece, A.~J. Tolley and S.-Y. Zhou, \emph{{Snowmass
  White Paper: UV Constraints on IR Physics}},  in \emph{{2022 Snowmass Summer
  Study}}, 3, 2022, \href{https://arxiv.org/abs/2203.06805}{{\ttfamily
  2203.06805}}.

\bibitem{Kruczenski:2022lot}
M.~Kruczenski, J.~Penedones and B.~C. van Rees, \emph{{Snowmass White Paper:
  S-matrix Bootstrap}},  \href{https://arxiv.org/abs/2203.02421}{{\ttfamily
  2203.02421}}.

\bibitem{Parke:1986gb}
S.~J. Parke and T.~R. Taylor, \emph{{An Amplitude for $n$ Gluon Scattering}},
  \href{http://dx.doi.org/10.1103/PhysRevLett.56.2459}{\emph{Phys. Rev. Lett.}
  {\bfseries 56} (1986) 2459}.

\bibitem{Witten:2003nn}
E.~Witten, \emph{{Perturbative gauge theory as a string theory in twistor
  space}}, \href{http://dx.doi.org/10.1007/s00220-004-1187-3}{\emph{Commun.
  Math. Phys.} {\bfseries 252} (2004) 189--258},
  [\href{https://arxiv.org/abs/hep-th/0312171}{{\ttfamily hep-th/0312171}}].

\bibitem{Cachazo:2004kj}
F.~Cachazo, P.~Svrcek and E.~Witten, \emph{{MHV vertices and tree amplitudes in
  gauge theory}},
  \href{http://dx.doi.org/10.1088/1126-6708/2004/09/006}{\emph{JHEP} {\bfseries
  09} (2004) 006}, [\href{https://arxiv.org/abs/hep-th/0403047}{{\ttfamily
  hep-th/0403047}}].

\bibitem{Roiban:2004yf}
R.~Roiban, M.~Spradlin and A.~Volovich, \emph{{On the tree level S matrix of
  Yang-Mills theory}},
  \href{http://dx.doi.org/10.1103/PhysRevD.70.026009}{\emph{Phys. Rev. D}
  {\bfseries 70} (2004) 026009},
  [\href{https://arxiv.org/abs/hep-th/0403190}{{\ttfamily hep-th/0403190}}].

\bibitem{Britto:2005fq}
R.~Britto, F.~Cachazo, B.~Feng and E.~Witten, \emph{{Direct proof of tree-level
  recursion relation in Yang-Mills theory}},
  \href{http://dx.doi.org/10.1103/PhysRevLett.94.181602}{\emph{Phys. Rev.
  Lett.} {\bfseries 94} (2005) 181602},
  [\href{https://arxiv.org/abs/hep-th/0501052}{{\ttfamily hep-th/0501052}}].

\bibitem{Penedones:2010ue}
J.~Penedones, \emph{{Writing CFT correlation functions as AdS scattering
  amplitudes}}, \href{http://dx.doi.org/10.1007/JHEP03(2011)025}{\emph{JHEP}
  {\bfseries 03} (2011) 025},
  [\href{https://arxiv.org/abs/1011.1485}{{\ttfamily 1011.1485}}].

\bibitem{Baumann:2022jpr}
D.~Baumann, D.~Green, A.~Joyce, E.~Pajer, G.~L. Pimentel, C.~Sleight et~al.,
  \emph{{Snowmass White Paper: The Cosmological Bootstrap}},
  \href{http://dx.doi.org/10.21468/SciPostPhysCommRep.1}{\emph{SciPost Phys.
  Comm. Rep.} {\bfseries 2024} (2024) 1},
  [\href{https://arxiv.org/abs/2203.08121}{{\ttfamily 2203.08121}}].

\bibitem{Melville:2023kgd}
S.~Melville and G.~L. Pimentel, \emph{{de Sitter S matrix for the masses}},
  \href{http://dx.doi.org/10.1103/PhysRevD.110.103530}{\emph{Phys. Rev. D}
  {\bfseries 110} (2024) 103530},
  [\href{https://arxiv.org/abs/2309.07092}{{\ttfamily 2309.07092}}].

\bibitem{Spradlin:2001nb}
M.~Spradlin and A.~Volovich, \emph{{Vacuum states and the S matrix in dS /
  CFT}}, \href{http://dx.doi.org/10.1103/PhysRevD.65.104037}{\emph{Phys. Rev.
  D} {\bfseries 65} (2002) 104037},
  [\href{https://arxiv.org/abs/hep-th/0112223}{{\ttfamily hep-th/0112223}}].

\bibitem{Cheung:2022pdk}
C.~Cheung, J.~Parra-Martinez and A.~Sivaramakrishnan, \emph{{On-shell
  correlators and color–kinematics duality in curved symmetric spacetimes}},
  \href{http://dx.doi.org/10.1007/JHEP05(2022)027}{\emph{JHEP} {\bfseries 05}
  (2022) 027}, [\href{https://arxiv.org/abs/2201.05147}{{\ttfamily
  2201.05147}}].

\bibitem{Grall:2020tqc}
T.~Grall and S.~Melville, \emph{{Inflation in motion: unitarity constraints in
  effe...e field theories with (spontaneously) broken Lorentz symmetry}},
  \href{http://dx.doi.org/10.1088/1475-7516/2020/09/017}{\emph{JCAP} {\bfseries
  09} (2020) 017}, [\href{https://arxiv.org/abs/2005.02366}{{\ttfamily
  2005.02366}}].

\bibitem{Melville:2021lst}
S.~Melville and E.~Pajer, \emph{{Cosmological Cutting Rules}},
  \href{http://dx.doi.org/10.1007/JHEP05(2021)249}{\emph{JHEP} {\bfseries 05}
  (2021) 249}, [\href{https://arxiv.org/abs/2103.09832}{{\ttfamily
  2103.09832}}].

\bibitem{Grall:2021xxm}
T.~Grall and S.~Melville, \emph{{Positivity bounds without boosts: New
  constraints on low energy effective field theories from the UV}},
  \href{http://dx.doi.org/10.1103/PhysRevD.105.L121301}{\emph{Phys. Rev. D}
  {\bfseries 105} (2022) L121301},
  [\href{https://arxiv.org/abs/2102.05683}{{\ttfamily 2102.05683}}].

\bibitem{Cespedes:2020xqq}
S.~C\'espedes, A.-C. Davis and S.~Melville, \emph{{On the time evolution of
  cosmological correlators}},
  \href{http://dx.doi.org/10.1007/JHEP02(2021)012}{\emph{JHEP} {\bfseries 02}
  (2021) 012}, [\href{https://arxiv.org/abs/2009.07874}{{\ttfamily
  2009.07874}}].

\bibitem{Goodhew:2020hob}
H.~Goodhew, S.~Jazayeri and E.~Pajer, \emph{{The Cosmological Optical
  Theorem}}, \href{http://dx.doi.org/10.1088/1475-7516/2021/04/021}{\emph{JCAP}
  {\bfseries 04} (2021) 021},
  [\href{https://arxiv.org/abs/2009.02898}{{\ttfamily 2009.02898}}].

\bibitem{Goodhew:2021oqg}
H.~Goodhew, S.~Jazayeri, M.~H.~G. Lee and E.~Pajer, \emph{{Cutting cosmological
  correlators}},
  \href{http://dx.doi.org/10.1088/1475-7516/2021/08/003}{\emph{JCAP} {\bfseries
  08} (2021) 003}, [\href{https://arxiv.org/abs/2104.06587}{{\ttfamily
  2104.06587}}].

\bibitem{Jazayeri:2021fvk}
S.~Jazayeri, E.~Pajer and D.~Stefanyszyn, \emph{{From locality and unitarity to
  cosmological correlators}},
  \href{http://dx.doi.org/10.1007/JHEP10(2021)065}{\emph{JHEP} {\bfseries 10}
  (2021) 065}, [\href{https://arxiv.org/abs/2103.08649}{{\ttfamily
  2103.08649}}].

\bibitem{Stefanyszyn:2024msm}
D.~Stefanyszyn, X.~Tong and Y.~Zhu, \emph{{There and Back Again: Mapping and
  Factorizing Cosmological Observables}},
  \href{http://dx.doi.org/10.1103/PhysRevLett.133.221501}{\emph{Phys. Rev.
  Lett.} {\bfseries 133} (2024) 221501},
  [\href{https://arxiv.org/abs/2406.00099}{{\ttfamily 2406.00099}}].

\bibitem{Albayrak:2023hie}
S.~Albayrak, P.~Benincasa and C.~Duaso~Pueyo, \emph{{Perturbative unitarity and
  the wavefunction of the Universe}},
  \href{http://dx.doi.org/10.21468/SciPostPhys.16.6.157}{\emph{SciPost Phys.}
  {\bfseries 16} (2024) 157},
  [\href{https://arxiv.org/abs/2305.19686}{{\ttfamily 2305.19686}}].

\bibitem{Goodhew:2024eup}
H.~Goodhew, A.~Thavanesan and A.~C. Wall, \emph{{The Cosmological CPT
  Theorem}},  \href{https://arxiv.org/abs/2408.17406}{{\ttfamily 2408.17406}}.

\bibitem{Thavanesan:2025kyc}
A.~Thavanesan, \emph{{No-go Theorem for Cosmological Parity Violation}},
  \href{https://arxiv.org/abs/2501.06383}{{\ttfamily 2501.06383}}.

\bibitem{Benincasa:2020uph}
P.~Benincasa and M.~Parisi, \emph{{Positive geometries and differential forms
  with non-logarithmic singularities. Part I}},
  \href{http://dx.doi.org/10.1007/JHEP08(2020)023}{\emph{JHEP} {\bfseries 08}
  (2020) 023}, [\href{https://arxiv.org/abs/2005.03612}{{\ttfamily
  2005.03612}}].

\bibitem{Melville:2024zjq}
S.~Melville, \emph{{Causality and quasi-normal modes in the GREFT}},
  \href{http://dx.doi.org/10.1140/epjp/s13360-024-05520-5}{\emph{Eur. Phys. J.
  Plus} {\bfseries 139} (2024) 725},
  [\href{https://arxiv.org/abs/2401.05524}{{\ttfamily 2401.05524}}].

\bibitem{Serra:2022pzl}
F.~Serra, J.~Serra, E.~Trincherini and L.~G. Trombetta, \emph{{Causality
  constraints on black holes beyond GR}},
  \href{http://dx.doi.org/10.1007/JHEP08(2022)157}{\emph{JHEP} {\bfseries 08}
  (2022) 157}, [\href{https://arxiv.org/abs/2205.08551}{{\ttfamily
  2205.08551}}].

\bibitem{Bittermann:2022hhy}
N.~Bittermann, D.~McLoughlin and R.~A. Rosen, \emph{{On causality conditions in
  de Sitter spacetime}},
  \href{http://dx.doi.org/10.1088/1361-6382/accc05}{\emph{Class. Quant. Grav.}
  {\bfseries 40} (2023) 115006},
  [\href{https://arxiv.org/abs/2212.02559}{{\ttfamily 2212.02559}}].

\bibitem{Dubovsky:2007ac}
S.~Dubovsky, A.~Nicolis, E.~Trincherini and G.~Villadoro, \emph{{Microcausality
  in curved space-time}},
  \href{http://dx.doi.org/10.1103/PhysRevD.77.084016}{\emph{Phys. Rev. D}
  {\bfseries 77} (2008) 084016},
  [\href{https://arxiv.org/abs/0709.1483}{{\ttfamily 0709.1483}}].

\bibitem{Baumgart:2020oby}
M.~Baumgart and R.~Sundrum, \emph{{Manifestly Causal In-In Perturbation Theory
  about the Interacting Vacuum}},
  \href{http://dx.doi.org/10.1007/JHEP03(2021)080}{\emph{JHEP} {\bfseries 03}
  (2021) 080}, [\href{https://arxiv.org/abs/2010.10785}{{\ttfamily
  2010.10785}}].

\bibitem{AguiSalcedo:2023nds}
S.~Agui~Salcedo and S.~Melville, \emph{{The cosmological tree theorem}},
  \href{http://dx.doi.org/10.1007/JHEP12(2023)076}{\emph{JHEP} {\bfseries 12}
  (2023) 076}, [\href{https://arxiv.org/abs/2308.00680}{{\ttfamily
  2308.00680}}].

\bibitem{Salcedo:2022aal}
S.~A. Salcedo, M.~H.~G. Lee, S.~Melville and E.~Pajer, \emph{{The Analytic
  Wavefunction}}, \href{http://dx.doi.org/10.1007/JHEP06(2023)020}{\emph{JHEP}
  {\bfseries 06} (2023) 020},
  [\href{https://arxiv.org/abs/2212.08009}{{\ttfamily 2212.08009}}].

\bibitem{Baumann:2021fxj}
D.~Baumann, W.-M. Chen, D.~Pueyo...s, A.~Joyce, H.~Lee and G.~L. Pimentel,
  \emph{{Linking the singularities of cosmological correlators}},
  \href{http://dx.doi.org/10.1007/JHEP09(2022)010}{\emph{JHEP} {\bfseries 09}
  (2022) 010}, [\href{https://arxiv.org/abs/2106.05294}{{\ttfamily
  2106.05294}}].

\bibitem{deRham:2020zyh}
C.~de~Rham and A.~J. Tolley, \emph{{Causality in curved spacetimes: The speed
  of light and gravity}},
  \href{http://dx.doi.org/10.1103/PhysRevD.102.084048}{\emph{Phys. Rev. D}
  {\bfseries 102} (2020) 084048},
  [\href{https://arxiv.org/abs/2007.01847}{{\ttfamily 2007.01847}}].

\bibitem{CarrilloGonzalez:2023emp}
M.~Carrillo~Gonz\'alez, \emph{{Bounds on EFT\textquoteright{}s in an expanding
  universe}}, \href{http://dx.doi.org/10.1103/PhysRevD.109.085008}{\emph{Phys.
  Rev. D} {\bfseries 109} (2024) 085008},
  [\href{https://arxiv.org/abs/2312.07651}{{\ttfamily 2312.07651}}].

\bibitem{CarrilloGonzalez:2025fqq}
M.~Carrillo~Gonz{\'a}lez and S.~C{\'e}spedes, \emph{{Causality bounds on the
  primordial power spectrum}},
  \href{http://dx.doi.org/10.1088/1475-7516/2025/08/071}{\emph{JCAP} {\bfseries
  08} (2025) 071}, [\href{https://arxiv.org/abs/2502.19477}{{\ttfamily
  2502.19477}}].

\bibitem{Creminelli:2011mw}
P.~Creminelli, \emph{{Conformal invariance of scalar perturbations in
  inflation}}, \href{http://dx.doi.org/10.1103/PhysRevD.85.041302}{\emph{Phys.
  Rev. D} {\bfseries 85} (2012) 041302},
  [\href{https://arxiv.org/abs/1108.0874}{{\ttfamily 1108.0874}}].

\bibitem{Assassi:2012zq}
V.~Assassi, D.~Baumann and D.~Green, \emph{{On Soft Limits of Inflationary
  Correlation Functions}},
  \href{http://dx.doi.org/10.1088/1475-7516/2012/11/047}{\emph{JCAP} {\bfseries
  11} (2012) 047}, [\href{https://arxiv.org/abs/1204.4207}{{\ttfamily
  1204.4207}}].

\bibitem{Maldacena:2002vr}
J.~M. Maldacena, \emph{{Non-Gaussian features of primordial fluctuations in
  single field inflationary models}},
  \href{http://dx.doi.org/10.1088/1126-6708/2003/05/013}{\emph{JHEP} {\bfseries
  05} (2003) 013}, [\href{https://arxiv.org/abs/astro-ph/0210603}{{\ttfamily
  astro-ph/0210603}}].

\bibitem{Hinterbichler:2012nm}
K.~Hinterbichler, L.~Hui and J.~Khoury, \emph{{Conformal Symmetries of
  Adiabatic Modes in Cosmology}},
  \href{http://dx.doi.org/10.1088/1475-7516/2012/08/017}{\emph{JCAP} {\bfseries
  08} (2012) 017}, [\href{https://arxiv.org/abs/1203.6351}{{\ttfamily
  1203.6351}}].

\bibitem{Creminelli:2012ed}
P.~Creminelli, J.~Nore{\~n}a and M.~Simonovi{\'c}, \emph{{Conformal consistency
  relations for single-field inflation}},
  \href{http://dx.doi.org/10.1088/1475-7516/2012/07/052}{\emph{JCAP} {\bfseries
  07} (2012) 052}, [\href{https://arxiv.org/abs/1203.4595}{{\ttfamily
  1203.4595}}].

\bibitem{Hinterbichler:2013dpa}
K.~Hinterbichler, L.~Hui and J.~Khoury, \emph{{An Infinite Set of Ward
  Identities for Adiabatic Modes in Cosmology}},
  \href{http://dx.doi.org/10.1088/1475-7516/2014/01/039}{\emph{JCAP} {\bfseries
  01} (2014) 039}, [\href{https://arxiv.org/abs/1304.5527}{{\ttfamily
  1304.5527}}].

\bibitem{Berezhiani:2013ewa}
L.~Berezhiani and J.~Khoury, \emph{{Slavnov-Taylor Identities for Primordial
  Perturbations}},
  \href{http://dx.doi.org/10.1088/1475-7516/2014/02/003}{\emph{JCAP} {\bfseries
  02} (2014) 003}, [\href{https://arxiv.org/abs/1309.4461}{{\ttfamily
  1309.4461}}].

\bibitem{Mirbabayi:2014zpa}
M.~Mirbabayi and M.~Zaldarriaga, \emph{{Double Soft Limits of Cosmological
  Correlations}},
  \href{http://dx.doi.org/10.1088/1475-7516/2015/03/025}{\emph{JCAP} {\bfseries
  03} (2015) 025}, [\href{https://arxiv.org/abs/1409.6317}{{\ttfamily
  1409.6317}}].

\bibitem{Avis:2019eav}
G.~Avis, S.~Jazayeri, E.~Pajer and J.~Supe{\l}, \emph{{Spatial Curvature at the
  Sound Horizon}},
  \href{http://dx.doi.org/10.1088/1475-7516/2020/02/034}{\emph{JCAP} {\bfseries
  02} (2020) 034}, [\href{https://arxiv.org/abs/1911.04454}{{\ttfamily
  1911.04454}}].

\bibitem{Jazayeri:2019nbi}
S.~Jazayeri, E.~Pajer and D.~van~der Woude, \emph{{Solid Soft Theorems}},
  \href{http://dx.doi.org/10.1088/1475-7516/2019/06/011}{\emph{JCAP} {\bfseries
  06} (2019) 011}, [\href{https://arxiv.org/abs/1902.09020}{{\ttfamily
  1902.09020}}].

\bibitem{Basile:2024ydc}
T.~Basile, E.~Joung, K.~Mkrtchyan and M.~Mojaza, \emph{{Spinor-helicity
  representations of particles of any mass in dS4 and AdS4 spacetimes}},
  \href{http://dx.doi.org/10.1103/PhysRevD.109.125003}{\emph{Phys. Rev. D}
  {\bfseries 109} (2024) 125003},
  [\href{https://arxiv.org/abs/2401.02007}{{\ttfamily 2401.02007}}].

\bibitem{Maldacena:2011nz}
J.~M. Maldacena and G.~L. Pimentel, \emph{{On graviton non-Gaussianities during
  inflation}}, \href{http://dx.doi.org/10.1007/JHEP09(2011)045}{\emph{JHEP}
  {\bfseries 09} (2011) 045},
  [\href{https://arxiv.org/abs/1104.2846}{{\ttfamily 1104.2846}}].

\bibitem{Nagaraj:2018nxq}
B.~Nagaraj and D.~Ponomarev, \emph{{Spinor-helicity formalism for massless
  fields in AdS$_{4}$ I: higher-spin symmetry and quartic interactions}},
  \href{http://dx.doi.org/10.1103/PhysRevLett.122.101602}{\emph{Phys. Rev.
  Lett.} {\bfseries 122} (2019) 101602},
  [\href{https://arxiv.org/abs/1811.08438}{{\ttfamily 1811.08438}}].

\bibitem{Nagaraj:2019zmk}
B.~Nagaraj and D.~Ponomarev, \emph{{Spinor-helicity formalism for massless
  fields in AdS$_{4}$. Part II. Potentials}},
  \href{http://dx.doi.org/10.1007/JHEP06(2020)068}{\emph{JHEP} {\bfseries 06}
  (2020) 068}, [\href{https://arxiv.org/abs/1912.07494}{{\ttfamily
  1912.07494}}].

\bibitem{Nagaraj:2020sji}
B.~Nagaraj and D.~Ponomarev, \emph{{Spinor-helicity formalism for massless
  fields in AdS$_{4}$ III: contact four-point amplitudes}},
  \href{http://dx.doi.org/10.1007/JHEP08(2020)012}{\emph{JHEP} {\bfseries 08}
  (2020) 012}, [\href{https://arxiv.org/abs/2004.07989}{{\ttfamily
  2004.07989}}].

\bibitem{Buchbinder:2018nkp}
E.~I. Buchbinder, J.~Hutomo and S.~M. Kuzenko, \emph{{Higher spin supercurrents
  in anti-de Sitter space}},
  \href{http://dx.doi.org/10.1007/JHEP09(2018)027}{\emph{JHEP} {\bfseries 09}
  (2018) 027}, [\href{https://arxiv.org/abs/1805.08055}{{\ttfamily
  1805.08055}}].

\bibitem{David:2019mos}
A.~David, N.~Fischer and Y.~Neiman, \emph{{Spinor-helicity variables for
  cosmological horizons in de Sitter space}},
  \href{http://dx.doi.org/10.1103/PhysRevD.100.045005}{\emph{Phys. Rev. D}
  {\bfseries 100} (2019) 045005},
  [\href{https://arxiv.org/abs/1906.01058}{{\ttfamily 1906.01058}}].

\bibitem{Binder:2020raz}
D.~J. Binder, D.~Z. Freedman and S.~S. Pufu, \emph{{A bispinor formalism for
  spinning Witten diagrams}},
  \href{http://dx.doi.org/10.1007/JHEP02(2022)040}{\emph{JHEP} {\bfseries 02}
  (2022) 040}, [\href{https://arxiv.org/abs/2003.07448}{{\ttfamily
  2003.07448}}].

\bibitem{Hitchin:1982gh}
N.~J. Hitchin, \emph{{MONOPOLES AND GEODESICS}},
  \href{http://dx.doi.org/10.1007/BF01208717}{\emph{Commun. Math. Phys.}
  {\bfseries 83} (1982) 579--602}.

\bibitem{Hitchin:1982vry}
N.~J. Hitchin, \emph{{Complex manifolds and Einstein\textquoteright{}s
  equations}}, \href{http://dx.doi.org/10.1007/BFb0066025}{\emph{Lect. Notes
  Math.} {\bfseries 970} (1982) 73--99}.

\bibitem{Jones:1985pla}
P.~Jones and K.~Tod, \emph{{Minitwistor spaces and Einstein-Weyl spaces}},
  \href{http://dx.doi.org/10.1088/0264-9381/2/4/021}{\emph{Class. Quant. Grav.}
  {\bfseries 2} (1985) 565--577}.

\bibitem{Ward:1990vs}
R.~S. Ward and R.~O. Wells, \emph{{Twistor geometry and field theory}}.
\newblock Cambridge Monographs on Mathematical Physics. Cambridge University
  Press, 8, 1991,
  \href{http://dx.doi.org/10.1017/CBO9780511524493}{10.1017/CBO9780511524493}.

\bibitem{Costa:2011mg}
M.~S. Costa, J.~Penedones, D.~Poland and S.~Rychkov, \emph{{Spinning Conformal
  Correlators}}, \href{http://dx.doi.org/10.1007/JHEP11(2011)071}{\emph{JHEP}
  {\bfseries 11} (2011) 071},
  [\href{https://arxiv.org/abs/1107.3554}{{\ttfamily 1107.3554}}].

\bibitem{Caron-Huot:2021kjy}
S.~Caron-Huot and Y.-Z. Li, \emph{{Helicity basis for three-dimensional
  conformal field theory}},
  \href{http://dx.doi.org/10.1007/JHEP06(2021)041}{\emph{JHEP} {\bfseries 06}
  (2021) 041}, [\href{https://arxiv.org/abs/2102.08160}{{\ttfamily
  2102.08160}}].

\bibitem{Kawai:1985xq}
H.~Kawai, D.~C. Lewellen and S.~H.~H. Tye, \emph{{A Relation Between Tree
  Amplitudes of Closed and Open Strings}},
  \href{http://dx.doi.org/10.1016/0550-3213(86)90362-7}{\emph{Nucl. Phys. B}
  {\bfseries 269} (1986) 1--23}.

\bibitem{Bern:2008qj}
Z.~Bern, J.~J.~M. Carrasco and H.~Johansson, \emph{{New Relations for
  Gauge-Theory Amplitudes}},
  \href{http://dx.doi.org/10.1103/PhysRevD.78.085011}{\emph{Phys. Rev. D}
  {\bfseries 78} (2008) 085011},
  [\href{https://arxiv.org/abs/0805.3993}{{\ttfamily 0805.3993}}].

\bibitem{Bern:2010ue}
Z.~Bern, J.~J.~M. Carrasco and H.~Johansson, \emph{{Perturbative Quantum
  Gravity as a Double Copy of Gauge Theory}},
  \href{http://dx.doi.org/10.1103/PhysRevLett.105.061602}{\emph{Phys. Rev.
  Lett.} {\bfseries 105} (2010) 061602},
  [\href{https://arxiv.org/abs/1004.0476}{{\ttfamily 1004.0476}}].

\bibitem{Adamo:2022dcm}
T.~Adamo, J.~J.~M. Carrasco, M.~Carrillo-Gonz\'alez, M.~Chiodaroli, H.~Elvang,
  H.~Johansson et~al., \emph{{Snowmass White Paper: the Double Copy and its
  Applications}},  in \emph{{Snowmass 2021}}, 4, 2022,
  \href{https://arxiv.org/abs/2204.06547}{{\ttfamily 2204.06547}}.

\bibitem{Bern:2022wqg}
Z.~Bern, J.~J. Carrasco, M.~Chiodaroli, H.~Johansson and R.~Roiban, \emph{{The
  SAGEX review on scattering amplitudes Chapter 2: An invitation to
  color-kinematics duality and the double copy}},
  \href{http://dx.doi.org/10.1088/1751-8121/ac93cf}{\emph{J. Phys. A}
  {\bfseries 55} (2022) 443003},
  [\href{https://arxiv.org/abs/2203.13013}{{\ttfamily 2203.13013}}].

\bibitem{carrillogonzalez2018classical}
M.~Carrillo-Gonzalez, R.~Penco and M.~Trodden, \emph{The classical double copy
  in maximally symmetric spacetimes},  2018.

\bibitem{Farnsworth:2023mff}
K.~Farnsworth, M.~L. Graesser and G.~Herczeg, \emph{{Double Kerr-Schild
  spacetimes and the Newman-Penrose map}},
  \href{http://dx.doi.org/10.1007/JHEP10(2023)010}{\emph{JHEP} {\bfseries 10}
  (2023) 010}, [\href{https://arxiv.org/abs/2306.16445}{{\ttfamily
  2306.16445}}].

\bibitem{Liang:2023zxo}
Q.~Liang and S.~Nagy, \emph{{Convolutional double copy in (anti) de Sitter
  space}}, \href{http://dx.doi.org/10.1007/JHEP04(2024)139}{\emph{JHEP}
  {\bfseries 04} (2024) 139},
  [\href{https://arxiv.org/abs/2311.14319}{{\ttfamily 2311.14319}}].

\bibitem{Alkac:2023glx}
G.~Alkac, M.~K. Gumus, O.~Kasikci, M.~A. Olpak and M.~Tek, \emph{{Regularized
  Weyl double copy}},
  \href{http://dx.doi.org/10.1103/PhysRevD.109.084047}{\emph{Phys. Rev. D}
  {\bfseries 109} (2024) 084047},
  [\href{https://arxiv.org/abs/2310.06048}{{\ttfamily 2310.06048}}].

\bibitem{Han:2022mze}
S.~Han, \emph{{The Weyl double copy in vacuum spacetimes with a cosmological
  constant}}, \href{http://dx.doi.org/10.1007/JHEP09(2022)238}{\emph{JHEP}
  {\bfseries 09} (2022) 238},
  [\href{https://arxiv.org/abs/2205.08654}{{\ttfamily 2205.08654}}].

\bibitem{Prabhu:2020avf}
S.~G. Prabhu, \emph{{The classical double copy in curved spacetimes:
  perturbative Yang-Mills from the bi-adjoint scalar}},
  \href{http://dx.doi.org/10.1007/JHEP05(2024)117}{\emph{JHEP} {\bfseries 05}
  (2024) 117}, [\href{https://arxiv.org/abs/2011.06588}{{\ttfamily
  2011.06588}}].

\bibitem{Bahjat-Abbas:2017htu}
N.~Bahjat-Abbas, A.~Luna and C.~D. White, \emph{{The Kerr-Schild double copy in
  curved spacetime}},
  \href{http://dx.doi.org/10.1007/JHEP12(2017)004}{\emph{JHEP} {\bfseries 12}
  (2017) 004}, [\href{https://arxiv.org/abs/1710.01953}{{\ttfamily
  1710.01953}}].

\bibitem{Ilderton:2024oly}
A.~Ilderton and W.~Lindved, \emph{{Toward double copy on arbitrary
  backgrounds}},  \href{https://arxiv.org/abs/2405.10016}{{\ttfamily
  2405.10016}}.

\bibitem{Garcia-Compean:2024uie}
H.~Garc\'\i{}a-Compe\'an and C.~Ramos, \emph{{Classical Kerr-Schild double copy
  in bigravity for maximally symmetric spacetimes}},
  \href{http://dx.doi.org/10.1007/JHEP07(2024)074}{\emph{JHEP} {\bfseries 07}
  (2024) 074}, [\href{https://arxiv.org/abs/2403.19608}{{\ttfamily
  2403.19608}}].

\bibitem{Chacon:2024qsq}
E.~Chac\'on, H.~Garc\'\i{}a-Compe\'an and G.~Robles, \emph{{Hyper-Hermitian
  Weyl Double Copy}},  \href{https://arxiv.org/abs/2410.21610}{{\ttfamily
  2410.21610}}.

\bibitem{Chawla:2022ogv}
S.~Chawla and C.~Keeler, \emph{{Aligned fields double copy to Kerr-NUT-(A)dS}},
  \href{http://dx.doi.org/10.1007/JHEP04(2023)005}{\emph{JHEP} {\bfseries 04}
  (2023) 005}, [\href{https://arxiv.org/abs/2209.09275}{{\ttfamily
  2209.09275}}].

\bibitem{White:2020sfn}
C.~D. White, \emph{{Twistorial Foundation for the Classical Double Copy}},
  \href{http://dx.doi.org/10.1103/PhysRevLett.126.061602}{\emph{Phys. Rev.
  Lett.} {\bfseries 126} (2021) 061602},
  [\href{https://arxiv.org/abs/2012.02479}{{\ttfamily 2012.02479}}].

\bibitem{Chacon:2021wbr}
E.~Chac\'on, S.~Nagy and C.~D. White, \emph{{The Weyl double copy from twistor
  space}}, \href{http://dx.doi.org/10.1007/JHEP05(2021)239}{\emph{JHEP}
  {\bfseries 05} (2021) 2239},
  [\href{https://arxiv.org/abs/2103.16441}{{\ttfamily 2103.16441}}].

\bibitem{Chacon:2021hfe}
E.~Chac{\'o}n, A.~Luna and C.~D. White, \emph{{Double copy of the multipole
  expansion}}, \href{http://dx.doi.org/10.1103/PhysRevD.106.086020}{\emph{Phys.
  Rev. D} {\bfseries 106} (2022) 086020},
  [\href{https://arxiv.org/abs/2108.07702}{{\ttfamily 2108.07702}}].

\bibitem{Chacon:2021lox}
E.~Chac\'on, S.~Nagy and C.~D. White, \emph{{Alternative formulations of the
  twistor double copy}},
  \href{http://dx.doi.org/10.1007/JHEP03(2022)180}{\emph{JHEP} {\bfseries 03}
  (2022) 180}, [\href{https://arxiv.org/abs/2112.06764}{{\ttfamily
  2112.06764}}].

\bibitem{Luna:2022dxo}
A.~Luna, N.~Moynihan and C.~D. White, \emph{{Why is the Weyl double copy local
  in position space?}},
  \href{http://dx.doi.org/10.1007/JHEP12(2022)046}{\emph{JHEP} {\bfseries 12}
  (2022) 046}, [\href{https://arxiv.org/abs/2208.08548}{{\ttfamily
  2208.08548}}].

\bibitem{Adamo:2021dfg}
T.~Adamo and U.~Kol, \emph{{Classical double copy at null infinity}},
  \href{http://dx.doi.org/10.1088/1361-6382/ac635e}{\emph{Class. Quant. Grav.}
  {\bfseries 39} (2022) 105007},
  [\href{https://arxiv.org/abs/2109.07832}{{\ttfamily 2109.07832}}].

\bibitem{Armstrong-Williams:2023ssz}
K.~Armstrong-Williams and C.~D. White, \emph{{A spinorial double copy for $
  \mathcal{N} $ = 0 supergravity}},
  \href{http://dx.doi.org/10.1007/JHEP05(2023)047}{\emph{JHEP} {\bfseries 05}
  (2023) 047}, [\href{https://arxiv.org/abs/2303.04631}{{\ttfamily
  2303.04631}}].

\bibitem{Guevara:2021yud}
A.~Guevara, \emph{{Reconstructing Classical Spacetimes from the S-Matrix in
  Twistor Space}},  \href{https://arxiv.org/abs/2112.05111}{{\ttfamily
  2112.05111}}.

\bibitem{CarrilloGonzalez:2022ggn}
M.~Carrillo~Gonz\'alez, W.~T. Emond, N.~Moynihan, J.~Rumbutis and C.~D. White,
  \emph{{Mini-twistors and the Cotton double copy}},
  \href{http://dx.doi.org/10.1007/JHEP03(2023)177}{\emph{JHEP} {\bfseries 03}
  (2023) 177}, [\href{https://arxiv.org/abs/2212.04783}{{\ttfamily
  2212.04783}}].

\bibitem{Beetar:2024ptv}
C.~Beetar, M.~Carrillo~Gonz\'alez, S.~Jaitly and T.~Keseman, \emph{{Double copy
  in AdS$_{3}$ from minitwistor space}},
  \href{http://dx.doi.org/10.1007/JHEP03(2025)125}{\emph{JHEP} {\bfseries 03}
  (2025) 125}, [\href{https://arxiv.org/abs/2410.23342}{{\ttfamily
  2410.23342}}].

\bibitem{Bu:2023cef}
W.~Bu and S.~Seet, \emph{{Celestial holography and AdS$_{3}$/CFT$_{2}$ from a
  scaling reduction of twistor space}},
  \href{http://dx.doi.org/10.1007/JHEP12(2023)168}{\emph{JHEP} {\bfseries 12}
  (2023) 168}, [\href{https://arxiv.org/abs/2306.11850}{{\ttfamily
  2306.11850}}].

\bibitem{Bu:2024cql}
W.~Bu and S.~Seet, \emph{{A systematic approach to celestial holography: a case
  study in Einstein gravity}},
  \href{https://arxiv.org/abs/2404.04637}{{\ttfamily 2404.04637}}.

\bibitem{Bu:2023vjt}
W.~Bu and S.~Seet, \emph{{A hidden 2d CFT for self-dual Yang-Mills on the
  celestial sphere}},
  \href{http://dx.doi.org/10.1007/JHEP08(2024)022}{\emph{JHEP} {\bfseries 08}
  (2024) 022}, [\href{https://arxiv.org/abs/2310.17457}{{\ttfamily
  2310.17457}}].

\bibitem{Seet:2024vmh}
S.~Seet, \emph{{Twistor Space and Celestial Holography}}, Ph.D. thesis,
  Cambridge U., DAMTP, 2024.
\newblock 10.17863/CAM.112793.

\bibitem{Adamo:2016rtr}
T.~Adamo, D.~Skinner and J.~Williams, \emph{{Twistor methods for AdS$_{5}$}},
  \href{http://dx.doi.org/10.1007/JHEP08(2016)167}{\emph{JHEP} {\bfseries 08}
  (2016) 167}, [\href{https://arxiv.org/abs/1607.03763}{{\ttfamily
  1607.03763}}].

\bibitem{Bala:2025qxr}
A.~Bala and D.~K. S, \emph{{An Ode to the Penrose and Witten transforms in
  Twistor space for 3D CFT}},
  \href{https://arxiv.org/abs/2505.14082}{{\ttfamily 2505.14082}}.

\bibitem{Bala:2025gmz}
A.~Bala, S.~Jain, D.~K. S., D.~Mazumdar and V.~Singh, \emph{{3D Conformal Field
  Theory in Twistor Space}},
  \href{https://arxiv.org/abs/2502.18562}{{\ttfamily 2502.18562}}.

\bibitem{Cheung:2009dc}
C.~Cheung and D.~O'Connell, \emph{{Amplitudes and Spinor-Helicity in Six
  Dimensions}},
  \href{http://dx.doi.org/10.1088/1126-6708/2009/07/075}{\emph{JHEP} {\bfseries
  07} (2009) 075}, [\href{https://arxiv.org/abs/0902.0981}{{\ttfamily
  0902.0981}}].

\bibitem{Simmons-Duffin:2012juh}
D.~Simmons-Duffin, \emph{{Projectors, Shadows, and Conformal Blocks}},
  \href{http://dx.doi.org/10.1007/JHEP04(2014)146}{\emph{JHEP} {\bfseries 04}
  (2014) 146}, [\href{https://arxiv.org/abs/1204.3894}{{\ttfamily 1204.3894}}].

\bibitem{Penrose:1986ca}
R.~Penrose and W.~Rindler, \emph{{SPINORS AND SPACE-TIME. VOL. 2: SPINOR AND
  TWISTOR METHODS IN SPACE-TIME GEOMETRY}}.
\newblock Cambridge Monographs on Mathematical Physics. Cambridge University
  Press, 4, 1988,
  \href{http://dx.doi.org/10.1017/CBO9780511524486}{10.1017/CBO9780511524486}.

\bibitem{Woodhouse:1985id}
N.~M.~J. Woodhouse, \emph{{REAL METHODS IN TWISTOR THEORY}},
  \href{http://dx.doi.org/10.1088/0264-9381/2/3/006}{\emph{Class. Quant. Grav.}
  {\bfseries 2} (1985) 257--291}.

\bibitem{Costa:2014kfa}
M.~S. Costa, V.~Gon\c{c}alves and J.~a. Penedones, \emph{{Spinning AdS
  Propagators}}, \href{http://dx.doi.org/10.1007/JHEP09(2014)064}{\emph{JHEP}
  {\bfseries 09} (2014) 064},
  [\href{https://arxiv.org/abs/1404.5625}{{\ttfamily 1404.5625}}].

\bibitem{Bena:1999py}
I.~Bena, \emph{{The Antisymmetric tensor propagator in AdS}},
  \href{http://dx.doi.org/10.1103/PhysRevD.62.127901}{\emph{Phys. Rev. D}
  {\bfseries 62} (2000) 127901},
  [\href{https://arxiv.org/abs/hep-th/9910059}{{\ttfamily hep-th/9910059}}].

\bibitem{Bena:1999be}
I.~Bena, \emph{{The Propagator for a general form field in AdS(d+1)}},
  \href{http://dx.doi.org/10.1103/PhysRevD.62.126008}{\emph{Phys. Rev. D}
  {\bfseries 62} (2000) 126008},
  [\href{https://arxiv.org/abs/hep-th/9911073}{{\ttfamily hep-th/9911073}}].

\bibitem{Anguelova:2003kf}
L.~Anguelova and P.~Langfelder, \emph{{Massive gravitino propagator in
  maximally symmetric spaces and fermions in dS / CFT}},
  \href{http://dx.doi.org/10.1088/1126-6708/2003/03/057}{\emph{JHEP} {\bfseries
  03} (2003) 057}, [\href{https://arxiv.org/abs/hep-th/0302087}{{\ttfamily
  hep-th/0302087}}].

\bibitem{Leonhardt:2003qu}
T.~Leonhardt, R.~Manvelyan and W.~Ruhl, \emph{{The Group approach to AdS space
  propagators}},
  \href{http://dx.doi.org/10.1016/j.nuclphysb.2003.07.007}{\emph{Nucl. Phys. B}
  {\bfseries 667} (2003) 413--434},
  [\href{https://arxiv.org/abs/hep-th/0305235}{{\ttfamily hep-th/0305235}}].

\bibitem{Basu:2006ti}
A.~Basu and L.~I. Uruchurtu, \emph{{Gravitino propagator in anti de Sitter
  space}}, \href{http://dx.doi.org/10.1088/0264-9381/23/20/023}{\emph{Class.
  Quant. Grav.} {\bfseries 23} (2006) 6059--6076},
  [\href{https://arxiv.org/abs/hep-th/0603089}{{\ttfamily hep-th/0603089}}].

\bibitem{Faizal:2011sa}
M.~Faizal, \emph{{Covariant Graviton Propagator in Anti-de Sitter Spacetime}},
  \href{http://dx.doi.org/10.1088/0264-9381/29/3/035007}{\emph{Class. Quant.
  Grav.} {\bfseries 29} (2012) 035007},
  [\href{https://arxiv.org/abs/1112.4369}{{\ttfamily 1112.4369}}].

\bibitem{Bailey:1985}
T.~N. Bailey, \emph{Twistors and fields with sources on worldlines},
  \href{http://dx.doi.org/10.1098/rspa.1985.0008}{\emph{Proceedings of the
  Royal Society of London A} {\bfseries 397} (1985) 143--155}.

\bibitem{Baumann:2020dch}
D.~Baumann, C.~Duaso~Pueyo, A.~Joyce, H.~Lee and G.~L. Pimentel, \emph{{The
  Cosmological Bootstrap: Spinning Correlators from Symmetries and
  Factorization}},
  \href{http://dx.doi.org/10.21468/SciPostPhys.11.3.071}{\emph{SciPost Phys.}
  {\bfseries 11} (2021) 071},
  [\href{https://arxiv.org/abs/2005.04234}{{\ttfamily 2005.04234}}].

\bibitem{Farrow:2018yni}
J.~A. Farrow, A.~E. Lipstein and P.~McFadden, \emph{{Double copy structure of
  CFT correlators}},
  \href{http://dx.doi.org/10.1007/JHEP02(2019)130}{\emph{JHEP} {\bfseries 02}
  (2019) 130}, [\href{https://arxiv.org/abs/1812.11129}{{\ttfamily
  1812.11129}}].

\bibitem{Lipstein:2019mpu}
A.~E. Lipstein and P.~McFadden, \emph{{Double copy structure and the flat space
  limit of conformal correlators in even dimensions}},
  \href{http://dx.doi.org/10.1103/PhysRevD.101.125006}{\emph{Phys. Rev. D}
  {\bfseries 101} (2020) 125006},
  [\href{https://arxiv.org/abs/1912.10046}{{\ttfamily 1912.10046}}].

\bibitem{Bzowski:2017poo}
A.~Bzowski, P.~McFadden and K.~Skenderis, \emph{{Renormalised 3-point functions
  of stress tensors and conserved currents in CFT}},
  \href{http://dx.doi.org/10.1007/JHEP11(2018)153}{\emph{JHEP} {\bfseries 11}
  (2018) 153}, [\href{https://arxiv.org/abs/1711.09105}{{\ttfamily
  1711.09105}}].

\bibitem{Albayrak:2020fyp}
S.~Albayrak, S.~Kharel and D.~Meltzer, \emph{{On duality of color and
  kinematics in (A)dS momentum space}},
  \href{http://dx.doi.org/10.1007/JHEP03(2021)249}{\emph{JHEP} {\bfseries 03}
  (2021) 249}, [\href{https://arxiv.org/abs/2012.10460}{{\ttfamily
  2012.10460}}].

\bibitem{Jain:2021qcl}
S.~Jain, R.~R. John, A.~Mehta, A.~A. Nizami and A.~Suresh, \emph{{Double copy
  structure of parity-violating CFT correlators}},
  \href{http://dx.doi.org/10.1007/JHEP07(2021)033}{\emph{JHEP} {\bfseries 07}
  (2021) 033}, [\href{https://arxiv.org/abs/2104.12803}{{\ttfamily
  2104.12803}}].

\bibitem{Alday:2021odx}
L.~F. Alday, C.~Behan, P.~Ferrero and X.~Zhou, \emph{{Gluon Scattering in AdS
  from CFT}}, \href{http://dx.doi.org/10.1007/JHEP06(2021)020}{\emph{JHEP}
  {\bfseries 06} (2021) 020},
  [\href{https://arxiv.org/abs/2103.15830}{{\ttfamily 2103.15830}}].

\bibitem{Lee:2022fgr}
H.~Lee and X.~Wang, \emph{{Cosmological double-copy relations}},
  \href{http://dx.doi.org/10.1103/PhysRevD.108.L061702}{\emph{Phys. Rev. D}
  {\bfseries 108} (2023) L061702},
  [\href{https://arxiv.org/abs/2212.11282}{{\ttfamily 2212.11282}}].

\bibitem{White:2024pve}
C.~D. White, \emph{{The Classical Double Copy}}.
\newblock World Scientific, 5, 2024,
  \href{http://dx.doi.org/10.1142/q0457}{10.1142/q0457}.

\bibitem{Bailey:1998zif}
T.~N. Bailey and E.~G. Dunne, \emph{{A twistor correspondence and Penrose
  transform for odd-dimensional hyperbolic space}},
  \href{http://dx.doi.org/10.1090/S0002-9939-98-04215-4}{\emph{Proc. Am. Math.
  Soc.} {\bfseries 126} (1998) 1245--1253}.

\bibitem{Adamo:2024hme}
T.~Adamo and S.~Klisch, \emph{{The KLT Kernel in Twistor Space}},
  \href{http://dx.doi.org/10.1007/s00220-025-05254-0}{\emph{Commun. Math.
  Phys.} {\bfseries 406} (2025) 79},
  [\href{https://arxiv.org/abs/2406.04539}{{\ttfamily 2406.04539}}].

\bibitem{Adamo:2015ina}
T.~Adamo, \emph{{Gravity with a cosmological constant from rational curves}},
  \href{http://dx.doi.org/10.1007/JHEP11(2015)098}{\emph{JHEP} {\bfseries 11}
  (2015) 098}, [\href{https://arxiv.org/abs/1508.02554}{{\ttfamily
  1508.02554}}].

\bibitem{Boels:2006ir}
R.~Boels, L.~J. Mason and D.~Skinner, \emph{{Supersymmetric Gauge Theories in
  Twistor Space}},
  \href{http://dx.doi.org/10.1088/1126-6708/2007/02/014}{\emph{JHEP} {\bfseries
  02} (2007) 014}, [\href{https://arxiv.org/abs/hep-th/0604040}{{\ttfamily
  hep-th/0604040}}].

\end{thebibliography}\endgroup
\bibliographystyle{JHEP}

\end{document}